\documentclass[fleqn,usenatbib]{mnras}

\usepackage{newtxtext,newtxmath}

\usepackage[T1]{fontenc}

\DeclareRobustCommand{\VAN}[3]{#2}\let\VANthebibliography\thebibliography\def\thebibliography{\DeclareRobustCommand{\VAN}[3]{##3}\VANthebibliography}

\usepackage{graphicx}	  
\usepackage{enumitem}     
\usepackage{subcaption}   
\usepackage{booktabs}     
\usepackage{amsmath}	  
\usepackage{xcolor}       
\usepackage{fontawesome5} 
\usepackage{orcidlink}    

\newcommand\Msol{\text{M}_{\odot}}
\newcommand\HI{\ion{H}{i}}
\newcommand\Hmolecular{\text{H}$_{2}$}
\newcommand{\Code}[1]{\texttt{#1}}
\newcommand{\siml}[1]{\textsc{#1}}
\newcommand\tng{\siml{IllustrisTNG}}
\newcommand{\telesc}[1]{\textsc{#1}}

\newcommand\dm{\textsc{dm}}
\newcommand\stars{\textsc{stars}}
\newcommand\gas{\textsc{gas}}
\newcommand\hi{\textsc{hi}}
\newcommand\hicm{\textsc{21cm}}
\newcommand\temp{\textsc{temp}}
\newcommand\bfield{\textsc{bfield}}

\makeatletter
\renewcommand{\theparagraph}{}
\makeatother

\definecolor{todocolor}{HTML}{E83A82}

\title[Galactic Alchemy]{Galactic Alchemy:\ Deep Learning Map-to-Map Translation in Hydrodynamical Simulations}

\author[P. Denzel et al.]{
  Philipp Denzel$^{\orcidlink{0000-0003-0126-0659},1}$\thanks{E-mail: philipp.denzel@zhaw.ch},
  Yann Billeter$^{\orcidlink{0009-0003-2783-9252},2}$,
  Frank-Peter Schilling$^{\orcidlink{0000-0002-7763-2140},1}$,
  and Elena Gavagnin$^{\orcidlink{0000-0001-5146-3759},1,3}$
\\
$^{1}$Centre for Artificial Intelligence, Zurich University of Applied Sciences ZHAW, Technikumstrasse 71, Winterthur 8400, Switzerland\\
$^{2}$Institute of Science Technology and Policy, ETH Zurich, Universitätstrasse 41, Zurich 8092, Switzerland\\
$^{3}$Institute of Business Information Technology, Zurich University of Applied Sciences ZHAW, Theaterstrasse 17, Winterthur 8400, Switzerland\\
}

\date{Accepted XXX.\ Received YYY;\ in original form ZZZ}

\pubyear{2025}

\begin{document}\label{firstpage}
\pagerange{\pageref{firstpage}--\pageref{lastpage}}
\maketitle

\begin{abstract}
  We present the first systematic study of multi-domain map-to-map translation in galaxy formation simulations, leveraging deep generative models to predict diverse galactic properties.
  Using high-resolution magneto-hydrodynamical simulation data, we compare conditional generative adversarial networks and diffusion models under unified preprocessing and evaluation, optimizing architectures and attention mechanisms for physical fidelity on galactic scales.
  Our approach jointly addresses seven astrophysical domains -- including dark matter, gas, neutral hydrogen, stellar mass, temperature, and magnetic field strength -- while introducing physics-aware evaluation metrics that quantify structural realism beyond standard computer vision measures.
  We demonstrate that translation difficulty correlates with physical coupling, achieving near-perfect fidelity for mappings from gas to dark matter and mappings involving astro-chemical components such as total gas to \ion{H}{i} content, while identifying fundamental challenges in weakly constrained tasks such as gas to stellar mass mappings.
  Our results establish GAN-based models as competitive counterparts to state-of-the-art diffusion approaches at a fraction of the computational cost (in training and inference), paving the way for scalable, physics-aware generative frameworks for forward modelling and observational reconstruction in the SKA era.
\end{abstract}

\begin{keywords}
 hydrodynamics --\ %
 galaxies: structure --\ %
 dark matter --\ %
 galaxies: stellar content --\ %
 software: machine learning
\end{keywords}



\section{Introduction}\label{sec:intro}

The spatial matter distribution of galaxies is the result of a complex and chaotic interaction between its individual components, such as dark matter (DM), stellar populations, (predominantly hydrogen) gas, super-massive black holes (SMBHs), and their surrounding environment. Interactions between these components are governed by their mutual gravitational and electromagnetic forces, and hydrodynamical processes which collectively shape the structure and properties of galaxies over cosmic time \citep{binney_tremaine_2011_galactic_dynamics, tinsley_2022_evolution_stars, conselice_2014_evolution_galaxy, donofrio_2016_physics_galaxy}. These interactions imprint subtle signatures in the phase-space distribution of a galaxy, retaining the various feedback mechanisms that have influenced its formation and evolution \citep{binney_2023_self_consistent_models, bassini_2024_inflow_outflow}. Thus, the physical components encode distinct aspects of galaxy evolution:

\begin{itemize}[leftmargin=*,topsep=1pt]
\item \textbf{DM haloes} of galaxies dominate the gravitational potential into which baryonic matter flows and forms visible substructure~\citep{white_1991_galaxy_formation, moore_1999_cold_collapse, frenk_2012_dark_matter}.

\item \textbf{Stellar mass} reflects the cumulative outcome of star formation and feedback, but its distribution is temporally highly non-local and entropic \citep{mckee_2007_theory_star, kennicutt_2012_star_formation, hwang_2019_evolution_star}, while star formation is spatially localized in \Hmolecular{} clouds within the interstellar medium \citep[ISM;][]{colman_2024_cloud_properties, schinnerer_2024_molecular_gas}.

\item \textbf{Gas} traces the baryonic backbone of galaxies, regulating cooling, heating, and star formation through radiative feedback cycles \citep{gavagnin_2017_star_cluster, luisi_2021_stellar_feedback} and turbulence induced by active galactic nuclei \citep[AGNs;][]{biernacki_2018_combined_effect, valentini_2019_impact_agn}, supernovae \citep{fielding_2017_how_supernovae, ibrahim_2023_impact_supernova}, stellar winds \citep{krumholz_2014_star_cluster, bally_2016_protostellar_outflows}, galaxy-galaxy mergers \citep{hopkins_2006_unified_merger_driven, cibinel_2019_early_late_stage}, or interaction with the intergalactic medium \citep[IGM;][]{muratov_2017_metal_flows, poggianti_2019_gasp_xxiii}.

\item \textbf{Neutral hydrogen} and \textbf{21cm brightness} are key observational tracers of the ISM for low-redshift galaxies, critical for radio surveys with \telesc{MeerKAT}, \telesc{ASKAP}, or the upcoming \telesc{SKA-Mid} \citep[such as WALLABY, MIGHTEE-\HI{}, MHONGOOSE, or the MeerKAT Fornax Survey;][]{maccagni_2024_neutral_atomic, obeirne_2025_wallaby_pilot_survey, maddox_2021_mightee_hi, blok_2024_mhongoose, maccagni_2025_meerkat_fornax_survey}.

\item \textbf{Temperature} captures thermodynamic states shaped by shocks, cooling, and AGN-driven outflows \citep{zubovas_2024_complex_effect, ward_2024_agn_driven}.

\item \textbf{Magnetic fields} emerge from turbulent amplification and influence gas dynamics \citep[e.g.,][]{beck_2015_magnetic_fields, rieder_2017_small_scale_dynamo}, yet remain poorly constrained observationally.
\end{itemize}

Recovering these domains from limited information is challenging; observationally due to technical limitations: for most instruments, signals beyond the local Universe -- especially from HI -- become too faint due to intrinsic dimming \citep{messias_2024_h_content}, foreground contamination \citep{meerklass_2025_l_band_deep_field}, and low-frequency RFI \citep{harper_2018_potential_impact, engelbrecht_2024_radio_frequency}.

Conversely, theoretical inference of these domains has computational challenges:
the understanding of the distribution of matter in the Universe remains largely driven by numerical simulations. Among these, (magneto-)hydrodynamical simulations present the most principled approach to model and capture the non-linear co-evolution of dark and baryonic matter fields across cosmological and astrophysical scales \citep[for recent reviews, see][]{crain_2023_hydrodynamical_simulations}. However, this quality comes at steep computational costs or forces detrimental trade-offs between resolution and volume.

To mitigate these challenges, simpler alternatives such as dark-matter-only (DMO) simulations \citep[e.g.,][]{potter_2017_pkdgrav3, ishiyama_2021_uchuu_simulations, cheng_2020_cube} reproduce large-scale structure and halo statistics at reduced cost but omit baryonic physics. Semi-analytical models (SAMs) attempt to compensate by applying post de facto prescriptions to approximate baryonic effects on top of DMO outputs \citep[e.g., ][]{berlind_2003_halo_occupation, somerville_2008_semi_analytic_model, schneider_2019_quantifying_baryon, obuljen_2023_modeling_hi}.
While these methods enable exploration of cosmological parameter space, they lack the fidelity needed to capture the full complexity of galaxy-scale feedback and morphology. In particular, their intrinsic post-hoc nature often ignores the gravitational back-reaction caused by the redistribution of baryons on the DM field.

With the proliferation of deep generative models, a complementary line of research has emerged that seeks to emulate aspects of these simulations rather than compute them from first principles. Recent efforts have explored enhancing simulations and augmenting galaxy models through scalable deep learning techniques in various ways. For instance, \cite{perraudin_2019_cosmological_n_body} use scalable GANs (for details see Section~\ref{sec:gans}) to produce entire N-body 3D cubes of the cosmic DM distribution in a multi-scale approach. Still, techniques aiming for the full 3D reconstruction of cosmological simulations often face challenges in scaling to resolutions where individual galaxies can be resolved. Alternatively, \cite{bernardini_2021_from_ember} employ Wasserstein-GANs to paint baryons onto thin slices of simulation boxes from the \siml{FIRE} simulation suite. \cite{li_2021_ai_assisted_superresolution, schanz_2024_stochastic_super_resolution} use StyleGAN and denoising diffusion models, respectively, to super-resolve cosmic large-scale structure predictions. \cite{thiele_2020_teaching_neural} applied a U-Net architecture (for details see Section~\ref{sec:architectures}) to infer observable thermal and kinematic Sunyaev-Zel'dovich maps of haloes from DMO simulations, explicitly linking theory to observations. Similarly, \cite{chadayammuri_2023_painting_baryons} use a U-Net for image-to-image translation of \tng{} galaxy cluster haloes to the corresponding baryonic fields.

Most studies focus on a single aspect of a simulation's galaxy formation or feedback model and do not fully reproduce (or harness) all physical modes of simulated galaxies (for details see Section~\ref{sec:methodology}, Equation~\ref{eq:galaxy_population}).

In this paper, we introduce a novel application of deep generative models for map-to-map translation across multiple astrophysical domains in cosmological simulations on the galaxy-scale level. In contrast to other works, we propose a more comprehensive representation of a formation scenario by fitting various permutations of galaxy properties, without explicit heuristics or phenomenological tuning. Using high-resolution magneto-hydrodynamical simulation data from the \tng{} suite \citep[\textsc{TNG50--1};][]{springel_2017_first_results, nelson_2017_first_results, pillepich_2017_first_results, marinacci_2018_first_results, naiman_2018_first_results}, we systematically compare conditional generative adversarial networks (GANs) and diffusion models under unified preprocessing and evaluation. Our approach goes beyond prior work by jointly addressing multiple domains and introducing physics-aware metrics -- such as asymmetry, clumpiness, concentration, and power spectra  -- that assess structural realism and astrophysical fidelity beyond standard computer vision measures. We show that GAN-based models can achieve performance comparable to diffusion models at a fraction of the computational cost (in training and inference), in particular for map-to-map translations involving astro-chemical components. Moreover, a set of deep generative models including all domain translations provides a comprehensive representation of a galaxy's formation scenario (for details see Section~\ref{sec:methodology} and Equation~\ref{eq:galaxy_population}). Finally, the generative models establish a bridge between theory and observation by incorporating domains that are directly observable, such as 21-cm brightness, into the translation process. This is particularly relevant for upcoming large-scale radio surveys with the Square Kilometre Array \citep[\telesc{SKA};][]{braun_2015_advancing_astrophysics, staveley-smith_2015_hi_science} telescopes, which will probe the cosmic distribution of \HI{} through 21cm emission. By enabling the reconstruction of astrophysical quantities from observational proxies and forward modelling of instrument-specific effects, our approach provides a scalable pathway to interpret SKA data within the context of galaxy formation scenarios.

The remainder of this paper is structured as follows: Section~\ref{sec:methodology} details our methodology, models, evaluation metrics, and data, Section~\ref{sec:results} presents the results, and Section~\ref{sec:conclusion} discusses implications and future directions.

\section{Data \& Methodology}\label{sec:methodology}

Our work aims to address the limitations identified above by leveraging high-resolution simulation data as the foundation for a generative modelling approach. To this end, we require a dataset that captures the full complexity of baryonic and DM interactions (i.e. magneto-hydrodynamics) at galaxy scales. In the following Section~\ref{sec:tng}, we detail the selection criteria and preprocessing steps applied to construct our dataset of galaxy maps.

\subsection{Dataset}\label{sec:tng}

The \tng{} project is a series of publicly released, cosmological magneto-hydrodynamical simulations of galaxy formation, run with the \Code{AREPO}~\citep{weinberger_2020_arepo_public} moving-mesh code~\citep{springel_2017_first_results, nelson_2017_first_results, pillepich_2017_first_results, marinacci_2018_first_results, naiman_2018_first_results}. Each simulation self-consistently solves the coupled evolution of DM, cosmic gas, luminous stars, and SMBHs.
The \siml{TNG50--1} simulation was run with a total of $2\times2160^{3}$ resolution elements, a DM mass resolution of $3.1\times10^{5}\,\Msol/h$, and a baryon mass resolution of $5.7\times10^{4}\,\Msol/h$, providing a rare combination of large volume and fine resolution in a simulation released to the public. Galaxies were selected from snapshots between $z=1$ and $z=0$ with a required minimum number of resolution elements of $10^{4}$, to ensure sufficient resolution even for larger satellite and dwarf galaxies.

Projections onto images for each selected galaxy were performed in multiple domains (galaxy properties $\zeta$):
\begin{itemize}[leftmargin=*,topsep=1pt]
\item DM mass (\dm{}; column density)
\item stellar mass (\stars{}; column density)
\item total gas mass (\gas{}; column density)
\item \HI{} gas mass (\hi{}; column density)
\item (mock) 21-cm brightness temperature (\hicm{})
\item gas temperature (\temp{})
\item magnetic field strength (\bfield{})
\end{itemize}

The projections extend to two half-mass radii of a galaxy's total gas mass, ensuring each domain image has the same spatial resolution for a given galaxy.  All but the 21-cm brightness temperature maps are directly simulated quantities; the former were generated following~\cite{villaescusa-navarro_2018_ingredients_for}.  The map projections were performed using an adapted version \Code{Pylians3} code~\citep{pylians_2018}.  The resulting dataset of multiple domains counts 504,000 512$\times$512 images in total (72,000 images per domain), produced from roughly 3000 galaxies per snapshot (6 in total), each galaxy randomly rotated (on all axes) four different ways before projection for data augmentation. Note that the first iteration of the dataset contained fewer samples with a slightly higher average total halo mass; this dataset was used where explicitly stated in Section~\ref{sec:experiments}~and~\ref{sec:results}.

In summary, the dataset contains a set of galaxy projections in multiple domains (different physical modes of a galaxy) which jointly approximate the fiducial \siml{TNG} model, i.e.\ \tng{}'s formation scenario.

Deep learning networks often work best on non-peaked data distributions, numerically standardized in intervals between $[0, 1]$ (for uniform priors) or $[-1, 1]$ (for Gaussian priors). Inspired by common transformation used in the high-energy physics domain \citep[see e.g.][]{finke_2021_autoencoders}, we use the following scaling for all maps
\begin{equation}\label{eq:hep_scaling}
  \tilde{x} = (b+1) \cdot \left( \frac{x}{c} \right)^{\frac{1}{\gamma}} - b
\end{equation}
where $c\neq0$ is the normalization constant (around the maximum of the data distribution), $\gamma \sim \mathcal{U}\{0, \mathcal{O}(10)\}$ the power scaling, and the Boolean parameter $b\in\{0, 1\}$ depending on whether the interval should map to $[0, 1]$ or $[-1, 1]$. The exact values for the $\gamma$ parameters were found via grid search (such that the median of the dataset distribution is $\ge0.3$ and $\le0.6$) per domain as listed in Table~\ref{tab:preprocessing}; $b$ was 0 for all models with a uniform prior (GANs) and 1 for all models with a Gaussian prior (diffusion models). This transformation normalizes the data ranges and stabilizes the variances in the data, making them more Gaussian-like.

\begin{table}
  \centering
  \caption{The preprocessing transformation parameters: $c$ is the normalization constant (in the respective units of the corresponding maps) and $\gamma$ the power scaling. The Boolean $b$ deciding whether the transformation maps to a symmetric or non-negative interval was always 1 for diffusion models and 0 for GANs.}\label{tab:preprocessing}
  \begin{tabular}{rccccccc}
  \toprule
  & \dm{} & \stars{} & \gas{} & \hi{} & \hicm{} & \temp{} & \bfield{} \\
  \midrule
  c & $2\times10^{10}$ & $8\times10^{11}$ & $10^{10}$ & $10^{8}$ & $165$ & $10^{8}$ & $10^{-1}$ \\
  $\gamma$ & 8 & 16 & 8 & 8 & 8 & 8 & 8 \\
  \bottomrule
  \end{tabular}
\end{table}

\subsection{Galaxy formation scenario}

Capturing the complex interplay between baryonic components and DM distributions at the galaxy scale is computationally the most expensive task in any numerical simulation.
Often, a trade-off between simulation size and resolution is required to make a hydrodynamical treatment even feasible.
Additionally, surrogate techniques, so-called sub-grid models, are employed to capture effects of baryonic components below the resolution limit.
The ill-constrained parameters of such sub-grid models are calibrated to match observed properties at the simulated scales, leading to degeneracy and difficulties in the interpretation of outcomes \citep[cf.][]{crain_2023_hydrodynamical_simulations}.

For this reason, different simulation suites produce similarly realistic galaxies with a wide variety of formation ``recipes''.
Notable, publicly available (and thus for this work relevant) examples of such suites are the \siml{EAGLE} \citep{schaye_2014_eagle_project, crain_2015_eagle_simulations, mcalpine_2016_eagle_simulations}, \siml{Horizon-AGN} \citep{dubois_2014_dancing_in}, \tng{} \citep{springel_2017_first_results, nelson_2017_first_results, pillepich_2017_first_results, marinacci_2018_first_results, naiman_2018_first_results}, and \siml{SIMBA} \citep{dave_2019_simba} suites.

The summary of all these physical effects characterizing a simulated population of galaxies, we will abstractly describe as a galaxy \emph{formation scenario} $\Phi$.
In Bayesian terms, a simulation describes galaxy samples from a \emph{population} $\Gamma_i$ by the marginalization

\begin{equation}\label{eq:galaxy_population}
  P(\Gamma | \Phi) = \sum_{\zeta\in\Omega}{P(\Gamma | \zeta) P(\zeta | \Phi)}
\end{equation}

where $\zeta \in \Omega$ represents a \emph{galaxy property} from the set of galaxy characteristics $\Omega$.
There will also be \emph{nuisance parameters} $\nu$ which lead to the expression of a galaxy distribution but are not related to any physical galaxy property, such as orientation

\begin{equation}\label{eq:nuisance_margin}
  P(\Gamma | \zeta) = \sum_\nu{P(\Gamma | \zeta, \nu) P(\nu)}\,.
\end{equation}

A major inconvenience of simulations is the impracticality of drawing new samples from the galaxy population $g \sim P(\Gamma | \Phi)$, as this would require re-running an entirely new simulation at repeated computational expense.
We pose that one or a set of deep generative models can properly encapsulate a simulation's formation scenario $\phi$ by learning individual galaxy properties $\zeta$, enabling in-painting a learnt formation scenario onto DMO simulations.

Recent advancements in deep learning techniques have demonstrated their efficacy in performing generative tasks that involve complex functional mappings between images.
Given that simulated galaxies are typically reduced to 2D for comparison with observational data, this study will focus on image-based deep learning techniques.

\subsection{Deep generative modelling}

As general function approximators, deep learning neural networks have proven extremely useful for data processing across various scientific disciplines~\citep{hornik_1989_multilayer_feedforward, goodfellow_2016_deep_learning}.
Their ability to beat the curse of dimensionality allows for extraction of subliminal signals from complex data, finding hidden patterns or concepts that are difficult to (manually) formalize.
Especially deep learning generative models have demonstrated unparalleled results, creating high-quality synthetic data, modelling complex systems and processes~\citep{whang_2023_applications, bengesi_2024_advancements_generative_ai}.
The goal of deep generative models is to learn an implicit (true) data distribution from which a finite number of samples is available for training~\citep[cf.][]{bond-taylor_2022_deep_generative_modelling}; this usually means fitting an over-parametrized model $p_\theta(x) \approx p(x)$ such that new samples $\hat{x} \sim p_\theta(x)$ can be drawn and/or the likelihood $p_\theta(x)$ be evaluated.
\emph{Conditional} generative models additionally include control variables $c$ which guide the generative process such that $p_\theta(x | c) \approx p(x | c)$.
\emph{Image-to-image translation} is a particular application of conditional generation, where an input image of one domain is transformed into a corresponding output image in a different domain~\citep{pang_2022_image_to}; in this study, the domains are defined by individual galaxy properties $c \equiv \zeta$ where $\zeta \in \Omega$ (see Section~\ref{sec:tng}).
Examples of image-to-image tasks include style transfer, image colourization, denoising, super-resolution, or semantic segmentation.
Within the sciences, such tasks have been adapted in and across many disciplines, showing impressive performance in modelling the distribution of atomistic systems, proteins, and biomolecules \citep[e.g.,][]{ingraham_2023_illuminating_protein, rives_2021_biological_structure, schneuing_2024_struture_based_drug, roenne_2024_generative_diffusion}, particle jets \citep[e.g.,][]{leigh_2024_pc_jedi, golling_2024_masked_particle}, or for medical imaging enhancements \citep[e.g.,][]{amirian_2024_artifact_reduction, bullock_2019_xnet}.

The conditional probability distributions approximated by these models directly correspond with the terms in Equation~(\ref{eq:galaxy_population}); thus such methods are particularly well-suited for this investigation.
We examined GAN and diffusion-based approaches, as detailed in Section~\ref{sec:gans} and~\ref{sec:diffusion}.
Both approaches are known to produce high-quality samples. While \emph{diffusion models} are considered state-of-the-art in scientific applications of image generation, they are intrinsically inefficient in their inference process, even when applied in latent space, even more so in pixel space~\citep[cf.][]{dhariwal_2021_diffusion_models}.
On the other hand, \emph{GANs} can efficiently generate samples with a single forward pass, but generally have poorer training stability and distribution coverage.
Therefore, we investigated both approaches for this work's use case and compared their results, advantages, and challenges.
As a secondary objective, we assess whether GAN-based models can achieve performance comparable to diffusion models, as this would substantially reduce computational costs and enable scalable deployment in large-scale simulation pipelines.
Demonstrating such parity would not only accelerate inference but also substantially reduce the time required to iterate over all image translation directions, enabling more comprehensive exploration of domain mappings within practical computational budgets.

\subsubsection{Generative Adversarial Networks}\label{sec:gans}

\emph{GANs} are two-component models where a generative network, the \emph{generator} $G$, and a discriminative network, the \emph{discriminator} $D$, compete in an adversarial game; first introduced by~\cite{goodfellow_2014_generative_adversarial}. $G$ aims to map\footnote{in a classical GAN the input $z$ is typically a noise variable $z \sim \mathcal{N}(0, \mathbb{I})$} out of an implicit distribution $p_{G}(z)$ to samples indistinguishable from the true data distribution $p(x)$, while at the same time $D$ is optimized to distinguish between generated samples from $p_{G}$ and real samples from the true data distribution. The adversarial game simultaneously invokes the minimization of the objective $\mathcal{L}_{\text{adversarial}}(G, D)$ by $G$ and the maximization of the same by $D$. These seemingly diametrical goals give rise to an efficient mechanism which optimizes $G \rightarrow G^\ast$ leading to plausible, high-quality samples
\begin{equation}\label{eq:minimax}
  G^\ast = \arg \min_{G}\max_{D} \mathcal{L}_{\text{adversarial}}(G, D)\,.
\end{equation}
This effectively eliminates the need to formulate an explicit loss function, as the discriminator will take that role; in other words, the loss function is learnt.

\cite{isola_2016_image_to_image} furthermore demonstrated a conditional version (cGAN) of this adversarial game as a general-purpose solution to image-to-image translation dubbed \emph{pix2pix}. Although very similar to the classical GAN formulation, both cGAN networks are additionally conditioned on an input image $x$. The generator learns a mapping from input to output image domain space $G: (x, z) \mapsto y$. The discriminator is also additionally shown the input image with the corresponding generator output $G(x, z)$. This subtly changes the interpretation of its task from originally judging the realness of generated images to a judgment on the plausibility of the domain mapping. Accordingly, the GAN optimization objective is constructed as follows
\begin{equation}\label{eq:cgan_objective}
  \mathcal{L}_{\text{cGAN}} = \mathbb{E}_{x, y}\left[\log{D(x, y)}\right] + \mathbb{E}_{x, z} \left[\log{(1-D(x, G(x, z)))}\right]
\end{equation}
where the first term is the average prediction strength of the discriminator when the images are sampled from the data distribution. The second term establishes the actual adversarial game, describing the average discriminator's prediction strength when the images are sampled from the generator.

Moreover,\ \cite{isola_2016_image_to_image} proposed to mix the GAN objective with a traditional $L_p$ loss term
\begin{equation}\label{eq:lp_objective}
  \mathcal{L}_{L_{p}} = \mathbb{E}_{x, y, z}\left[ ||y-G(x, z)||_p \right]
\end{equation}
where $p=1$ was found to be optimal by the authors whereas $p=2$ lead to blurriness in the predicted images.

The final adversarial objective is then given by
\begin{equation}\label{eq:adversarial_objective}
  \mathcal{L}_{\text{adversarial}}(G, D) = \mathcal{L}_{L1}(G) + \lambda \cdot \mathcal{L}_{\text{cGAN}}(G, D)
\end{equation}
where the objective weighting factor $\lambda$ can be treated as a fixed hyper-parameter or adaptively tuned similar to~\cite{esser_2020_taming_transformers}.

Finally, note that the noise variable $z$ is necessary to learn a stochastic mapping, matching a distribution other than a delta function. However,\ \cite{isola_2016_image_to_image} have found noise input ineffective as cGAN models tend to simply ignore the noise and suggested to use dropout at test time instead to capture the full entropy of the modelled conditional distributions.

In practice, GANs are notoriously difficult to train despite their proven ability to generate high-quality samples. Two major challenges are \textit{vanishing gradients} and \textit{mode collapse}, which can be mitigated through architectural and objective modifications. Architectural strategies include residual skip connections to improve radient flow \citep{he_2015_deep_residual}, experimenting with normalization layers (batch \cite{ioffe_2015_batch_normalization}, group \cite{wu_2018_group_normalization}, layer \cite{ba_2016_layer_normalization}, or none), and refining deconvolution operations near the generator output \citep{odena_2016_deconvolution}. Objective-based approaches involve alternative loss formulations for $\mathcal{L}_{\text{cGAN}}(G, D)$, such as DCGAN \citep{radford_2015_unsupervised_representation}, LSGAN \citep{mao_2016_least_squares}, or Wasserstein-GAN variants \citep[WGAN, WGAN-GP;][]{arjovsky_2017_wasserstein_gan,gulrajani_2017_improved_training}. Due to the minimax nature of GANs, losses often oscillate rather than converge, making diagnosis difficult. Overall, balancing generator and discriminator remains inherently unstable \citep[cf.][]{arjovsky_2017_towards_principled}, requiring alternating gradient updates or separately scheduled learning rates.

In this study, we closely followed the implementation of the Pix2Pix model by~\cite{isola_2016_image_to_image}, including the aforementioned techniques and best practices. The generator is implemented as a standard \emph{U-Net} (\cite{ronneberger_2015_u_net}; architecture modifications are detailed in Section~\ref{sec:architectures}), paired with a \emph{PatchGAN} discriminator which evaluates the plausibility of an image in sub-regions rather than a classical full-image discrimination. The discriminator can be restricted to enforce the correctness in local patches because the L1 loss in Equation~(\ref{eq:adversarial_objective}) motivates the model to correctly predict low-frequency features in images, ultimately leading to more details in generated samples.

\subsubsection{Diffusion-based models}\label{sec:diffusion}

Diffusion models have emerged as the de facto state of the art in computer vision (CV), surpassing in stability, distribution coverage, and arguably in sample quality models like GANs, normalizing flows or variational auto-encoders. They, colloquially speaking, learn to iteratively denoise a corrupted version of the data. More precisely speaking, diffusion models include a \emph{forward} (noising) process which is designed to push samples off the data manifold and a \emph{backward} (denoising) process for which a model is trained to produce trajectories back to that data manifold, generating plausible samples. There are various framings for diffusion models, leading to slightly different expressions for these forward and backward processes. Here, we give a high-level overview of the formalisms relevant to this study.

Following~\cite{ho_2020_denoising_diffusion}'s description of Denoising Diffusion Probabilistic Models (DDPMs), the forward and backward processes take the form of Markov chains. The forward process starts from the input $x_0$ and step-wise transitions to latent variables $\{x_1, \ldots, x_{T}\}$ (and vice versa for the backward process). Each forward transition at a particular time step only depends on the previous step and its probability is parametrized as a diagonal Gaussian
\begin{equation}
  q(x_t | x_{t-1}) \coloneq \mathcal{N}(x_t;\sqrt{1-\beta_t}x_{t-1},\beta_t\mathbb{I})
\end{equation}
where the variance is $\beta_t\in(0, 1)$ and typically scheduled as $\beta_{t-1}<\beta_t$. In the limit of infinitesimal step sizes, the true reverse process has the same functional form as the forward process, a well-known fact from Brownian diffusion in physics \citep[see Equations 76 and 77 in][]{feller_1949_stochastic_processes}. Thus, learning to approximate the backward process for small (enough) step sizes becomes feasible and can be analogously parametrized as
\begin{equation}
  p_\theta(x_{t-1} | x_t) \coloneq \mathcal{N}(x_{t-1};\mu_{\theta}(x_{t}, t),\Sigma_{\theta}(x_t, t))\,.
\end{equation}

Like the forward process, the backward process is a Markov chain for which its joint probability is given by the product of individual step conditionals
\begin{equation}
  p_\theta(x_{0:T}) \coloneq p(x_{T}) \prod_{t=1}^{T}p_\theta(x_{t-1}|x_t)
\end{equation}
where the marginal probability is a pure Gaussian $p(x_{T}) = q(x_{T}) = \mathcal{N}(x_{T};0,\mathbb{I})$.

The actual objective of the diffusion process, the sample probability $p_\theta(x_0)$, is generally intractable, as it would require marginalization over all possible trajectories. However, akin to latent space models such as VAEs \citep{kingma_2013_auto_encoding}, an \emph{evidence lower variational bound} (ELBO) can be estimated \citep[see][and references therein]{kingma_2021_variational_diffusion, sohl-dickstein_2015_deep_unsupervised}
\begin{align}
  &\log\,{p_\theta}(x_0) \nonumber\\
  &\geq \mathbb{E}_{q}\left[ \log{p(x_{T})} + \sum_{t\geq1}\log{\frac{p_{\theta}(x_{t-1}|x_t)}{q(x_{t}|x_{t-1})}} \right] \label{eq:ELBO_unstable}\\
  &\geq \mathbb{E}_{q}\left[ \log{p_\theta(x_{0}|x_{1})}\right] - \mathbb{E}_{q}\left[ D_{\text{KL}}(q(x_{T}|x_0) || p(x_{T})) \right] \nonumber\\
  &\quad - \sum_{t>1}\mathbb{E}_{q}\left[D_{\text{KL}}(q(x_{t-1}|x_{t},x_0) || p_{\theta}(x_{t-1}|x_t)) \right] \label{eq:ELBO_stable}
\end{align}
where $D_{\text{KL}}$ denotes the Kullback-Leibler divergence. The lower bound (\ref{eq:ELBO_unstable}) can have high variance and hence limited training efficiency and stability, compared to (\ref{eq:ELBO_stable}). Note that the first term in (\ref{eq:ELBO_stable}) ensures sample reconstruction quality while the second term matches the priors (assumed Gaussian), analogous to a classical VAE.\ The third summation term is known as the diffusion loss $\mathcal{L}_{\text{diffusion}}$ and can be calculated in closed form given $x_0$ is known (as it is during training).

The reverse step $p_\theta(x_{t-1}|x_t)$, that is the neural network, has various implementations.\ \cite{ho_2020_denoising_diffusion} observed more stable training when the network only predicted $\mu_{\theta}$ and assumed the variances to be time-dependent constants $\Sigma_{\theta}(t)=\beta_t\mathbb{I}$. Through the \emph{reparametrization trick}, it is also possible to predict the added noise $\epsilon$ through a neural network $\epsilon_\theta$ rather than the mean of the Gaussian. Other forms of diffusion models directly predict the original data point $x_0$, or some combination of both \citep{salimans_2022_progressive_distillation}.

In any case, the diffusion loss in Equation~(\ref{eq:ELBO_stable}) can be shown to generally reduce to
\begin{equation}
  \mathcal{L}_{\text{diffusion}} = \mathbb{E}_{\epsilon\sim\mathcal{N}(0,\mathbb{I}), t\sim\mathcal{U}_{[0, T]}}\left[\gamma'_\eta(t)||\epsilon-\epsilon_\theta(x_t,t)||^2\right]
\end{equation}
where $\gamma'_\eta(t)$ is an optional weighting pre-factor (with learnable bounds) evaluated via automatic differentiation. The noise schedule $\gamma_\eta(t)$ also has various forms with the simplest schedule linearly increasing between two extremal bounds $\eta = \{\gamma_{\text{min}}, \gamma_{\text{max}}\}$.

For conditional generative tasks, conditioning variables $c$ (here images of the original domain) are fed as additional inputs to the network during training $\epsilon_\theta(x_t, t, c)$. The conditioning can be further enforced by guiding the diffusion process, pushing the backward process in the direction of the gradient of the target condition probability \citep{ho_2022_classifier_free_guidance}.\ \emph{Classifier-free diffusion guidance} achieves this through a modified training procedure by linearly combining null-labelled $\emptyset$ diffusion and conditioned diffusion $\tilde{\epsilon}_\theta(x_t, t, c) = \epsilon_\theta(x_t, t, \emptyset) + s (\epsilon_\theta(x_t, t, c) - \epsilon_\theta(x_t, t, \emptyset))$ given a guidance strength $s$. At inference time, samples can be artificially pushed towards the conditional direction by increasing the guidance strength $s\geq1$. 

In this study, various noise schedules and objective variations have been investigated, see Section~\ref{sec:experiments} for details.

\subsection{Neural Network Architectures}\label{sec:architectures}

All network implementations can be found in our \Code{chuchichaestli} package\footnote{release version \Code{v0.2.13}} published on \href{https://pypi.org/project/chuchichaestli/}{PyPI} and publicly available on \href{https://github.com/CAIIVS/chuchichaestli}{\faIcon{github} GitHub}. Here, we give an overview of their architecture, but for details we refer to the correspondingly listed sources.

\paragraph{U-Net.}
GAN as well as diffusion models implement their generative networks using the U-Net convolutional architecture, first introduced by~\cite{ronneberger_2015_u_net}. It was initially designed for segmentation of biomedical images, but has since been adapted to generative tasks for many other scientific fields \citep[e.g.,][]{bianco_2025_deep_learning, andersson_2019_separation_water, yao_2018_pixel_wise_regression}. Its basic structure consists of a contracting (encoder) and an expansive (decoder) path, resulting in characteristically U-shaped graphs. While the architectural blocks in a U-Net have seen various updates since its inception, the basic encoder level follows the typical convolutional network structure with repeated $3x3$ convolutional layers each followed by activations (LeakyReLU or ReLU) and a downsampling layer (a convolutional layer with stride 2); more recent versions additionally include residual block connections to improve gradient flow \citep{he_2015_deep_residual}. With multiple levels, this leads to image compression, feature extraction, and ultimately representational learning. For image-to-image domain translation tasks, the structure of the decoder blocks is typically mirrored using deconvolutional layers to recover the input image resolution. Due to the repeated application of downsampling convolutional operations, spatial information is lost in deeper levels of the encoder. To this end, U-Nets additionally include skip connections between the corresponding levels which directly pass the encoder output information, concatenated to the output from lower decoder levels, and effectively integrate spatial information in the expansive path of the U-Net.

The basic U-Net structure in this work resembles the implementation by~\cite{isola_2016_image_to_image} with a few notable updates:
\begin{itemize}[leftmargin=*,topsep=1pt]
\item we opted for Swish activation functions \citep{ramachandran_2017_searching_for} instead of ReLU and LeakyReLU
\item each block optionally includes a self-attention \citep{vaswani_2017_attention_is} or convolutional self-attention layer \citep{yang_2019_convolutional_self_attention}
\item dropout regularization in hidden layers (with a probability of 0.2)
\end{itemize}

Self-attention enables the handling of global interactions between pixels regardless of their relative position in the image and nicely complements the inherently local convolutional pixel treatment. Originally applied to language tasks, it quickly became an essential ingredient of any state-of-the-art neural network for image processing. However, since attention increases the computational complexity quadratically with sequence length, transformer networks become quickly infeasible, especially for high-dimensional data like images.\ \cite{parmar_2018_image_transformer, weissenborn_2019_scaling_autoregressive, ho_2019_axial_attention} proposed various solutions to this problem which usually entail reducing the receptive field and long-range interactions as compromise. We tested such \textit{convolutional self-attention layers}\footnote{where key, query, and value representations are mapped using two-dimensional convolutions instead of fully connected linear layers} in our U-Nets, but have not noticed any significant improvements in performance or efficiency over classical self-attention (see Section~\ref{sec:results}).

Moreover, for the use in diffusion models the U-Net additionally contains a sinusoidal time embedding (aka positional embedding) to keep track of the time step in the diffusion process.
The time embedding is injected in all residual blocks via linear projection layers whose outputs are added to the blocks' first convolutions.

\paragraph{PatchGAN.}
As introduced in Section~\ref{sec:gans} a PatchGAN is a patch-based discriminator which models an image as a Markovian field where each probability depends on neighbouring patches within a patch diameter. This concept was initially explored by \cite{li_2016_precomputed_real_time} in the context of texture synthesis and later implemented for general image-to-image translation by \cite{isola_2016_image_to_image}. Our discriminator networks for adversarial training were adapted from \cite{isola_2016_image_to_image} with a 70x70 pixel receptive field. The implementation follows a simple convolutional block pattern consisting of batch normalization, activation (LeakyReLU), and two-dimensional downsampling convolutional layers.

\subsection{Image-based evaluation metrics}\label{sec:cv_metrics}

To evaluate the similarity and quality of generated galaxy maps during and after training these neural networks, we first employ a set of widely used metrics from the CV domain.
These metrics provide a baseline for assessing pixel-level accuracy (distortion), perceptual fidelity, and statistical realism in image synthesis tasks.
While they are not tailored to astrophysical data, they offer valuable insights into the generative performance of deep learning models and can be used for initial hyper-parameter tuning.

\paragraph{Mean Squared Error (MSE)}
quantifies the average squared difference between corresponding pixels in two images
\begin{equation}
  \text{MSE}\left(x, \hat{x}\right) = \frac{1}{N} \sum_{i=1}^{N} \left(x_i - \hat{x}_i\right)^2
\end{equation}
where $x_i$ and $\hat{x}_i$ are pixel values in the reference and generated images, respectively, and $N$ is the total number of pixels.

It is sensitive to small pixel-level deviations and is often used to measure reconstruction accuracy. However, it does not account for perceptual or structural similarity.

\paragraph{Peak Signal-to-Noise Ratio (PSNR)}
expresses the ratio between the maximum possible pixel value and the power of the error signal
\begin{equation}
  \text{PSNR}\left(x, \hat{x}\right) = 10 \cdot \log_{10} \left( \frac{\text{c}^2}{\text{MSE}\left(x, \hat{x}\right)} \right)
\end{equation}
where $c$ is the maximum pixel value range (typically 1 for normalized images, or 2 if the data range from -1 to 1).

Higher PSNR values indicate better fidelity.
It is commonly used in image compression and denoising tasks.

\paragraph{Structural Similarity Index (SSIM)}
evaluates perceptual similarity by comparing luminance, contrast, and structural information between two images \citep{wang_2004_image_quality}:
\begin{equation}
  \text{SSIM}\left(x, \hat{x}\right) = \frac{\left(2\mu_x\mu_{\hat{x}} + k_1\right)\left(2\sigma_{x\hat{x}} + k_2\right)}{\left(\mu_x^2 + \mu_{\hat{x}}^2 + k_1\right)\left(\sigma_x^2 + \sigma_{\hat{x}}^2 + k_2\right)}
\end{equation}
where $\mu$, $\sigma$, and $\sigma_{x\hat{x}}$ are the means, variances, and covariances of the images, and $k_1$, $k_2$ are stabilizing constants.

SSIM ranges from 0 to 1, with higher values indicating greater structural similarity. It is more aligned with human visual perception than MSE or PSNR.

\paragraph{Fr\'echet Inception Distance (FID)}
measures the distance between the distributions of real and generated images in a feature space extracted by a pre-trained neural network \citep[such as][]{szegedy_2015_rethinking_inception}:
\begin{equation}
  \text{FID} = \|\mu_r - \mu_g\|^2 + \text{Tr}\left(\Sigma_r + \Sigma_g - 2(\Sigma_r \Sigma_g)^{1/2}\right)
\end{equation}
where $(\mu_r, \Sigma_r)$ and $(\mu_g, \Sigma_g)$ are the mean and covariance of real and generated image set features.

Lower FID scores indicate that the generated images are statistically similar to real ones in terms of feature distribution.
FID is widely used to evaluate generative models such as GANs and diffusion models.
However, since it is evaluated with model backbones typically pre-trained on ImageNet \citep[][an extensive dataset consisting of 3-channel, natural images]{imagenet_2009}, its application on scientific maps may be problematic.
In our use case, each map is replicated on 3-channels before its features are extracted and thus does not exhibit the same colour variation as for natural images.
Moreover, critics argue that FID’s reliance on ImageNet-trained embeddings and its assumption of Gaussianity in high-level feature space render it ill-suited for domains with drastically different image statistics, such as scientific or medical imaging \citep{kynkaeaenniemi_2022_role_of, jayasumana_2023_rethinking_fid}.
In contrast, some studies have shown that using ImageNet-trained features can still correlate better with human perception than domain-specific feature extractors, even in, e.g., medical image synthesis tasks \citep{woodland_2024_lecture_notes}.
In any case, these findings caution against uncritical use of FID in non-natural image domains and suggest careful validation supported also by alternative metrics.

The metrics mentioned above serve as a foundational measures for evaluating image-to-image translation tasks.
While they offer general-purpose assessments of distortion and perceptual image quality, they do not capture the domain-specific physical properties of galaxies.
To address this, we complement them with a set of astrophysical metrics tailored to the structural and morphological characteristics of galaxy maps.

\subsection{Astrophysical evaluation metrics}\label{sec:astroph_metrics}

To assess the physical plausibility of the generated galaxy maps beyond pixel-wise similarity, we introduce a set of astrophysically motivated metrics.
These metrics are designed to quantify structural, morphological, and distributional properties of galaxies, enabling a more rigorous comparison between generated and ground truth samples.
Each metric captures a distinct aspect of galaxy morphology and mass distribution, reflecting the underlying formation scenario.

\paragraph{Asymmetry Error (AE).}
Asymmetry evaluates the rotational symmetry of a galaxy map by comparing it to its $180^\circ$ rotated counterpart (centred on the galaxy).
This is a standard morphological indicator in observational astronomy \citep[first introduced by][]{schade_1995_cfrs}, and often used to identify signs of mergers, tidal interactions, or structural disturbances.
It is typically defined as part of the concentration-asymmetry-smoothness parameter system \citep[CAS;][]{conselice_2000_asymmetry_of, conselice_2003_relationship_between}. Here, we define the AE in a slightly simplified adaptation as the difference between the normalized asymmetry of a ground truth $I_{r}$ and generated map $I_{g}$
\begin{equation}
  \text{AE}(I_{r}, I_{g}) = \frac{\sum_{ij}|I_{r,ij} - I^{180^\circ}_{r,ij}|}{\sum_{ij}|I_{r,ij}|} - \frac{\sum_{ij}|I_{g,ij} - I^{180^\circ}_{g,ij}|}{\sum_{ij}|I_{g,ij}|}
\end{equation}
where $I^{180^\circ}$ are the $180^{\circ}$-rotated map correspondents.
Higher asymmetry errors with respect to generated maps indicate discrepancies in structural symmetry which may indicate unrealistic morphology or artefacts.

\paragraph{Smoothness/Clumpiness Error (SCE).}
The so-called clumpiness quantifies the presence of small-scale structures such as star-forming regions or dense gas clumps~\citep[cf. CAS parameter system;][]{conselice_2000_asymmetry_of, conselice_2003_relationship_between}.
We calculate a proxy by subtracting a smoothed version of the map from the original and measuring the positive residuals, to avoid biasing the result through smoothing artefacts and removal of diffuse regions.
\begin{equation}
  \text{SCE}(I_{r}, I_{g}) = \frac{\sum_{ij}\max(|I_{r,ij} - S_{r,ij}|, 0)}{\sum_{ij}|I_{r,ij}|} - \frac{\sum_{ij}\max(|I_{g,ij} - S_{g,ij}|, 0)}{\sum_{ij}|I_{g,ij}|}
\end{equation}
where $S$ is a smoothed (Gaussian blurred) version of the corresponding map $I$.
A high SCE value may indicate excessive noise or unrealistic fragmentation, while a low error suggests smooth, well-resolved distributions.
This metric is particularly relevant for evaluating the realism of baryonic, frictional components like gas and stars and substructure in DM haloes.

\paragraph{Centre-Of-Mass Distance (COMD)}
measures the Euclidean distance between the centre of mass of the generated map and that of the ground truth.
The center of mass of a galaxy reflects the spatial alignment of its component distribution.
\begin{equation}
  \text{COMD}(I_{r}, I_{g}) = \left\Vert\frac{\sum_{ij}x_{ij}I_{r,ij}}{\sum_{ij}I_{r,ij}} - \frac{\sum_{ij}x_{ij}I_{g,ij}}{\sum_{ij}I_{g,ij}}\right\Vert_{2}
\end{equation}
where $x_{ij}$ are the spatial coordinates of distribution elements (pixels).
Misalignment may indicate translation artefacts, structural inconsistencies, or failure to preserve spatial coherence.
This metric is particularly important for tasks involving domain translation where positional accuracy is critical.

\paragraph{(Cumulative) Radial Curve Errors (CRCE/RCE).}
This metric compares the radial intensity or mass profile of the generated map to that of the ground truth.
The radial profile is computed by averaging pixel values in concentric radial bins centred on the galaxy’s centre of mass.
The \textit{Radial Curve Errors} capture deviations in the spatial distribution of matter, such as incorrect central concentration or scale mismatch.
It is essential for validating the structural integrity of generated galaxies.
As a complement, the analogous comparison of cumulative radial distributions of generated and ground truth maps measures cumulative content deviations at a given radius.
Errors from cumulative profiles are sensitive to global mass conservation and spatial allocation.

\paragraph{Power Spectrum Errors (PSE)}
compares the radially averaged 2D power spectra (i.e. squared magnitude of the Fourier coefficients at each frequency) of two maps.
For each map, we compute the two-dimensional discrete Fourier transform and derive the power spectrum as the squared modulus of the Fourier coefficients.
The resulting two-dimensional power spectrum is then radially averaged in Fourier space to obtain a one-dimensional power spectrum curve $P(k)$ which characterizes the distribution of power as a function of spatial frequency.
Finally, the normalized power spectrum curve residuals can be reduced by means of summation or averaging.
This approach assesses similarity in spatial structure, texture, and particularly characteristic, second-order (filamentary) scales, independent to normalization, translation, or rotation.
The PSE is especially useful when validating a model's accurate reproduction of multi-scale spatial features.

The aforementioned metrics collectively provide a robust framework for evaluating the astrophysical realism of generated galaxy maps.
They complement traditional image similarity metrics from Section~\ref{sec:cv_metrics} by incorporating domain-specific knowledge and physical constraints, thereby enabling a more meaningful assessment of generated samples.

Finally, beyond pixel-level and morphological assessments, we also perform an inter-model consistency analysis based on integrated physical quantities (such as e.g. average magnetic field strength or total mass content).
By substituting individual components with model predictions while keeping others at ground truth, we quantify biases and scatter, as well as cross-source disagreement for a fixed target domain.
These metrics reveal whether different translation models preserve masses and energy globally and maintain physically plausible component fractions, independent of local image fidelity.
Furthermore, they test whether models can be chained in cycles, potentially avoiding the need to train all model translation permutations if the goal is to complete a physical model from an arbitrary galaxy property.
This approach provides a complementary, physically grounded perspective on model performance, ensuring that generated maps respect fundamental conservation principles and astrophysical scaling relations.

\subsection{Experiments}\label{sec:experiments}
Given that both generative methodologies described in Sections~\ref{sec:gans} and~\ref{sec:diffusion} employ U-Net architectures as their backbone, it is essential to optimize the architectural hyper-parameters for the specific characteristics of the dataset.
However, exhaustive hyper-parameter searches across all possible configurations are computationally prohibitive, especially for generative models.
We therefore constrain our ablation studies to architectural components that have demonstrated the most significant impact on conditional image generation performance.
For these ablations, a coarse grid search across diverse optimizers, learning rates, and loss term weights have been carried out beforehand to find good values/choices.
The hyper-parameters for the final network architectures used in Section~\ref{sec:domain_translations} have been optimized using the \emph{Optuna} and \emph{Ray Tune} frameworks.

\paragraph{Training.}
All models were implemented in PyTorch. The experiments were conducted on Nvidia V100/A100/H100/H200 GPUs depending on the specific VRAM requirements.
Adam optimizers (separate ones in the case of GAN-based models) with $\beta_1=0.9$ and $\beta_2=0.999$, and weight decay of $10^{-5}$ were used.
Unless otherwise stated, the maximum learning rate was set to $5\times10^{-5}$ for generators and $1\times10^{-5}$ for discriminators (where used), with a one-cycle policy schedule \citep[following][]{smith_2017_super_convergence}. It provides a smoother warm-up phase at lower learning rates, a ramp up to the maximum learning rate, and a cosine-annealing phase to $10^{-4}$ of the maximum value. When attention layers are included, this schedule was found to lead to less instabilities during GAN training.
All model experiments were trained for 30 epochs with a batch size of 8.
The datasets were split into 85\% training, 10\% validation, and 5\% test sets, ensuring that galaxies from the same halo did not appear in multiple splits.
All reported metrics for experiments in Section~\ref{sec:experiments} were evaluated on the validation set whereas astrophysical validation was performed on the test set, and reported after the final epoch.

\paragraph{U-Net size experiments.}
Previous work has identified model capacity, defined by depth (number of U-Net levels) and width (number of hidden feature channels) as a key factors of generative fidelity \citep{ronneberger_2015_u_net, ho_2020_denoising_diffusion, isola_2016_image_to_image}.
Additionally, the choice of normalization layers \citep{dhariwal_2021_diffusion_models}, and the inclusion of residual and attention mechanisms \citep{zhang_2018_self_attention_generative, dhariwal_2021_diffusion_models}, have consistently shown to enhance both training stability and output quality.
In contrast, other design choices, such as the specific up-sampling scheme or minor variations in skip connections, tend to yield marginal improvements and diminishing returns.
To systematically assess these factors, we first examine the impact of model capacity on generative performance, limiting experiments to high-impact components to keep computational costs tolerable.

Table~\ref{tab:size_experiments} summarizes the U-Net configurations tested in this initial set of targeted experiments.
Each experiment varies only the architectural parameters under investigation, while all other training settings are held constant.
For bench-marking, we selected the \gas{}$\rightarrow$\dm{} translation task, which exhibited intermediate difficulty across all domain pairs in preliminary tests.

All models were trained with adversarial loss for 30 epochs using the standard discriminator configuration (as described in Section~\ref{sec:architectures}), with a warm restart technique for stochastic gradient descent and a cosine annealing learning rate schedule \citep{loshchilov_2016_sgdr}.
Initial learning rates were set to $10^{-4}$ for the generator and $5\times10^{-5}$ for the discriminator.

The results of the U-Net size ablation study are summarized in Table~\ref{tab:size_experiments_results}.
Among the tested configurations, \textsc{mediumU} (64 channels, 4 levels) consistently achieved the best overall performance across most evaluation metrics.
Notably, the SSIM metric seemed to saturate in all tests quickly, indicating most U-Net configurations yield structurally similar outputs to the ground truth, but may lack sensitivity with smooth, high-resolution distributions like those from simulations and may not capture subtle differences fine-grained textures and localized features.

Deeper and larger U-Net variants started exhibiting artefacts and over-fitting that degraded perceptual quality of generated samples evident by higher PSNR values, but at the cost of increased FID.
Moreover, larger models exhibited signs of mode collapse, with unreliable metric results.
Conversely, the shallower and smaller performed comparably or worse in PSNR and SSIM but suffered from a substantially worse FID, suggesting insufficient capacity to model the full complexity of the domain mapping.

Based on these findings, we identified the \textsc{mediumU} configuration as the optimal U-Net configuration for this task.
It offers a favourable trade-off between performance and computational cost, avoids over-fitting, and maintains stable training dynamics.
This configuration is therefore used as the default architecture in all subsequent experiments unless stated otherwise.

Note that for diffusion models, spot tests resulted in comparable distortion metrics, however, to offset the substantial increase in computational cost, the U-Net width was reduced to 32 channels, prioritizing the inclusion of additional attention layers.

\begin{table}
  \centering
  \caption{
    Various U-Net configurations with varying sizes in depth and width that were tested.
    ``Width'' refers to the base number of feature channel in the first U-Net layer, whereas ``Depth'' is the number of levels between down- and up-sampling layers.
    The ``encoder'' consists of base ``DownBlocks'' including  and ``decoder''.
    ``\# Params'' is the total number of trainable parameters in the U-Net.}
  \label{tab:size_experiments}
\small
\begin{tabular}{lrrrr}
  \hline
  Designation & Width & Depth & Levels & \# Params \\
  \hline
  \textsc{tinyU}        & 16  & 4 & 4 &   4,010,369 \\
  \textsc{smallU}       & 32  & 4 & 4 &  16,024,833 \\
  \textsc{mediumU}      & 64  & 4 & 4 &  64,066,049 \\
  \textsc{mediumU\_L3}  & 64  & 3 & 3 &  15,814,657 \\
  \textsc{mediumU\_L5}  & 64  & 5 & 5 & 257,037,825 \\
  \textsc{largeU}       & 128 & 4 & 4 & 256,197,633 \\
  \hline
\end{tabular}
\end{table}

\begin{table}
  \centering
  \caption{
    Evaluation results of various U-Net size configurations from Table~\ref{tab:size_experiments} after training for 30 epochs.
    All experiments are based on the translation \gas{}$\rightarrow$\dm{}, adversarially trained with the same discriminator configuration.
    Model results in bold are optimal values, and those marked with $^{\dagger}$ exhibit mode collapse and are not reliable.}
  \label{tab:size_experiments_results}
\small
\begin{tabular}{lrrrr}
  \hline
  Designation & PSNR ↑ & SSIM ↑ & MSE ↓ & FID ↓ \\
  \hline
  \textsc{tinyU}        & 35.31    & 0.9954    & $5.5\times10^{-4}$    & 12.02 \\
  \textsc{smallU}       & 39.12    & 0.9966    & $6.6\times10^{-4}$    & 12.75 \\
  \textsc{mediumU}      & 39.76    & \bf0.9977 & $4.2\times10^{-4}$ & \bf9.71 \\
  \textsc{mediumU\_L3}  & 39.57    & 0.9967    & $6.8\times10^{-4}$    & 30.44 \\
  \textsc{mediumU\_L5}  & \bf48.01 & 0.9972    & $\bf2.9\times10^{-4}$    & 18.31 \\
  \textsc{largeU}       & $^{\dagger}$65.28 & $^{\dagger}$0.9978 & $^{\dagger}3.1\times10^{-3}$ & $^{\dagger}$270.0 \\
  \hline
\end{tabular}
\end{table}

\paragraph{Attention layer placement experiments.}
Having established an optimal baseline configuration, we investigated the impact of attention layer placement within the U-Net architecture in a subsequent series of experiments.
While prior work suggests that attention mechanisms can enhance global context modelling \citep{vaswani_2017_attention_is}, their effectiveness and efficiency may depend on the resolution level at which they are applied.
To this end, we varied the position of self-attention blocks across encoder and decoder stages, including configurations with attention in early layers (high-resolution features), late layers (low-resolution, high-semantic features), and hybrid placements spanning multiple levels.
While attention layers are expected to yield superior results no matter the placement, the main purpose of these experiments was to assess the relative performance differences of less computationally demanding placement in late layers to those in high-resolution features.
All other architectural and training settings were kept identical to those in the optimal configuration from the U-Net size experiments.
Only the learning rate update schedule was changed to a one-cycle policy due to instabilities in the generator-discriminator dynamics and to keep learning rate comparably high.
The evaluation focused on image-based metrics to determine whether attention placement influences fine-grained structural fidelity. These experiments aim to identify the most effective strategy for leveraging attention without incurring unnecessary computational overhead.

The results of these experiments (Table~\ref{tab:attn_experiments_results}) reveal that the benefits of self-attention layers in U-Net blocks are indeed dependent on both the number and placement within the network.
While adding attention universally across all levels (\textsc{attnAll}) improved pixel-wise metrics such as PSNR, it comparatively degraded distributional consistency as measured by FID, indicating over-parametrization.
Conversely, a moderate number of self-attention layers in the deepest levels seems to generally improve distributional, perceptual fidelity compared to the previous experiments, in trade for distortion \citep[cf.][]{blau_2018_perception_distortion}.
In particular, adding attention layers near the bottleneck (\textsc{attn3xU3}) yielded the best FID scores while maintaining competitive PSNR and the other distortion metrics.
These findings suggest that for this dataset global interactions are most effectively modelled when attention is applied to low-resolution, high-semantic feature maps, whereas attention in high-resolution layers may lead to equally or better performance but introduces unnecessary inefficiency and instability.
Moreover, all experiments have been repeated using the convolutional attention variant, with nearly identical results in each run, and minimally shorter forward pass timings.
Based on these results, we adopt a configuration with three deep convolutional attention layers for all subsequent experiments, as it offers the best balance of generative fidelity, stability, and efficiency.

\begin{table}
  \centering
  \caption{
    U-Net configurations with various attention blocks positioning (encoder levels numbered top to bottom, continuing in the decoder bottom up).
    The second column ``\# Encoder'' indicates how many attention layers are included in the encoder, the third ``\# Decoder'', how many in the decoder.
    ``\# Params'' is the total number of trainable parameters in the U-Net.
    The right-most column is the average time for a forward pass with a single batch.}
  \label{tab:attn_experiments}
\small
\begin{tabular}{lrrrr}
  \hline
  Designation & \# Encoder & \# Decoder & \# Params & Forward pass \\
  \hline
  \textsc{attnU1}    & 1 & 0 & 64,082,817 & 15.3765 s \\
  \textsc{attnU3}    & 1 & 0 & 64,329,729 &  2.0411 s \\
  \textsc{attnU4}    & 1 & 0 & 65,117,697 &  1.9542 s \\
  \textsc{attnUMid}  & 1 & 0 & 68,266,497 &  1.9796 s \\
  \textsc{attnU5}    & 0 & 1 & 68,266,497 &  1.9904 s \\
  \textsc{attnU6}    & 0 & 1 & 65,117,697 &  2.2751 s \\
  \textsc{attnU8}    & 0 & 1 & 64,132,353 & 27.8642 s \\
  \textsc{attn3xU3}  & 2 & 1 & 69,581,825 &  5.7335 s \\
  \textsc{attnAll}   & 4 & 4 & 71,046,529 & 48.2022 s \\
  \hline
\end{tabular}
\end{table}

\begin{table}
  \centering
  \caption{
    Evaluation results of U-Net attention layer placement experiments from Table~\ref{tab:attn_experiments} after training for 30 epochs.
    All experiments are based on the translation \gas{}$\rightarrow$\dm{}, adversarially trained with the same discriminator configuration.
    Model results in bold are optimal values, and those marked with $^{\dagger}$ exhibit mode collapse and are not reliable.}
  \label{tab:attn_experiments_results}
\small
\begin{tabular}{lrrrr}
\hline
Designation & PSNR ↑ & SSIM ↑ & MSE ↓ & FID ↓ \\
  \hline
  \textsc{attnU1}    &    37.23    &    0.9968  & $3.4\times10^{-4}$    &    6.93 \\ 
  \textsc{attnU3}    &    37.10    &    0.9972  & $3.5\times10^{-4}$    &    6.64 \\
  \textsc{attnU4}    &    36.70    &    0.9968  & $3.6\times10^{-4}$    &    6.59 \\
  \textsc{attnUMid}  &    35.27    &    0.9983  & $3.7\times10^{-4}$    &    7.47 \\
  \textsc{attnU5}    &    35.76    & \bf0.9983  & $3.6\times10^{-4}$    &    6.60 \\
  \textsc{attnU6}    &    36.26    & 0.9974     & $3.6\times10^{-4}$    &    4.57 \\
  \textsc{attnU8}    & $^{\dagger}$22.55 & $^{\dagger}$0.9  & $^{\dagger}1.2\times10^{-2}$& $^{\dagger}$253.2 \\
  \textsc{attn3xU3}  &    37.71    &    0.9970  & $3.4\times10^{-4}$ & \bf4.04 \\
  \textsc{attnAll}   & \bf42.55    &    0.9934  & $\bf3.2\times10^{-4}$    &    8.70 \\
\hline
\end{tabular}
\end{table}

\paragraph{Model specifics}

With the attention configuration fixed, we proceed to model-specific refinements. In this stage, we tuned discriminator architectures for GAN-based models and evaluate noise scheduling strategies for diffusion models.

For GAN-based models, the PatchGAN discriminator's width, depth, and number of hidden layer configurations were spot tested (see Table~\ref{tab:patchgan_experiments} for configurations).
Based on the results in Table~\ref{tab:patchgan_experiments_results}, the \textsc{pgan\_medium} configuration was used for the final model training.
While there were no clear differences between the various configurations, smaller networks had the tendency to impose checker-board artefacts in the generator outputs and larger, wider ones lead to instabilities during training due to mismatched sizes between discriminator and generator.

\begin{table}
  \centering
  \caption{
    PatchGAN configurations of various sizes.
    ``Width'' refers to the base number of feature channel in the first hidden layer, whereas ``Depth'' is the number of hidden layers in the network.
    ``\# Params'' is the total number of trainable parameters in the PatchGAN network.}
  \label{tab:patchgan_experiments}
\small
\begin{tabular}{lrrr}
  \hline
  Designation & Width & Depth & \# Params \\
  \hline
  \textsc{pgan\_small}    & 64 & 3 & 2,765,505 \\  
  \textsc{pgan\_medium}   & 64 & 4 & 11,165,377 \\  
  \textsc{pgan\_large}    & 64 & 5 & 44,742,337 \\  
  \textsc{pgan\_wide}     & 128 & 3 & 178,875,777 \\  
  \textsc{pgan\_narrow}   & 32 & 3 &  11,197,281 \\  
  \hline
\end{tabular}
\end{table}

\begin{table}
  \centering
  \caption{
    Evaluation results of PatchGAN size experiments from Table~\ref{tab:patchgan_experiments} after training for 30 epochs.
    All experiments are based on the translation \gas{}$\rightarrow$\dm{}, adversarially trained with the same generator configuration.
    Model results in bold are optimal values.}
  \label{tab:patchgan_experiments_results}
\small
\begin{tabular}{lrrrr}
  \hline
  Designation & PSNR ↑ & SSIM ↑ & MSE ↓ & FID ↓ \\
  \hline
  \textsc{pgan\_small}    &    32.46 &    0.9894 &    $8.3\times10^{-4}$ &   18.04 \\  
  \textsc{pgan\_medium}   & \bf37.16 &    0.9901 &    $\bf4.3\times10^{-4}$ & \bf4.62 \\  
  \textsc{pgan\_large}    &    36.56 & \bf0.9960 &    $5.7\times10^{-4}$ &    8.47 \\  
  \textsc{pgan\_wide}     &    34.77 &    0.9952 &    $7.1\times10^{-4}$ &    8.65 \\  
  \textsc{pgan\_narrow}   &    35.16 &    0.9909 & $6.0\times10^{-4}$ &    9.75 \\  
  \hline
\end{tabular}
\end{table}

For diffusion models, the U-Net includes a sinusoidal time-embedding with 32 channels in each block (as described in Section~\ref{sec:architectures}).
Moreover, linear, quadratic, and cosine noise schedules have been tested, and cosine clearly improved image quality (with a consistent 2-3 dB improvement in PSNR, 0.05-0.1 difference in SSIM), convergence, and provided smoother denoising transitions.

\subsection{Domain translations}\label{sec:domain_translations}
With all model components conservatively optimized, the final stage of experiments extended the map-to-map translation task to encompass all available domains.
Given the combinatorial nature of the dataset, exhaustively exploring all 5040 possible domain translations is infeasible.
However, it is reasonable to expect that the complexity of translations tasks varies between the astrophysical interactions between the components.
For instance, domain translations such as \gas{}$\rightarrow$\hi{} or \hicm{}$\rightarrow$\gas{} are likely to be less complex, as they represent information completion or reduction.
On the other hand, mappings like \stars{}$\rightarrow$\dm{} are inherently more challenging due to the ``weak'' coupling of these components in the simulation.

To capture this diversity, we selected a representative subset of domain pairs that span a broad range of translation difficulties; the translations are centred around \gas{} due to its close relation to observable quantities and, thus, astronomical relevance (as mentioned in Section~\ref{sec:intro}).
These included the following mappings:
\begin{itemize}[leftmargin=*,noitemsep]
\item within baryonic components
  \begin{itemize}[noitemsep,topsep=0pt]
  \item \gas{}$\rightarrow$\hi{},
  \item \gas{}$\rightarrow$\hicm{},
  \item \hicm{}$\rightarrow$\gas{},
  \item \gas$\rightarrow$\stars{}
  \end{itemize}
\item baryonic-to-DM translations
  \begin{itemize}[noitemsep,topsep=0pt]
  \item \gas{}$\rightarrow$\dm{},
  \item \dm{}$\rightarrow$\gas{}
  \end{itemize}
\item thermodynamic transformations
  \begin{itemize}[noitemsep,topsep=0pt]
  \item \gas{}$\rightarrow$\temp{}
  \end{itemize}
\item magnetic field strength reconstructions
  \begin{itemize}[noitemsep,topsep=0pt]
  \item \gas{}$\rightarrow$\bfield{}.
  \end{itemize}
\end{itemize}

For each selected pair, models were trained using the optimized U-Net configuration identified in previous experiments, with attention layers placed near the bottleneck.

Both GAN-based and diffusion-based models were evaluated, and their outputs compared using the full suite of CV and astrophysical metrics.
This strategy allowed us to assess not only the fidelity of individual translations but also the consistency of physical quantities across domains.
In particular, we investigated whether certain domain pairs exhibit systematic biases or structural artefacts, and whether translation difficulty correlates with the intrinsic entropy or sparsity of the source domain.
The results of these experiments are summarized in the following Section~\ref{sec:results}.

\section{Results}\label{sec:results}

\begin{table*}
  \centering
  \caption{
    Extensive results for the entire suite of map-to-map translation models with image-based metrics (see Section~\ref{sec:cv_metrics}).
    The values listed are mean $\pm$ standard deviation of the respective metrics from the last 5 epochs (duration chosen as patience parameter when testing for convergence), as metric values for GANs fluctuate more.
    Note that the data ranges differ for GAN and DDPM models, which inherently biases the metric towards DDPMs by $\sim$6.02 dB for the same MSE value.
    Thus, the PSNR values for DDPM models were implicitly unbiased in the discrimination analysis.
  }
  \label{tab:map_to_map_results_pixelwise}
\small
\begin{tabular*}{0.65\textwidth}{llrrrr}
  \hline
  Translation & Model & PSNR ↑ & SSIM ↑ & MSE ($\times10^{-4}$) ↓ & FID ↓  \\
  \hline
  \gas{}$\rightarrow$\dm{}     & GAN  &    35.31 $\pm$ 0.11 & \bf0.9974 $\pm$ 0.0002 &    3.82 $\pm$ \ 0.00 & \bf1.56 $\pm$ \ 0.36 \\
  \gas{}$\rightarrow$\dm{}     & DDPM &    41.17 $\pm$ 0.07 &    0.9970 $\pm$ 0.0001 &    4.24 $\pm$ \ 0.10 &    2.03 $\pm$ \ 0.08 \\
  \gas{}$\rightarrow$\stars{}  & GAN  &    18.55 $\pm$ 0.36 &    0.5738 $\pm$ 0.0154 &  324.12 $\pm$ \ 3.44 &   60.56 $\pm$ 15.98 \\
  \gas{}$\rightarrow$\stars{}  & DDPM &    23.34 $\pm$ 0.05 &    0.5577 $\pm$ 0.0046 &  324.60 $\pm$  10.58 &   56.17 $\pm$ 14.33 \\
  \gas{}$\rightarrow$\hi{}     & GAN  &    33.27 $\pm$ 0.16 &    0.9739 $\pm$ 0.0011 &   15.31 $\pm$ \ 0.80 &    4.57 $\pm$ \ 0.99 \\
  \gas{}$\rightarrow$\hi{}     & DDPM &    39.99 $\pm$ 0.14 &    0.9749 $\pm$ 0.0009 &   17.11 $\pm$ \ 0.77 &    5.86 $\pm$ \ 0.05 \\
  \gas{}$\rightarrow$\hicm{}   & GAN  &    31.60 $\pm$ 0.48 &    0.7958 $\pm$ 0.0115 &   17.90 $\pm$ \ 0.78 &    3.57 $\pm$ \ 1.03 \\
  \gas{}$\rightarrow$\hicm{}   & DDPM &    38.55 $\pm$ 0.05 &    0.8133 $\pm$ 0.0013 &   17.95 $\pm$ \ 0.12 &    5.78 $\pm$ \ 0.07 \\
  \gas{}$\rightarrow$\temp{}   & GAN  &    37.05 $\pm$ 0.27 &    0.9973 $\pm$ 0.0002 &    5.04 $\pm$ \ 0.19 &    9.91 $\pm$ \ 3.18 \\
  \gas{}$\rightarrow$\temp{}   & DDPM &    41.56 $\pm$ 0.06 &    0.9967 $\pm$ 0.0001 &    3.99 $\pm$ \ 0.32 &    7.86 $\pm$ \ 0.13 \\
  \gas{}$\rightarrow$\bfield{} & GAN  & \bf38.76 $\pm$ 0.62 &    0.9964 $\pm$ 0.0003 & \bf2.75 $\pm$ \ 0.51 &    9.80 $\pm$ \ 1.67 \\
  \gas{}$\rightarrow$\bfield{} & DDPM &    43.39 $\pm$ 0.26 &    0.9955 $\pm$ 0.0007 &    3.60 $\pm$ \ 0.28 &    8.38 $\pm$ \ 0.33 \\
  \dm{}$\rightarrow$\gas{}     & GAN  &    31.28 $\pm$ 0.12 &    0.9853 $\pm$ 0.0003 &   12.18 $\pm$ \ 0.79 &   36.36 $\pm$ \ 9.58 \\
  \dm{}$\rightarrow$\gas{}     & DDPM &    36.96 $\pm$ 0.03 &    0.9845 $\pm$ 0.0008 &   10.62 $\pm$ \ 0.40 &   22.87 $\pm$ \ 0.73 \\
  \hicm{}$\rightarrow$\gas{}   & GAN  &    35.95 $\pm$ 0.56 &    0.9904 $\pm$ 0.0013 &    4.46 $\pm$ \ 0.55 &    7.60 $\pm$ \ 2.24 \\
  \hicm{}$\rightarrow$\gas{}   & DDPM &    42.08 $\pm$ 0.07 &    0.9900 $\pm$ 0.0003 &    3.75 $\pm$ \ 0.10 &    5.63 $\pm$ \ 0.90 \\
  \hline
\end{tabular*}
\end{table*}

\begin{table*}
  \centering
  \caption{
    Extensive results for the entire suite of map-to-map translation models with astrophysical metrics (see Section~\ref{sec:astroph_metrics}).
    The values listed are mean $\pm$ standard deviation of the respective metrics from the last 5 epochs.
  }
  \label{tab:map_to_map_results_astroph}
\small
\begin{tabular*}{0.82\textwidth}{llrrrcr}
  \hline
  Translation & Model & AE ↓ & SCE ↓ & COMD ↓ & CRCE (at R$_{50}$) ↓ & PSE ↓ \\
  \hline
  \gas{}$\rightarrow$\dm{}     & GAN  &    0.0655 $\pm$ 0.0005 &    0.0027 $\pm$ 0.0000 &    0.0211 $\pm$ 0.0173 &    0.2132 $\pm$ 0.1982 &    0.0788 $\pm$ 0.0041 \\
  \gas{}$\rightarrow$\dm{}     & DDPM &    0.0746 $\pm$ 0.0005 &    0.0032 $\pm$ 0.0000 &    0.0215 $\pm$ 0.0168 &    0.2196 $\pm$ 0.1998 &    0.0856 $\pm$ 0.0042 \\
  \gas{}$\rightarrow$\stars{}  & GAN  &    0.7460 $\pm$ 0.0529 &    0.0975 $\pm$ 0.0265 &    0.0657 $\pm$ 0.0446 &    1.3772 $\pm$ 5.7906 &    0.0690 $\pm$ 0.0046 \\
  \gas{}$\rightarrow$\stars{}  & DDPM &    0.4466 $\pm$ 0.0494 &    0.0812 $\pm$ 0.0235 &    0.0297 $\pm$ 0.0393 &    1.2875 $\pm$ 3.4577 &    0.0596 $\pm$ 0.0042 \\
  \gas{}$\rightarrow$\hi{}     & GAN  &    0.0839 $\pm$ 0.0028 &    0.0207 $\pm$ 0.0013 &    0.0128 $\pm$ 0.0178 &    0.2684 $\pm$ 0.3197 & \bf0.0307 $\pm$ 0.0024 \\
  \gas{}$\rightarrow$\hi{}     & DDPM &    0.0885 $\pm$ 0.0031 &    0.0219 $\pm$ 0.0014 &    0.0136 $\pm$ 0.0191 &    0.2948 $\pm$ 0.3595 &    0.0363 $\pm$ 0.0028 \\
  \gas{}$\rightarrow$\hicm{}   & GAN  &    0.0713 $\pm$ 0.0027 &    0.0186 $\pm$ 0.0012 & \bf0.0109 $\pm$ 0.0155 &    0.2192 $\pm$ 0.3051 &    0.0452 $\pm$ 0.0270 \\
  \gas{}$\rightarrow$\hicm{}   & DDPM &    0.0813 $\pm$ 0.0029 &    0.0210 $\pm$ 0.0013 &    0.0120 $\pm$ 0.0167 &    0.2765 $\pm$ 0.2648 &    0.0524 $\pm$ 0.0308 \\
  \gas{}$\rightarrow$\temp{}   & GAN  &    0.0901 $\pm$ 0.0001 &    0.0024 $\pm$ 0.0000 &    0.0561 $\pm$ 0.0367 &    0.1754 $\pm$ 0.1909 &    0.0568 $\pm$ 0.0047 \\
  \gas{}$\rightarrow$\temp{}   & DDPM &    0.0793 $\pm$ 0.0001 & \bf0.0019 $\pm$ 0.0000 &    0.0484 $\pm$ 0.0332 & \bf0.1605 $\pm$ 0.1725 &    0.0597 $\pm$ 0.0041 \\
  \gas{}$\rightarrow$\bfield{} & GAN  &    0.0822 $\pm$ 0.0012 &    0.0093 $\pm$ 0.0002 &    0.0371 $\pm$ 0.0213 &    0.2209 $\pm$ 0.1843 &    0.1000 $\pm$ 0.0554 \\
  \gas{}$\rightarrow$\bfield{} & DDPM & \bf0.0647 $\pm$ 0.0010 &    0.0072 $\pm$ 0.0002 &    0.0294 $\pm$ 0.0192 &    0.1928 $\pm$ 0.1739 &    0.0875 $\pm$ 0.0497 \\
  \dm{}$\rightarrow$\gas{}     & GAN  &    0.1093 $\pm$ 0.0022 &    0.0224 $\pm$ 0.0006 &    0.0367 $\pm$ 0.0242 &    0.3143 $\pm$ 0.4294 &    0.0328 $\pm$ 0.0030 \\
  \dm{}$\rightarrow$\gas{}     & DDPM &    0.1085 $\pm$ 0.0021 &    0.0184 $\pm$ 0.0006 &    0.0357 $\pm$ 0.0246 &    0.2946 $\pm$ 0.3300 &    0.0333 $\pm$ 0.0030 \\
  \hicm{}$\rightarrow$\gas{}   & GAN  &    0.0891 $\pm$ 0.0019 &    0.0161 $\pm$ 0.0004 &    0.0148 $\pm$ 0.0141 &    0.3483 $\pm$ 0.3448 &    0.0641 $\pm$ 0.0038 \\
  \hicm{}$\rightarrow$\gas{}   & DDPM &    0.0705 $\pm$ 0.0018 &    0.0131 $\pm$ 0.0004 &    0.0124 $\pm$ 0.0115 &    0.3231 $\pm$ 0.3867 &    0.0621 $\pm$ 0.0037 \\
  \hline
\end{tabular*}
\end{table*}

\subsection{Qualitative assessment of samples}

Figure~\ref{fig:samples} shows representative samples of map-to-map translations across the (unseen) test set of domain pairs.
Each triplet shows the input map (left), the ground truth target (middle), and the model prediction (right).
For strongly coupled domains such as \gas{}$\rightarrow$\dm{}, both GAN and DDPM reproduce global morphology and substructures with high fidelity across various scales and mass ranges.
In some cases, smaller satellite haloes are either missing or were generated without any counterpart in the ground truth maps. When present, they are typically plausible domain translations of the input map.

Also, the translations \gas{}$\rightarrow$\hi{}, \gas{}$\rightarrow$\hicm{}, and \hicm{}$\rightarrow$\gas{} are consistently in excellent agreement for both models, with only mild over- or underestimation in some systems.

For thermodynamic and field-like targets (\gas{}$\rightarrow$\temp{}, \gas{}$\rightarrow$\bfield{}), DDPM predictions better preserve global gradients, whereas GANs sometimes sharpen local contrast and slightly overemphasize smaller map features.

The arguably most challenging inverse mappings (e.g., \dm{}$\rightarrow$\gas{}) reveal residual artefacts and misaligned substructures for both models, underscoring the difficulty of inferring baryonic components from DM alone.

Similarly, both models struggle to faithfully reproduce translations involving the weakly correlated components \gas{}$\rightarrow$\stars{}.
Samples from this task exhibit noticeable deviations: predicted stellar maps fail to capture the clumpy, centrally concentrated structures, reflecting the intrinsic (temporal) non-locality and higher entropy of the stellar distribution.

Overall, these examples illustrate that translation quality correlates strongly with the physical coupling between source and target domain, and that GAN models and DDPM reproduce very similar samples and differ mostly in the details and high-frequency features: GANs often excel in structural sharpness for tightly coupled mappings, whereas DDPMs better maintain global coherence in more weakly constrained tasks.

\begin{figure*}
  \caption{
    Samples from various models and tasks.
    Each panel shows a model input map on the left, the corresponding ground truth in the middle, and prediction on the right.
    Qualitative comparison confirms the alignment of astrophysical plausibility and human perception with astrophysical metrics and FID (see Tables~\ref{tab:map_to_map_results_pixelwise}~and~\ref{tab:map_to_map_results_astroph}).
  }\label{fig:samples}
  \begin{center}

    \begin{subfigure}[t]{0.48\textwidth}
      \centering
      \caption{
        \centering\gas{}$\rightarrow$\dm{}: GAN-inferred samples.\\
        \hspace{4.2em} input \hfill ground truth \hfill prediction \hspace{3.8em}
      }\label{fig:gas_dm_gan_samples}
      \includegraphics[width=\textwidth]{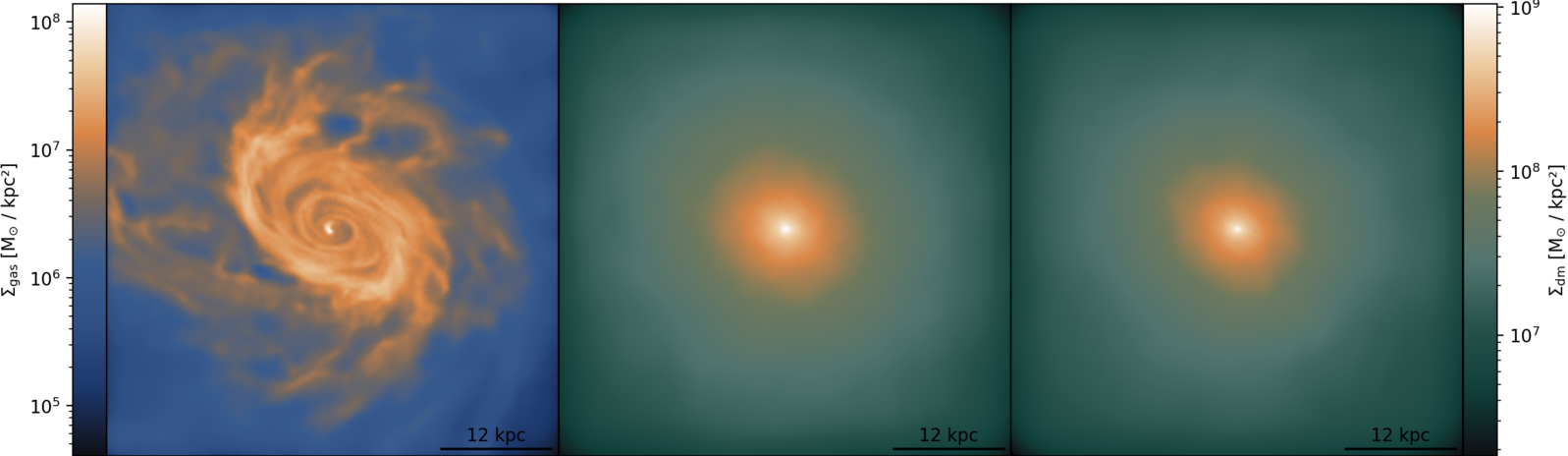}\\
      \includegraphics[width=\textwidth]{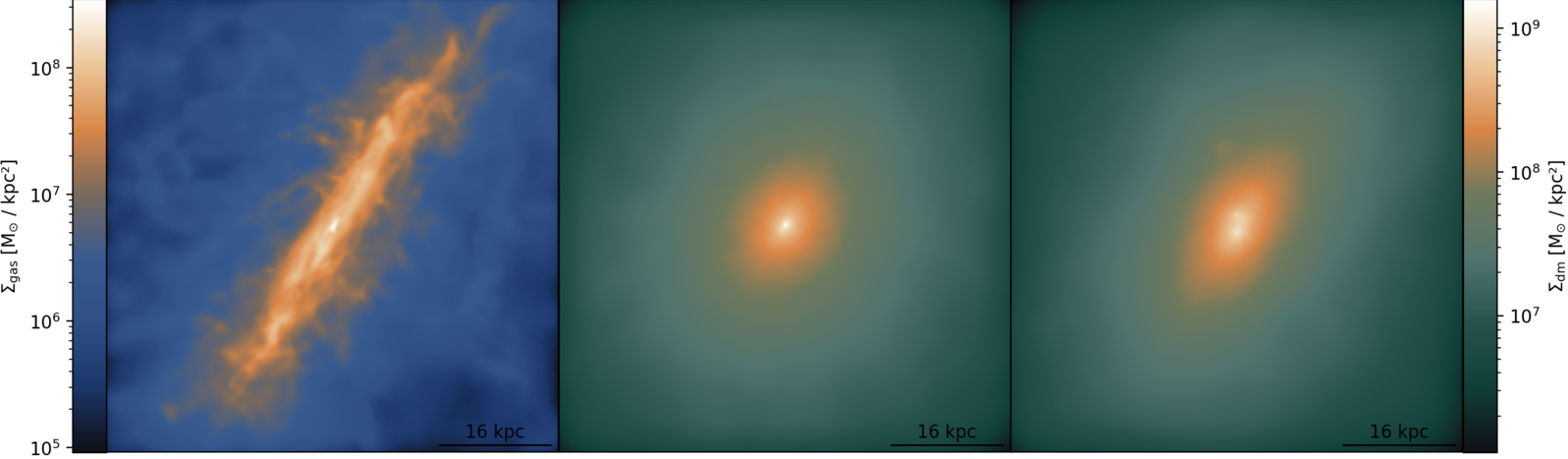}\\
      \includegraphics[width=\textwidth]{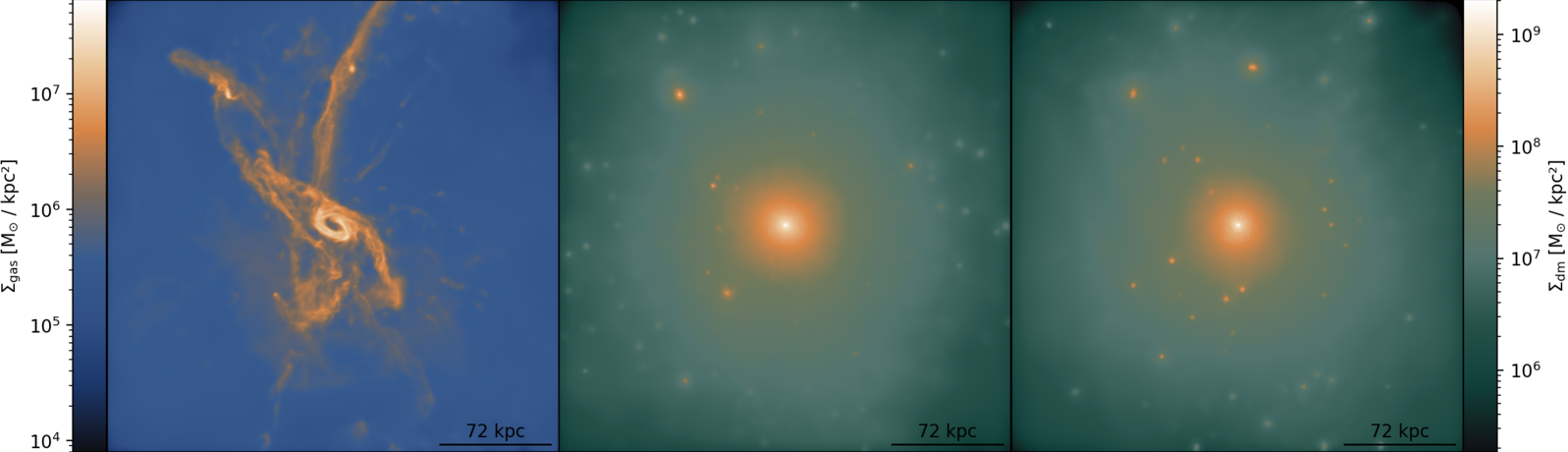}\\
      \includegraphics[width=\textwidth]{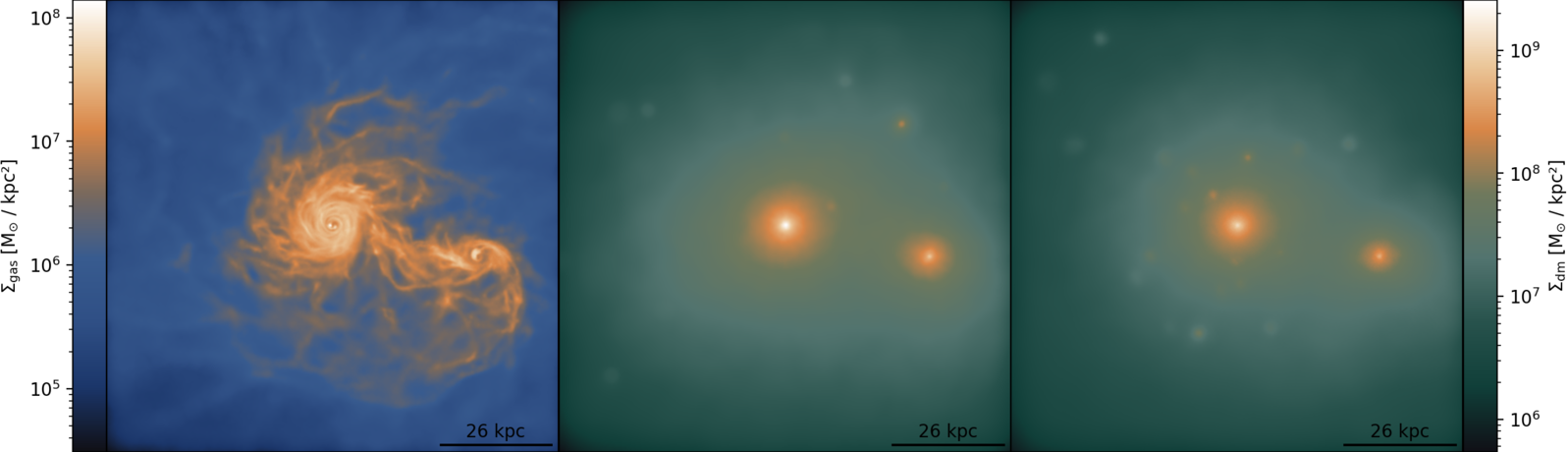}
    \end{subfigure}
    \begin{subfigure}[t]{0.48\textwidth}
      \centering
      \caption{\centering\gas{}$\rightarrow$\dm{}: DDPM-inferred samples.\\
        \hspace{4.2em} input \hfill ground truth \hfill prediction \hspace{3.8em}
      }\label{fig:gas_dm_ddpm_samples}
      \includegraphics[width=\textwidth]{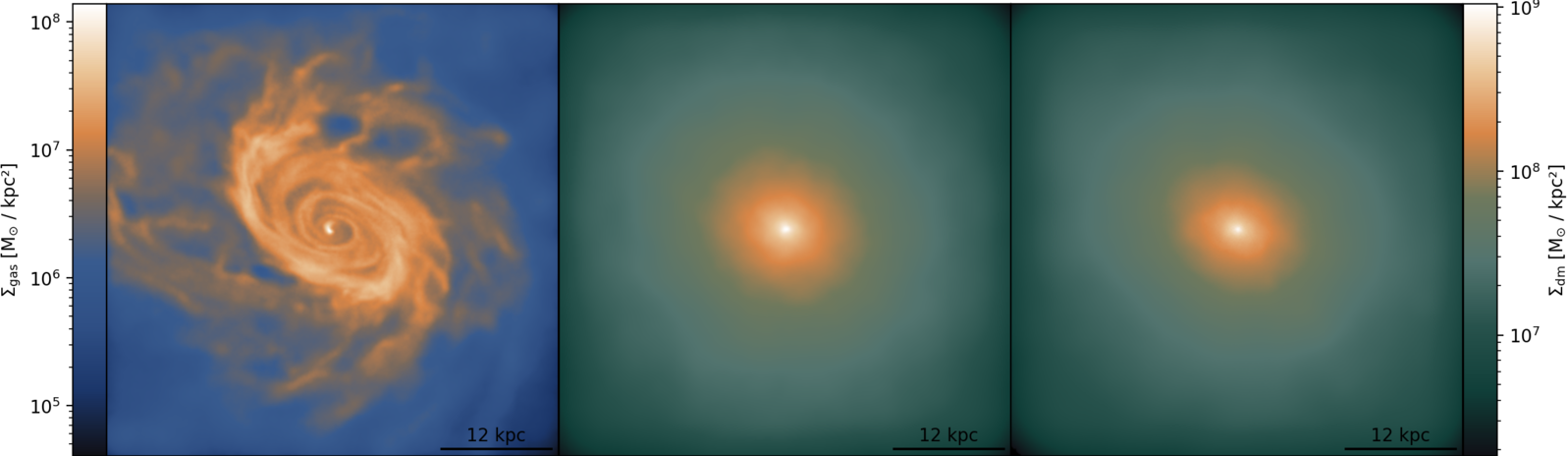}\\
      \includegraphics[width=\textwidth]{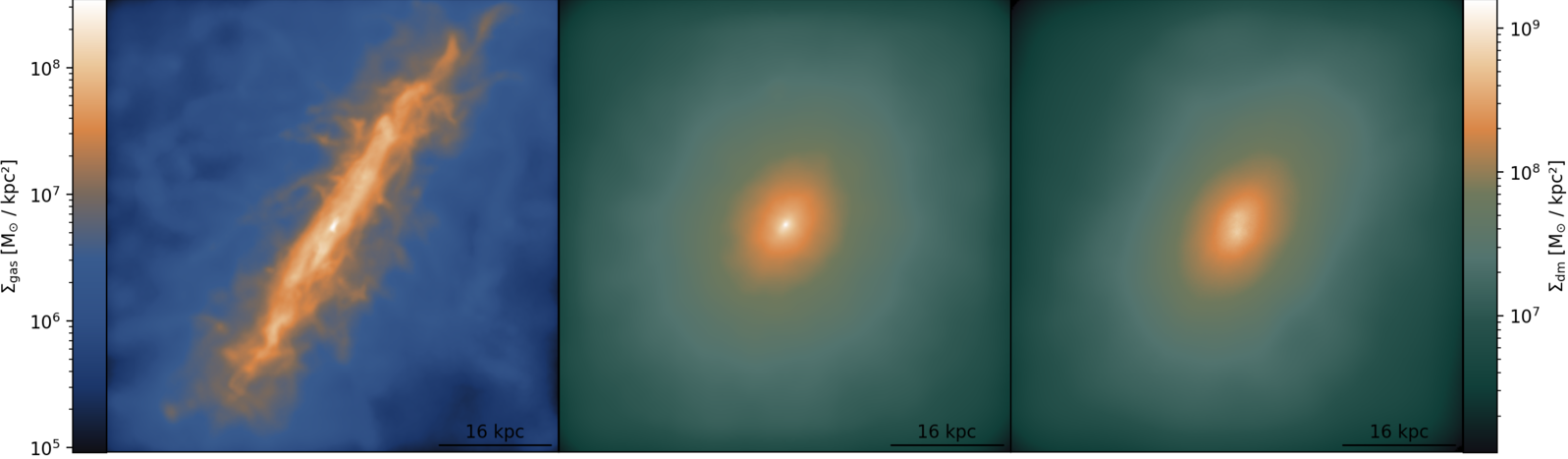}\\
      \includegraphics[width=\textwidth]{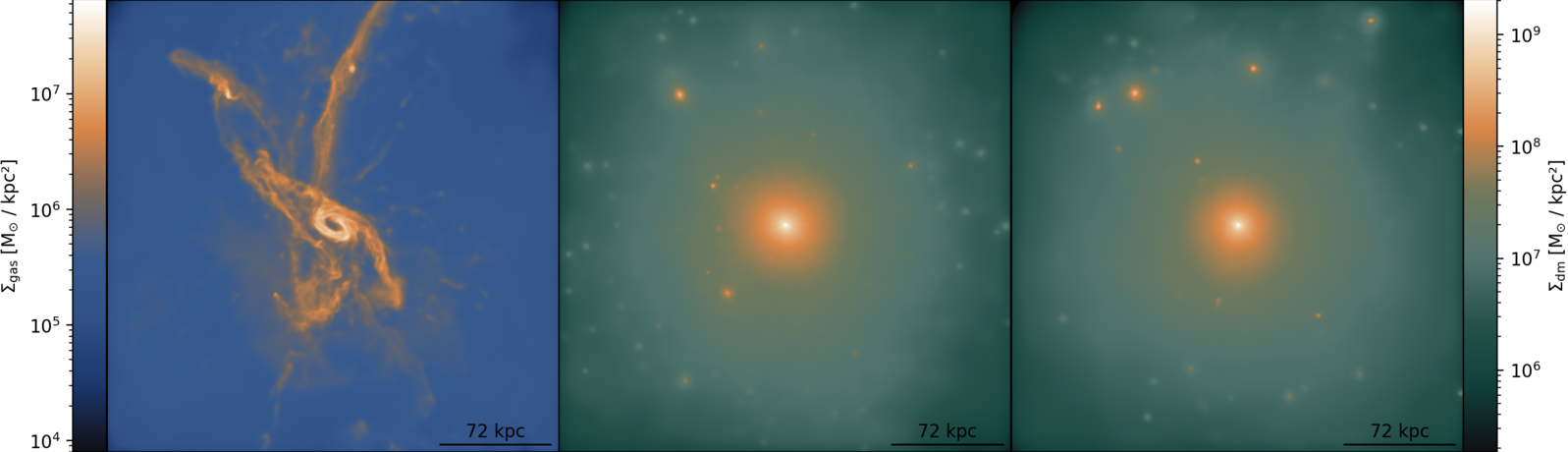}\\
      \includegraphics[width=\textwidth]{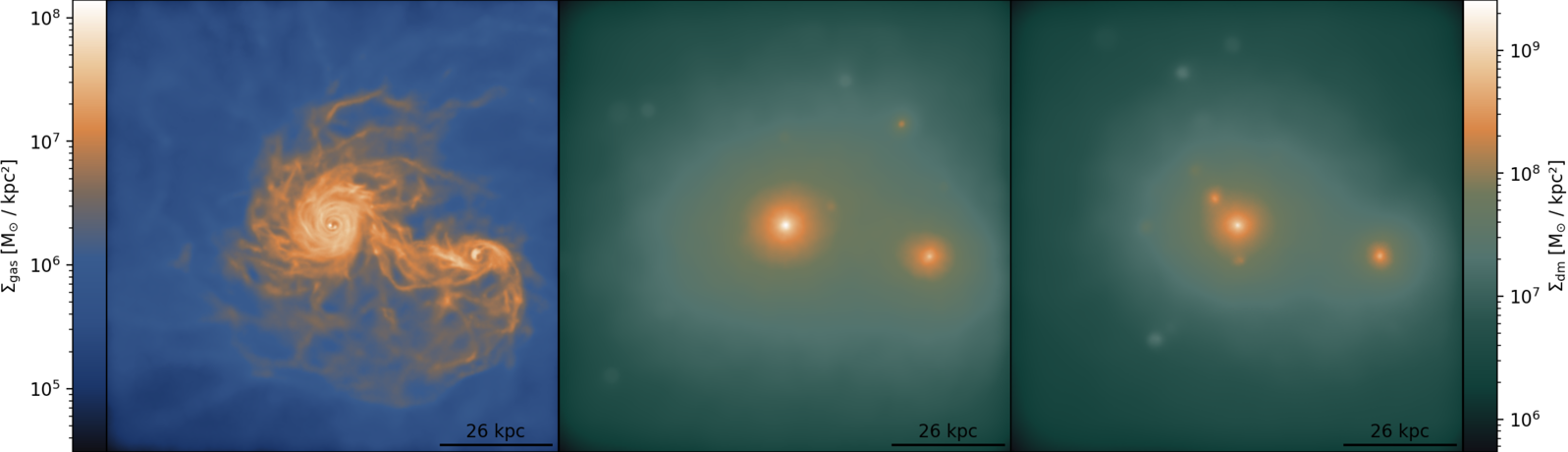}
    \end{subfigure}

    \vspace{2em}

    \begin{subfigure}[t]{0.48\textwidth}
      \centering
      \caption{\centering\gas{}$\rightarrow$\stars{}: GAN-inferred samples.\\
        \hspace{4.2em} input \hfill ground truth \hfill prediction \hspace{3.8em}
      }\label{fig:gas_star_gan_samples}
      \includegraphics[width=\textwidth]{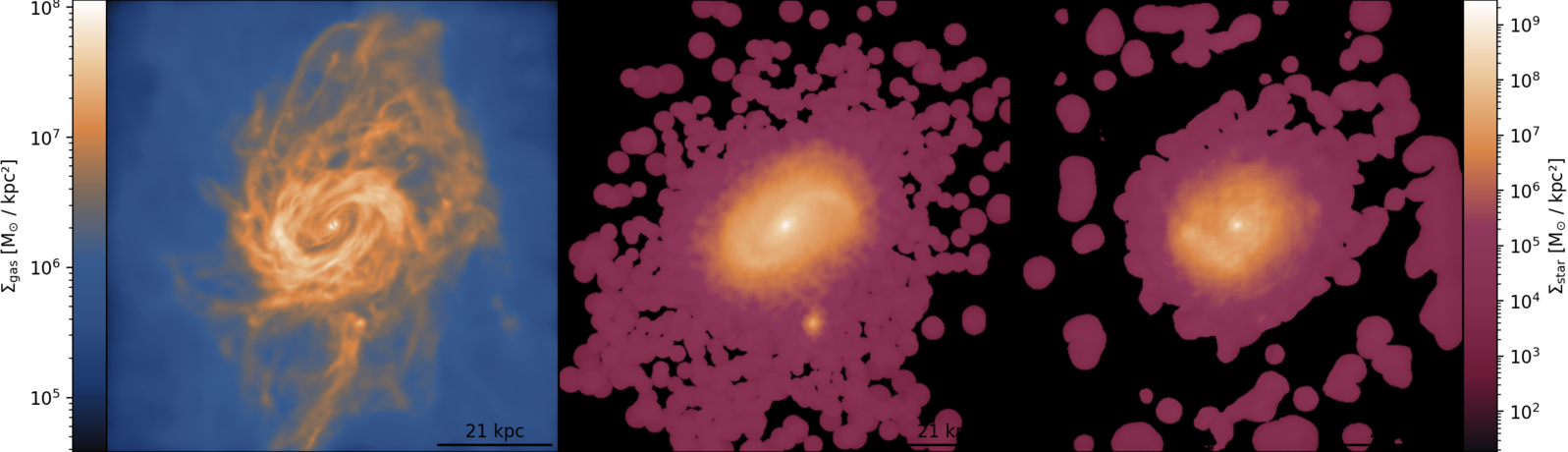}\\
      \includegraphics[width=\textwidth]{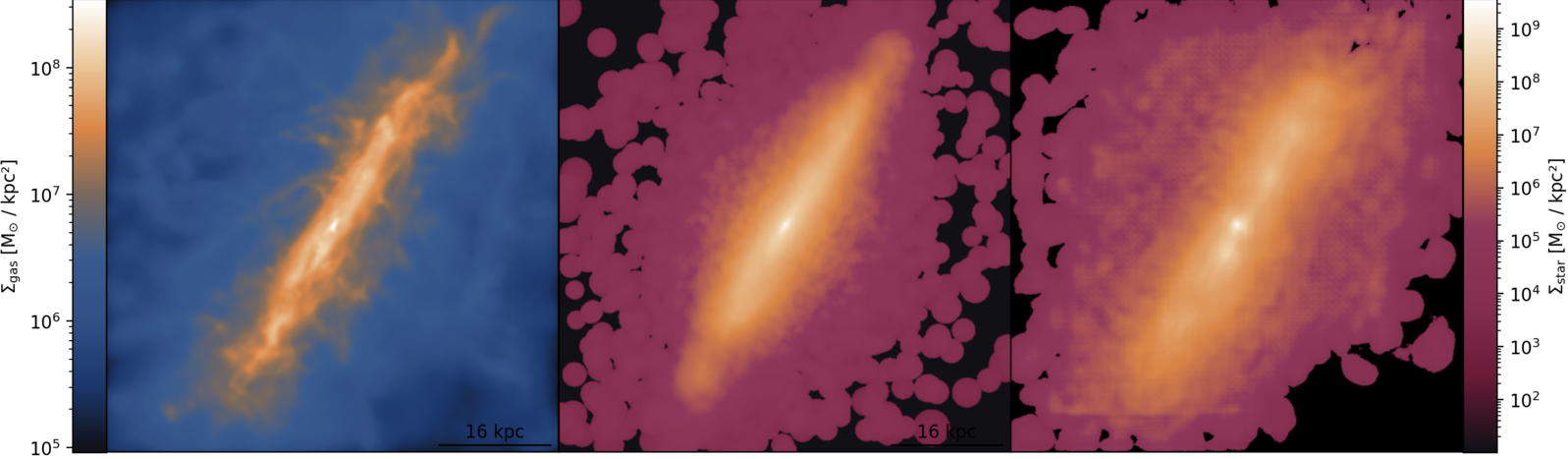}\\
      \includegraphics[width=\textwidth]{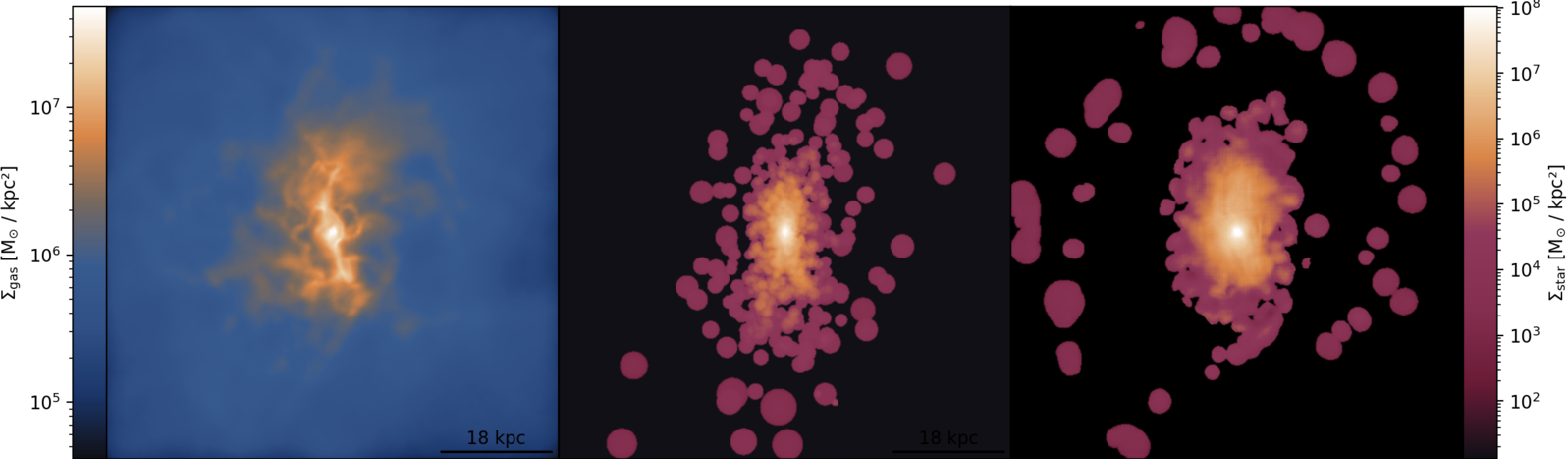}\\
      \includegraphics[width=\textwidth]{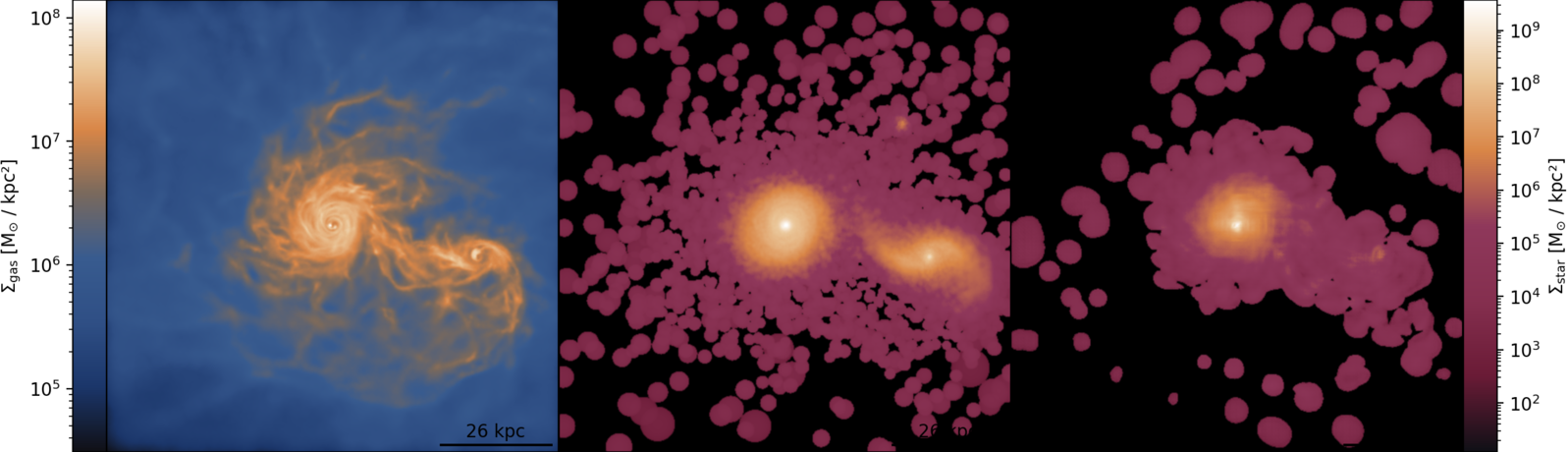}
    \end{subfigure}
    \begin{subfigure}[t]{0.48\textwidth}
      \centering
      \caption{\centering\gas{}$\rightarrow$\stars{}: DDPM-inferred samples.\\
        \hspace{4.2em} input \hfill ground truth \hfill prediction \hspace{3.8em}
      }\label{fig:gas_star_ddpm_samples}
      \includegraphics[width=\textwidth]{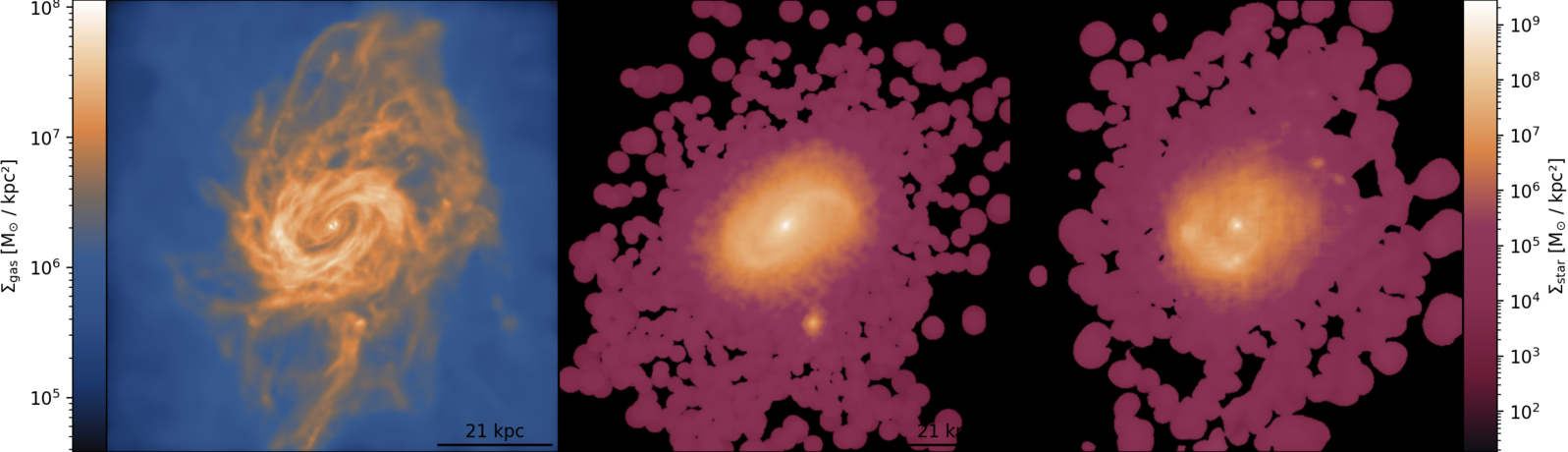}\\
      \includegraphics[width=\textwidth]{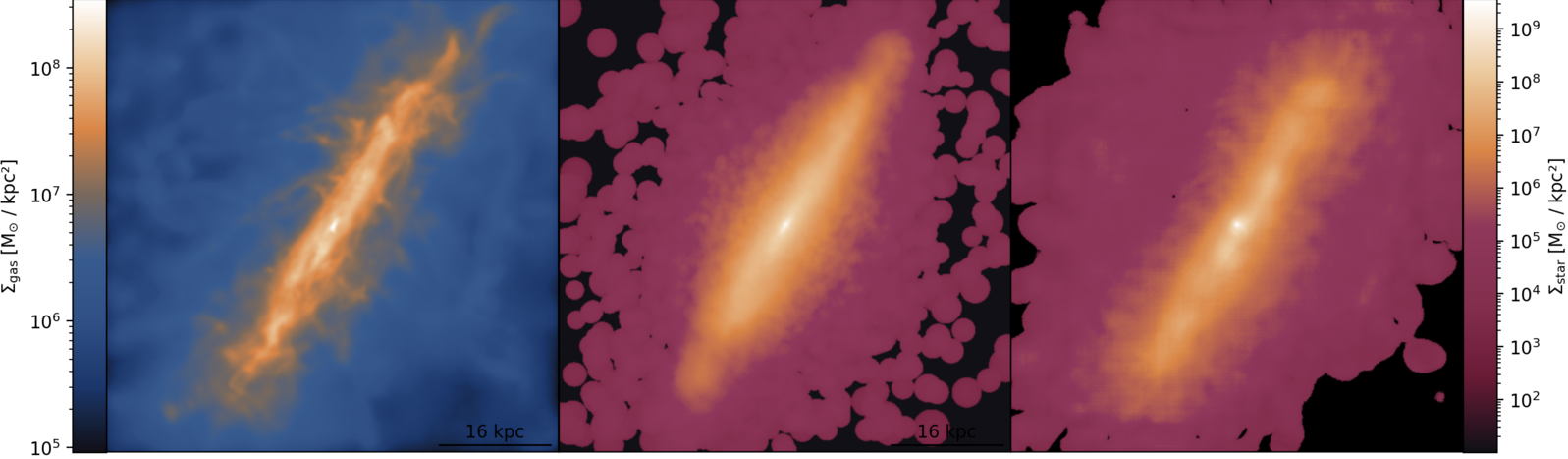}\\
      \includegraphics[width=\textwidth]{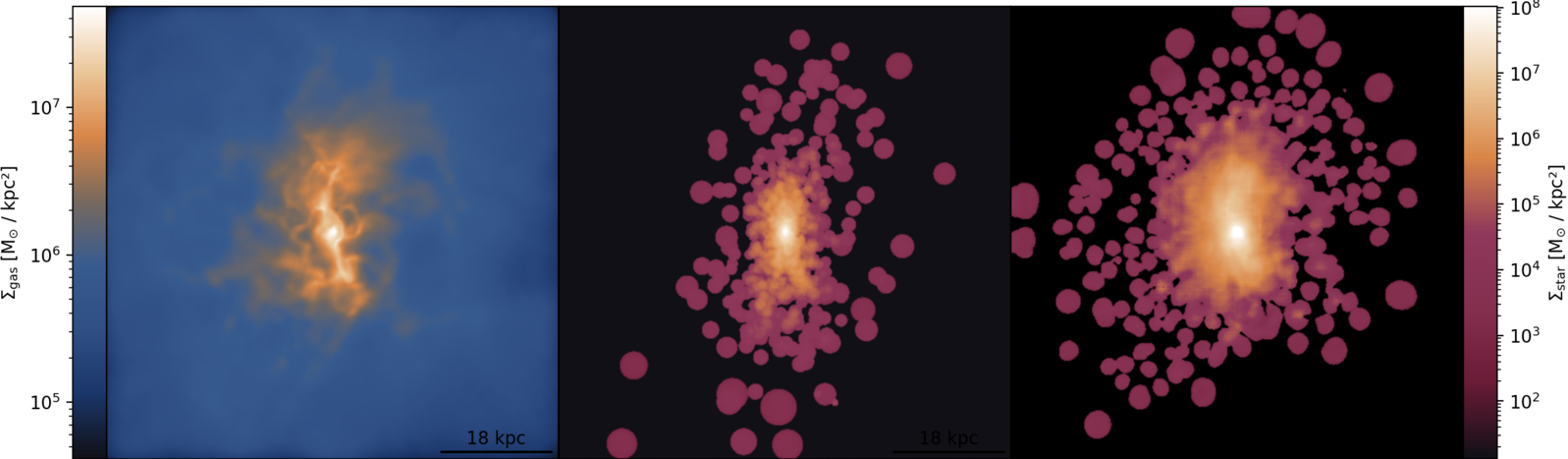}\\
      \includegraphics[width=\textwidth]{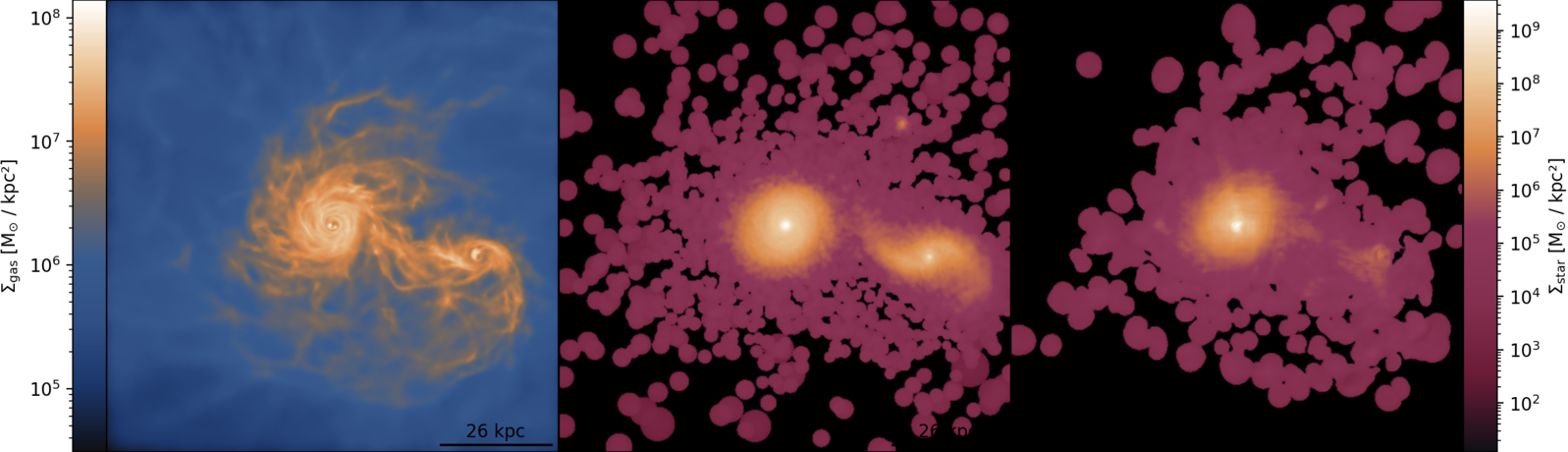}
    \end{subfigure}

  \end{center}
\end{figure*}

\begin{figure*}
  \ContinuedFloat
  \begin{center}
  
    \begin{subfigure}[t]{0.48\textwidth}
      \centering
      \caption{\centering\gas{}$\rightarrow$\hi{}: GAN-inferred samples.\\
        \hspace{4.2em} input \hfill ground truth \hfill prediction \hspace{3.8em}
      }\label{fig:gas_hi_gan_samples}
      \includegraphics[width=\textwidth]{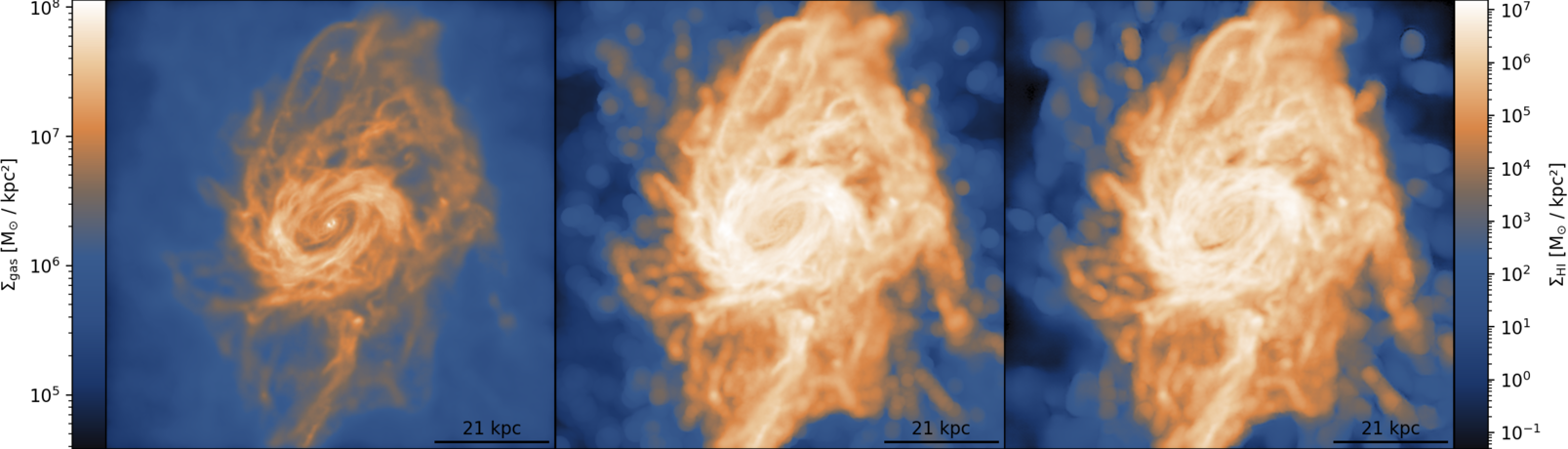}\\
      \includegraphics[width=\textwidth]{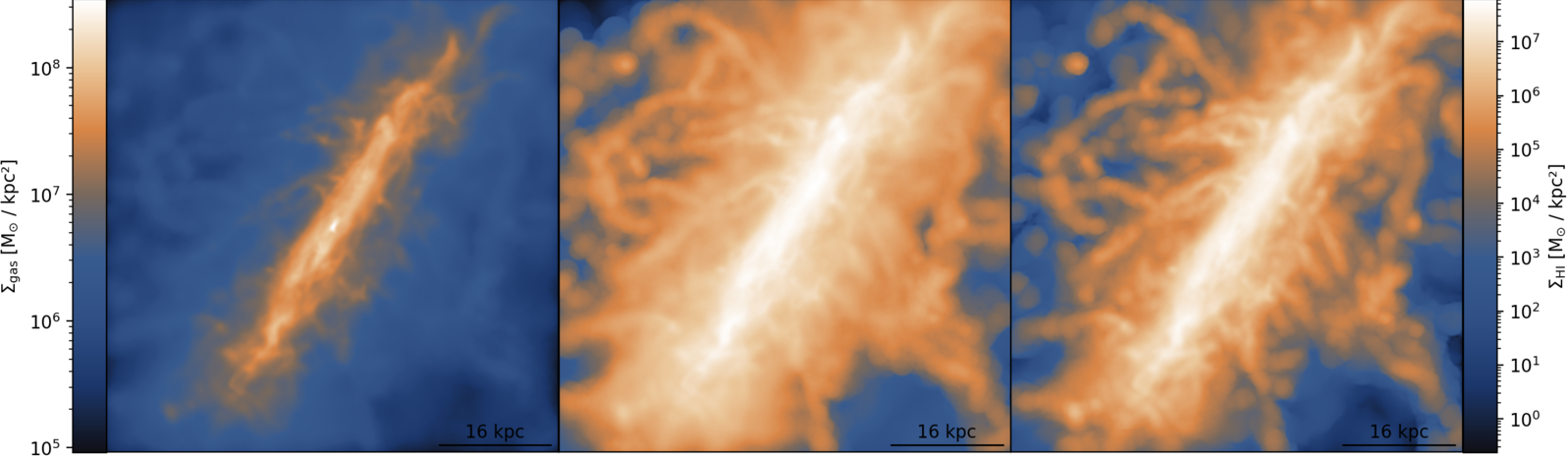}\\
      \includegraphics[width=\textwidth]{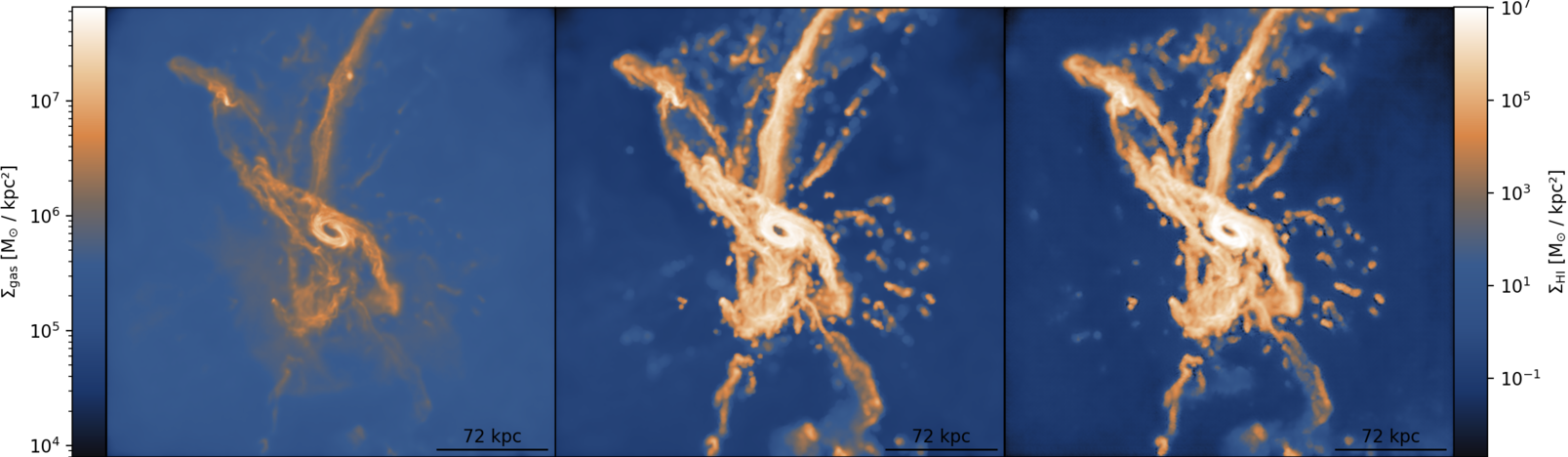}\\
      \includegraphics[width=\textwidth]{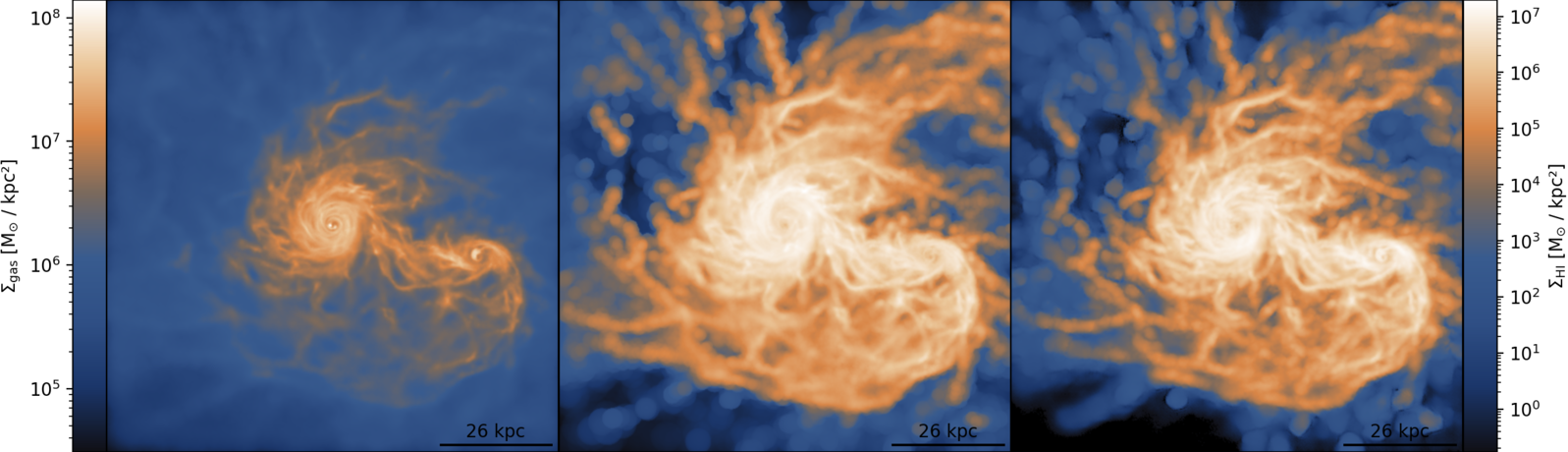}
    \end{subfigure}
    \begin{subfigure}[t]{0.48\textwidth}
      \centering
      \caption{\centering\gas{}$\rightarrow$\hi{}: DDPM-inferred samples.\\
        \hspace{4.2em} input \hfill ground truth \hfill prediction \hspace{3.8em}
      }\label{fig:gas_hi_ddpm_samples}
      \includegraphics[width=\textwidth]{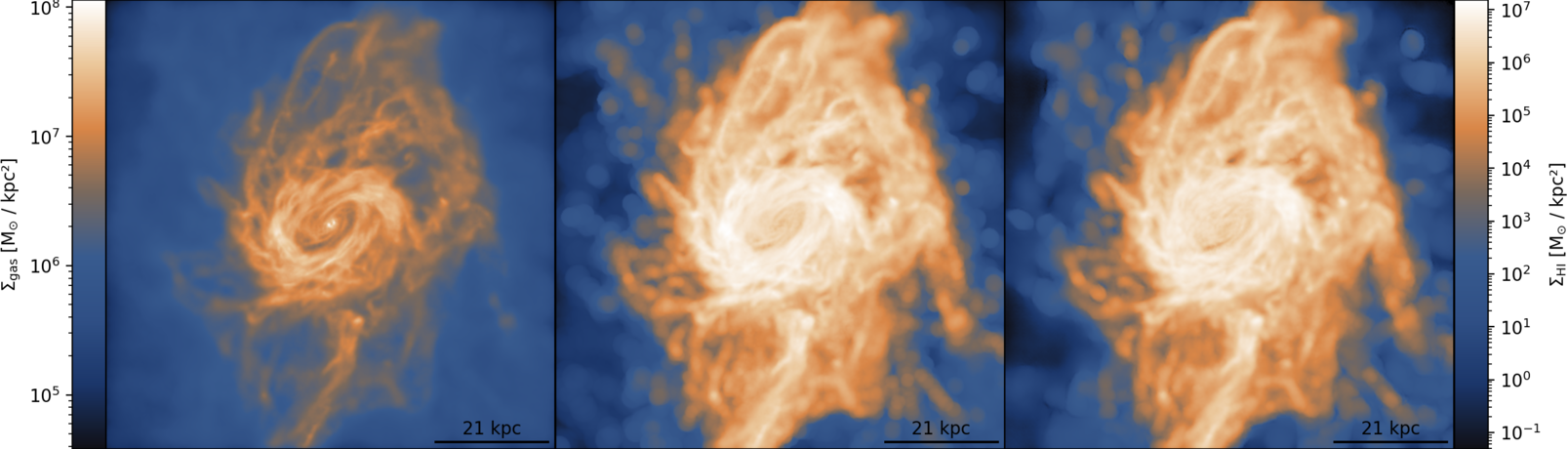}\\
      \includegraphics[width=\textwidth]{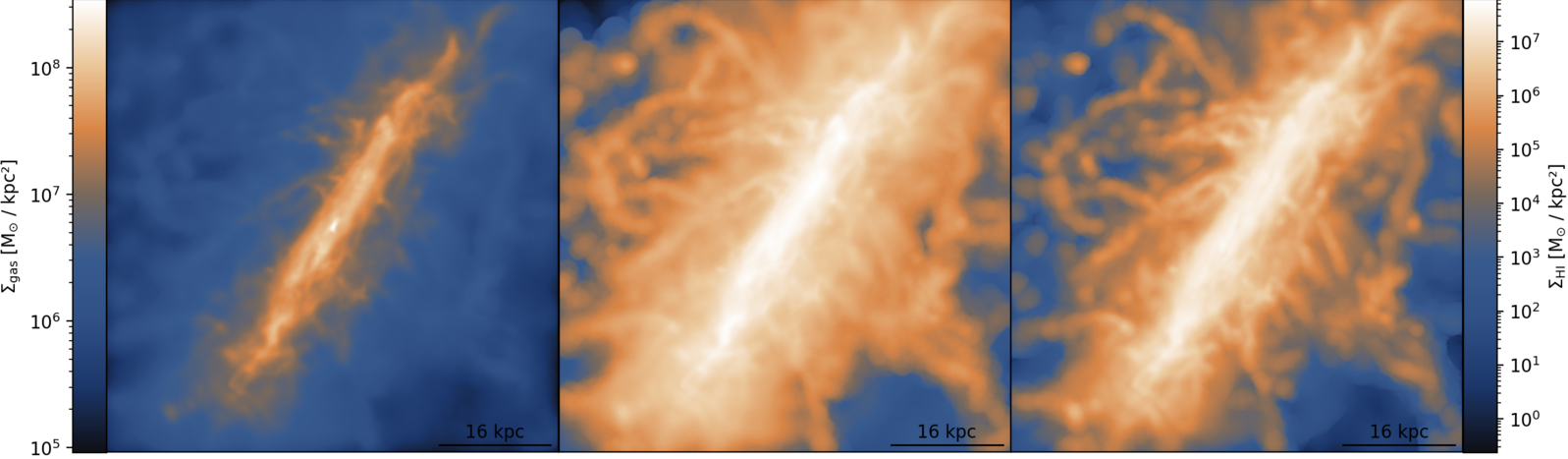}\\
      \includegraphics[width=\textwidth]{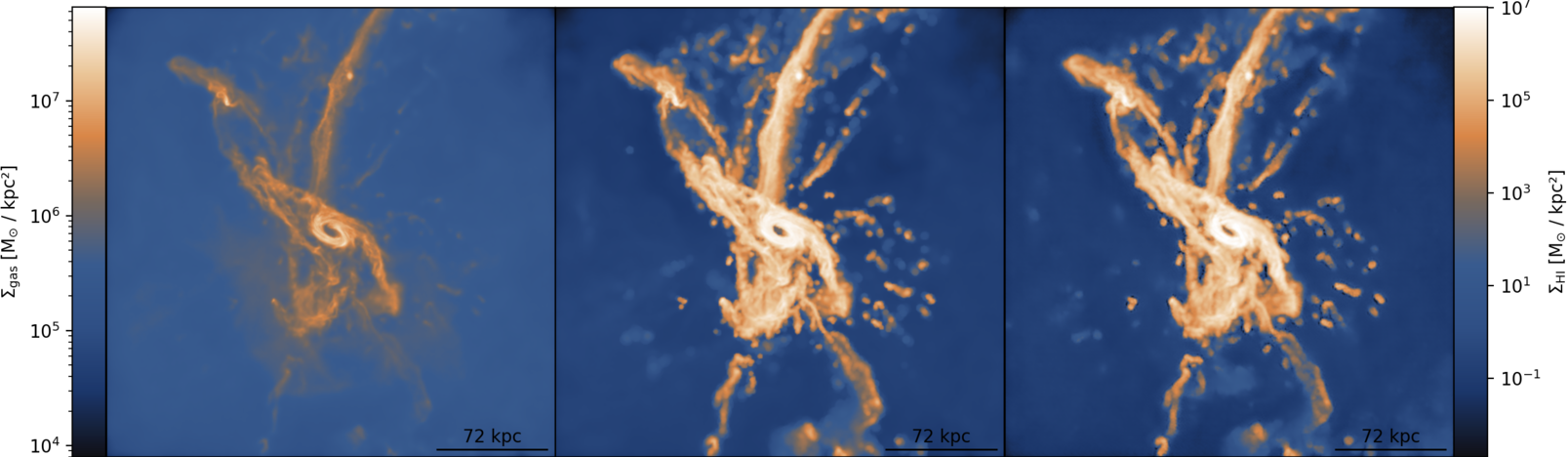}\\
      \includegraphics[width=\textwidth]{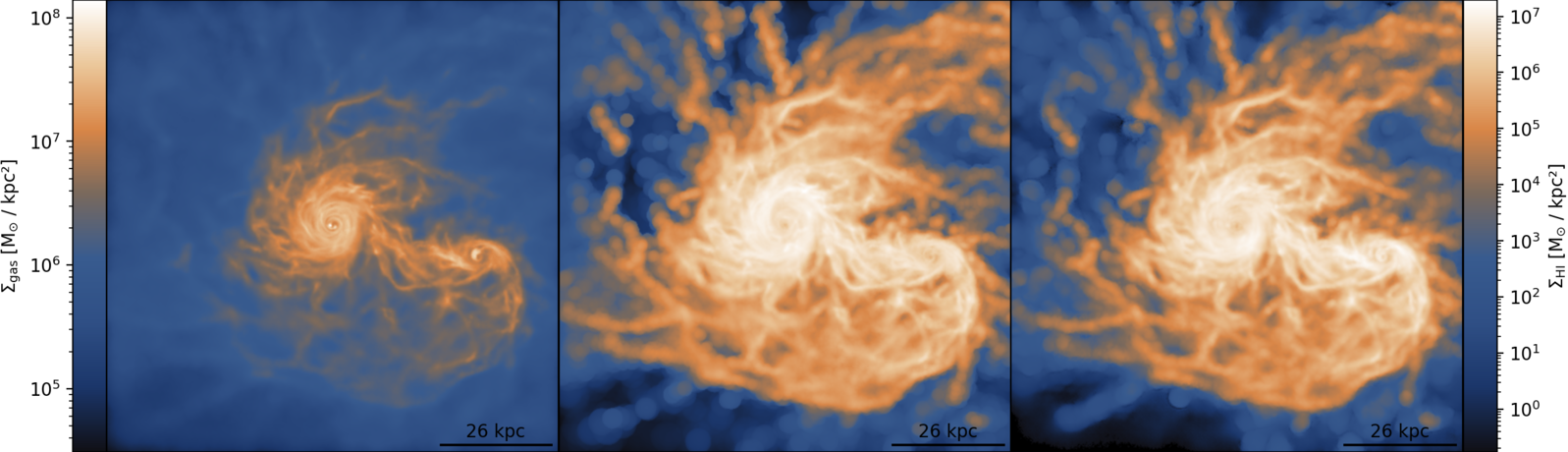}
    \end{subfigure}

    \vspace{2em}

    \begin{subfigure}[t]{0.48\textwidth}
      \centering
      \caption{\centering\gas{}$\rightarrow$\hicm{}: GAN-inferred samples.\\
        \hspace{4.2em} input \hfill ground truth \hfill prediction \hspace{3.8em}
      }\label{fig:gas_hicm_gan_samples}
      \includegraphics[width=\textwidth]{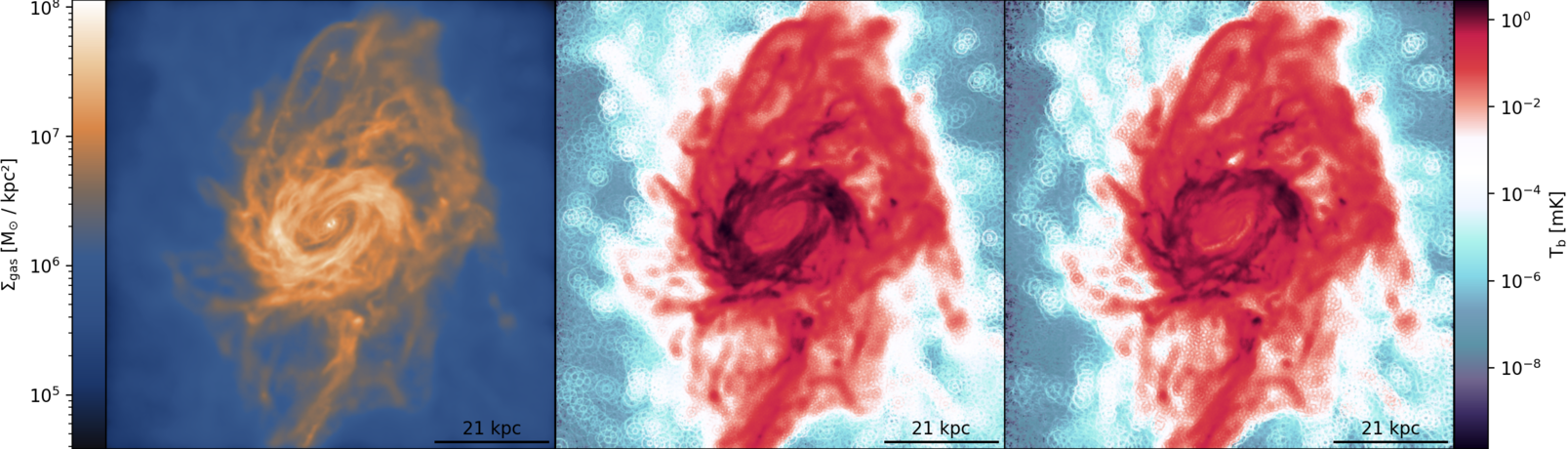}\\
      \includegraphics[width=\textwidth]{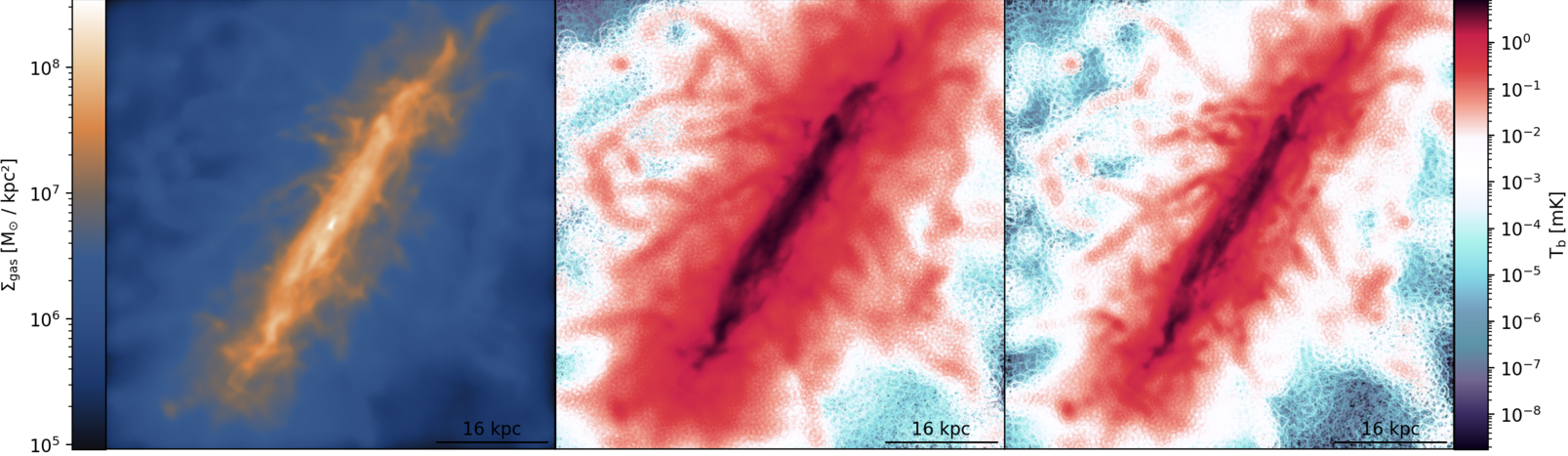}\\
      \includegraphics[width=\textwidth]{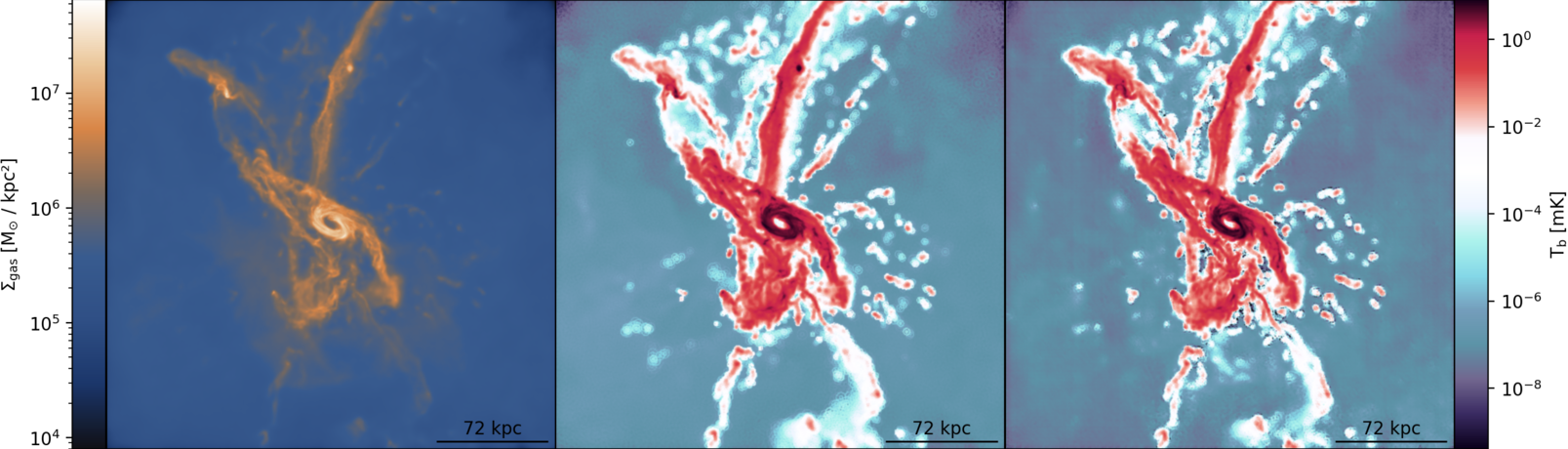}\\
      \includegraphics[width=\textwidth]{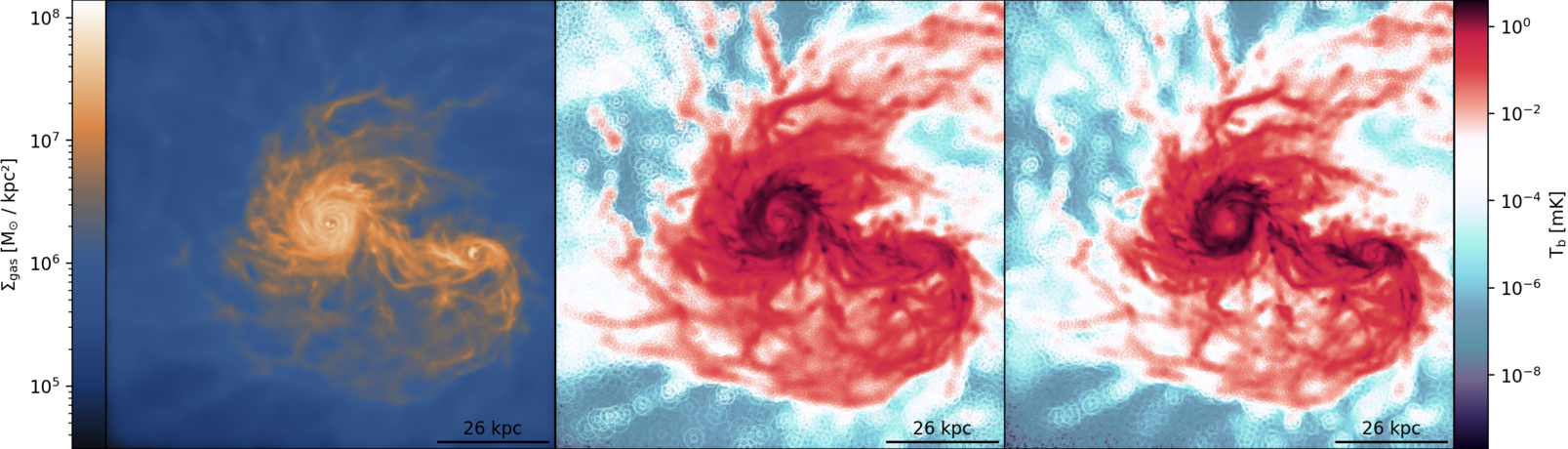}
    \end{subfigure}
    \begin{subfigure}[t]{0.48\textwidth}
      \centering
      \caption{\centering\gas{}$\rightarrow$\hicm{}: DDPM-inferred samples.\\
        \hspace{4.2em} input \hfill ground truth \hfill prediction \hspace{3.8em}
      }\label{fig:gas_hicm_ddpm_samples}
      \includegraphics[width=\textwidth]{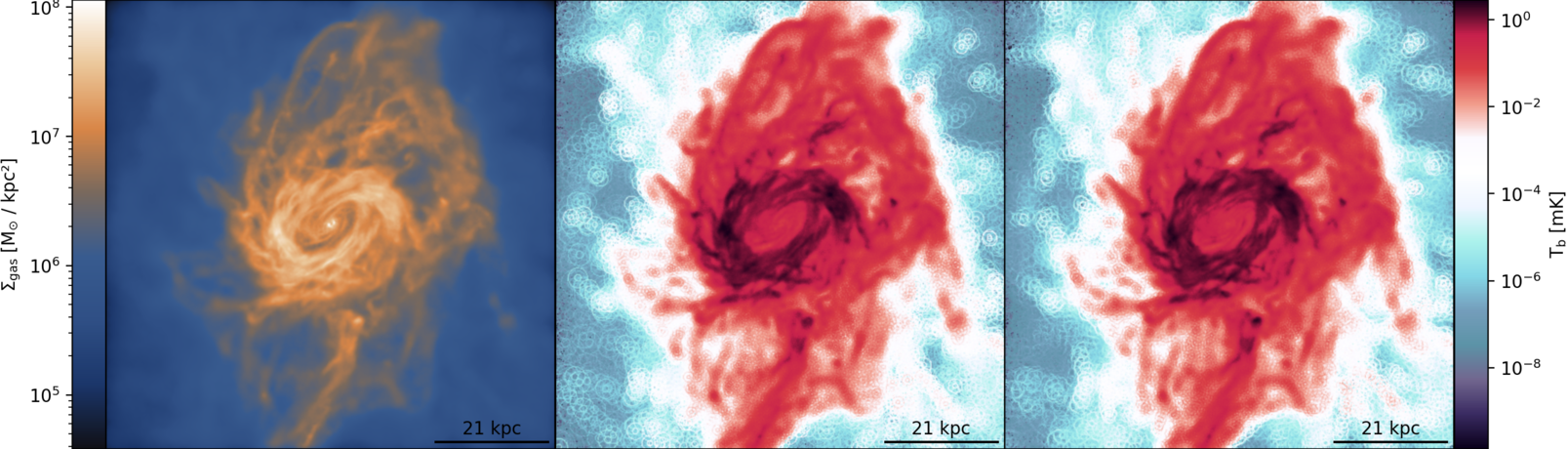}\\
      \includegraphics[width=\textwidth]{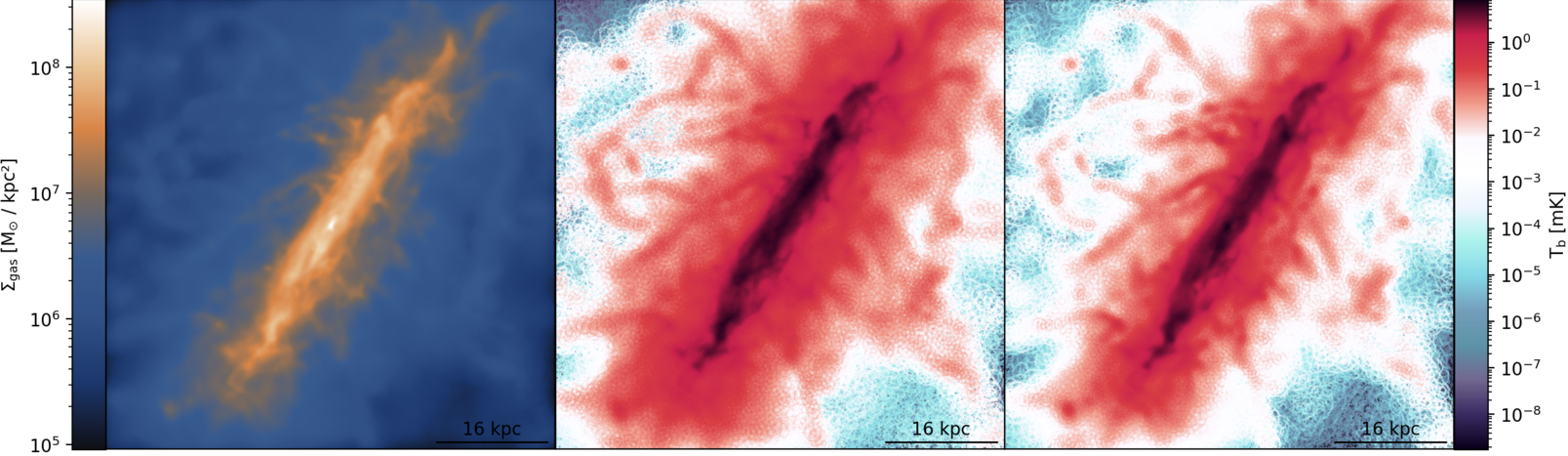}\\
      \includegraphics[width=\textwidth]{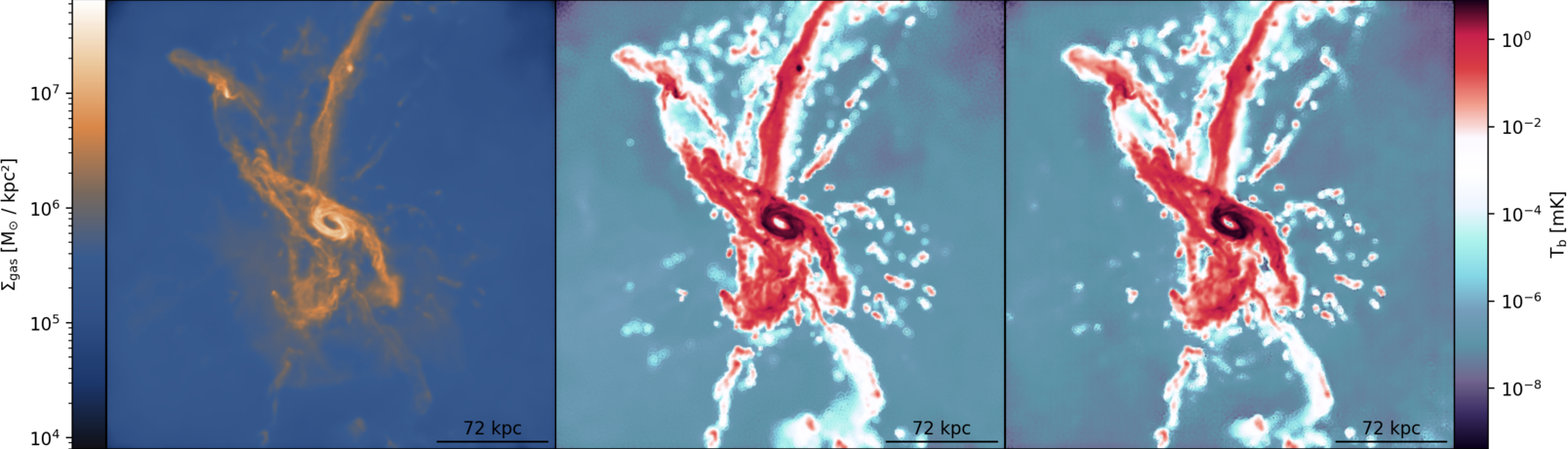}\\
      \includegraphics[width=\textwidth]{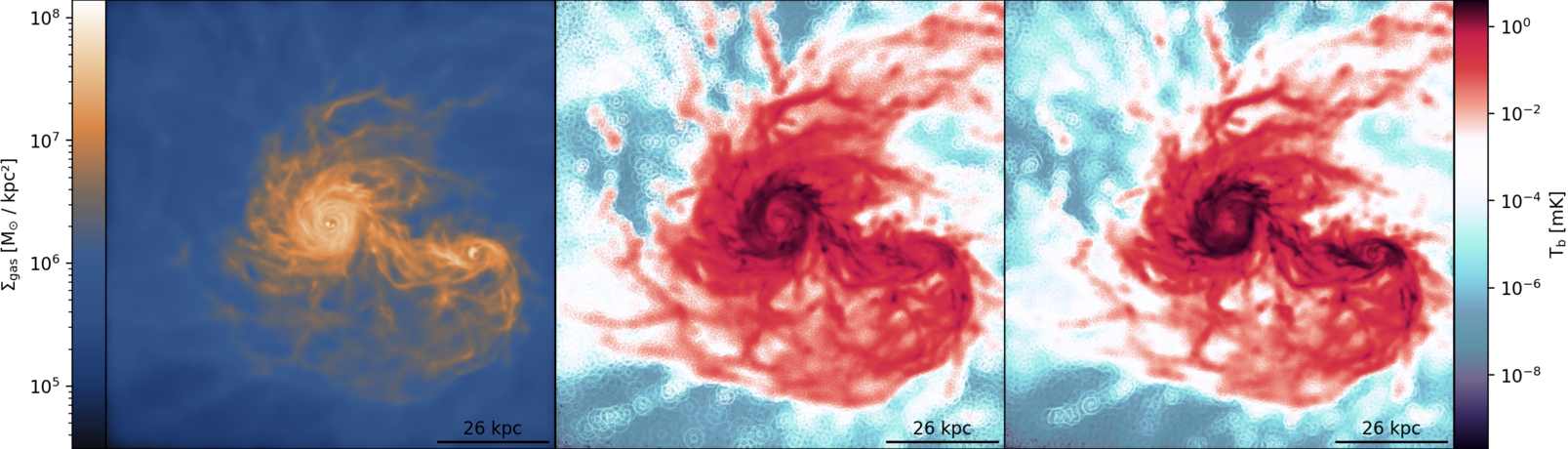}
    \end{subfigure}
    
  \end{center}
\end{figure*}

\begin{figure*}
  \ContinuedFloat
  \begin{center}

    \begin{subfigure}[t]{0.48\textwidth}
      \centering
      \caption{\centering\gas{}$\rightarrow$\temp{}: GAN-inferred samples.\\
        \hspace{4.2em} input \hfill ground truth \hfill prediction \hspace{3.8em}
      }\label{fig:gas_temp_gan_samples}
      \includegraphics[width=\textwidth]{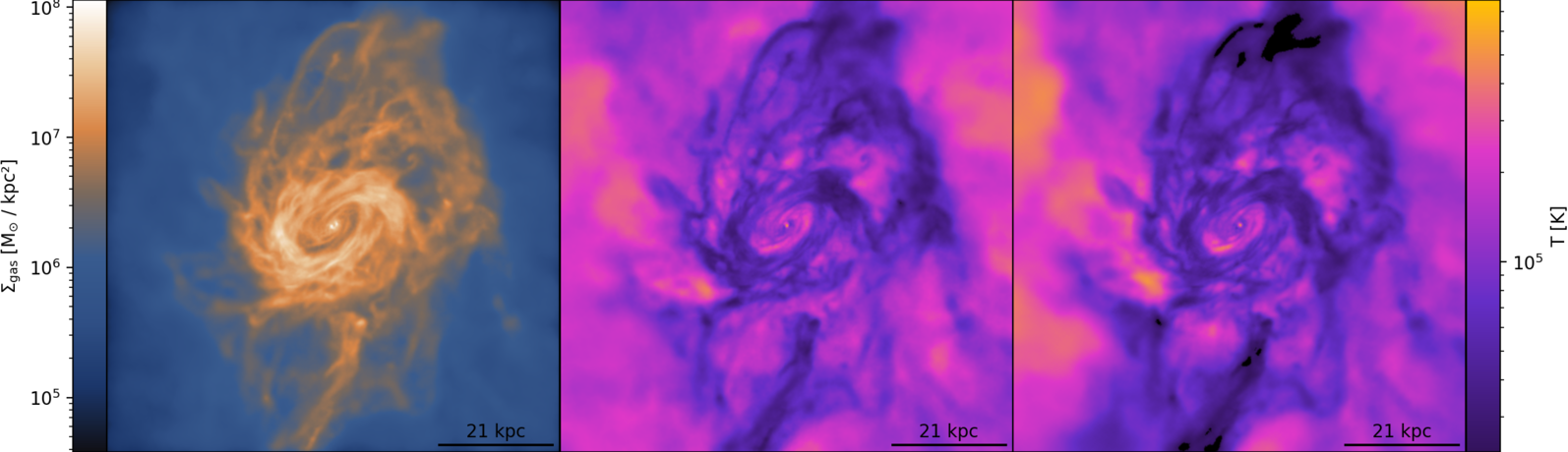}\\
      \includegraphics[width=\textwidth]{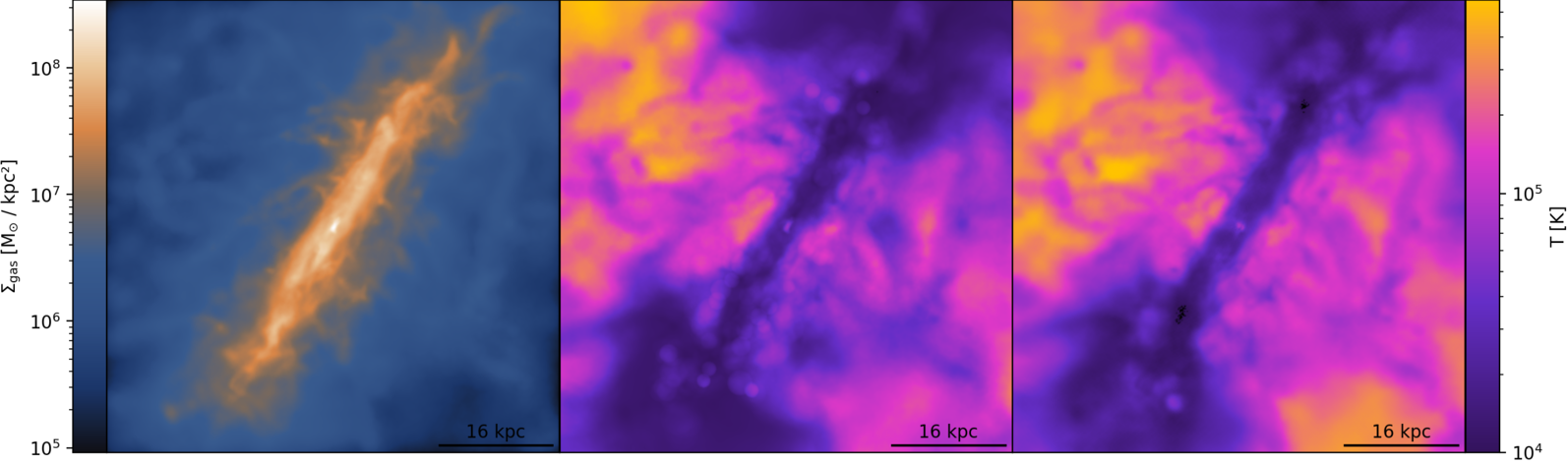}\\
      \includegraphics[width=\textwidth]{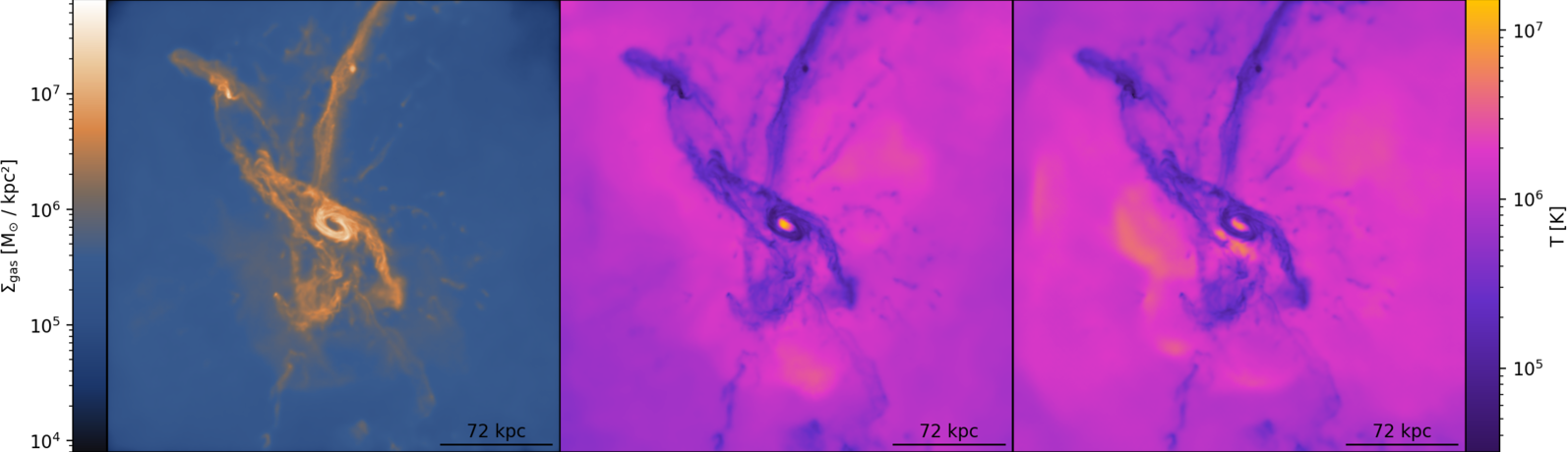}\\
      \includegraphics[width=\textwidth]{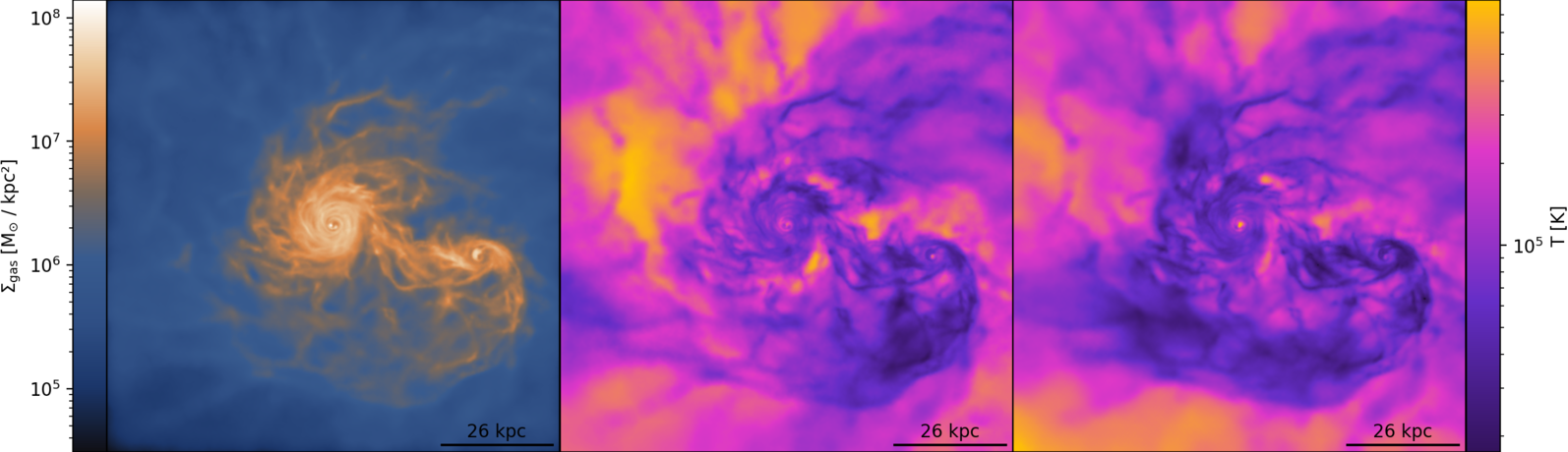}
    \end{subfigure}
    \begin{subfigure}[t]{0.48\textwidth}
      \centering
      \caption{\centering\gas{}$\rightarrow$\temp{}: DDPM-inferred samples.\\
        \hspace{4.2em} input \hfill ground truth \hfill prediction \hspace{3.8em}
      }\label{fig:gas_temp_ddpm_samples}
      \includegraphics[width=\textwidth]{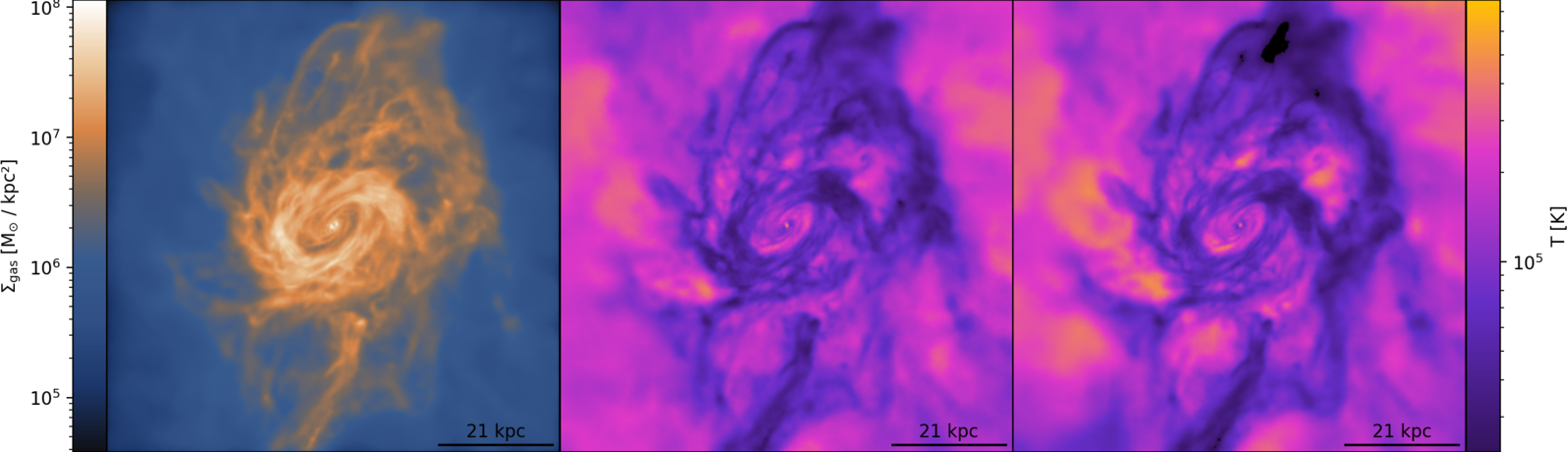}\\
      \includegraphics[width=\textwidth]{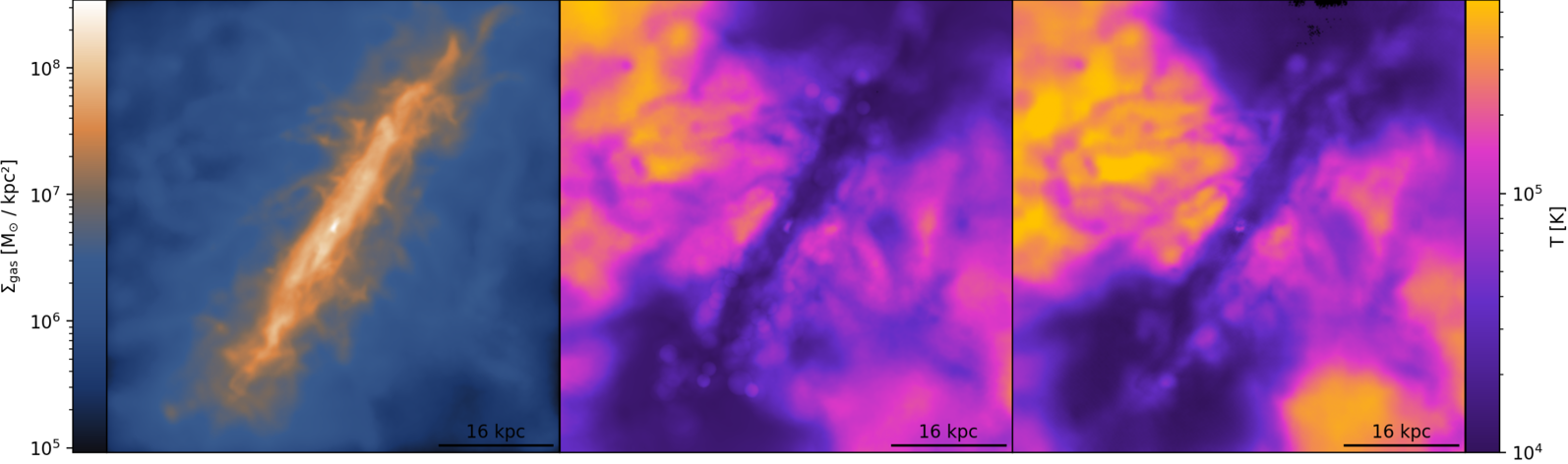}\\
      \includegraphics[width=\textwidth]{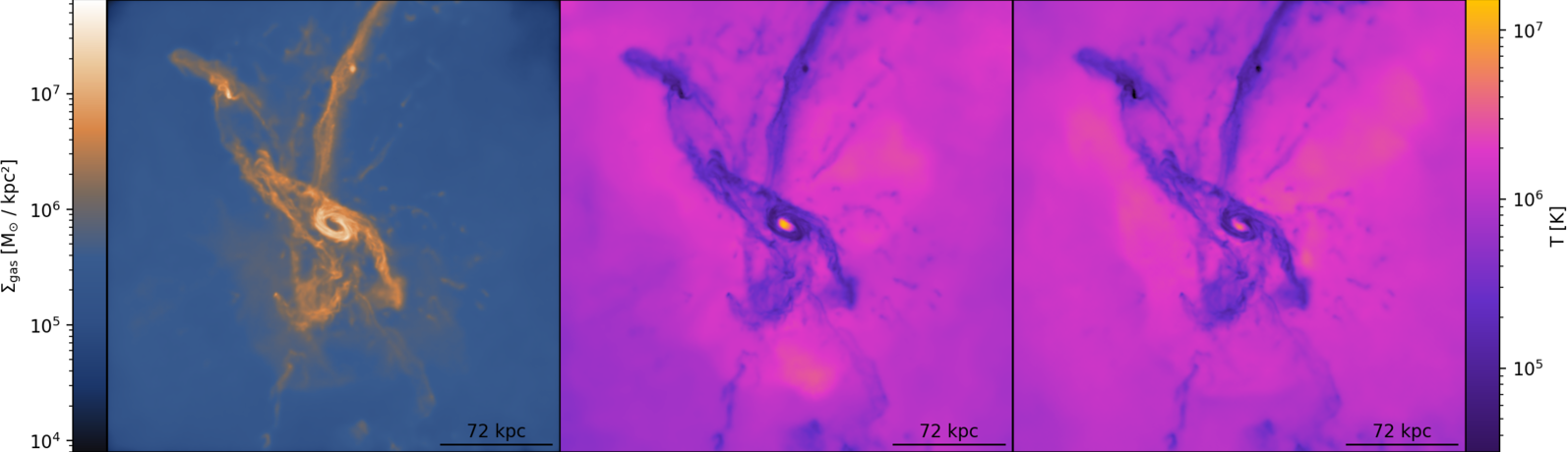}\\
      \includegraphics[width=\textwidth]{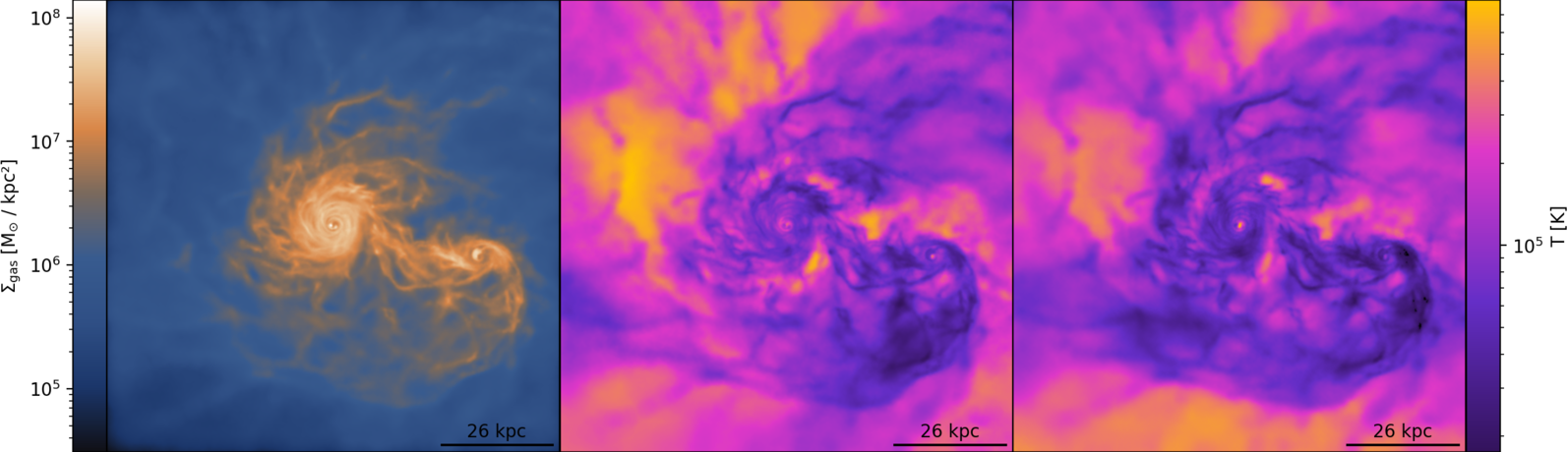}
    \end{subfigure}

    \vspace{2em}

    \begin{subfigure}[t]{0.48\textwidth}
      \centering
      \caption{\centering\gas{}$\rightarrow$\bfield{}: GAN-inferred samples.\\
        \hspace{4.2em} input \hfill ground truth \hfill prediction \hspace{3.8em}
      }\label{fig:gas_bfield_gan_samples}
      \includegraphics[width=\textwidth]{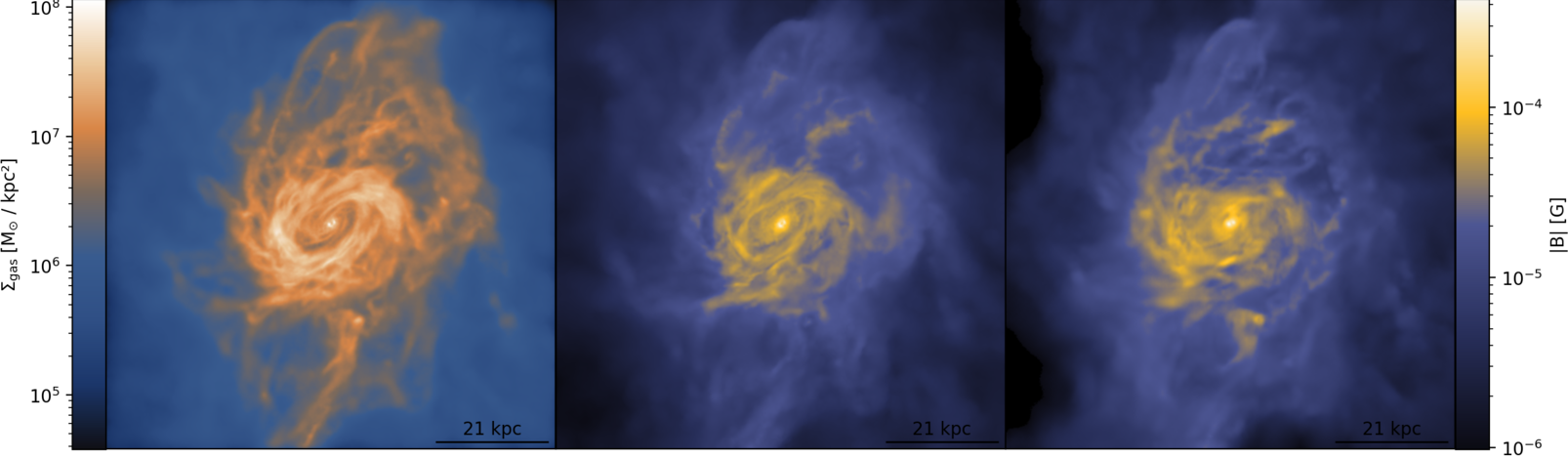}\\
      \includegraphics[width=\textwidth]{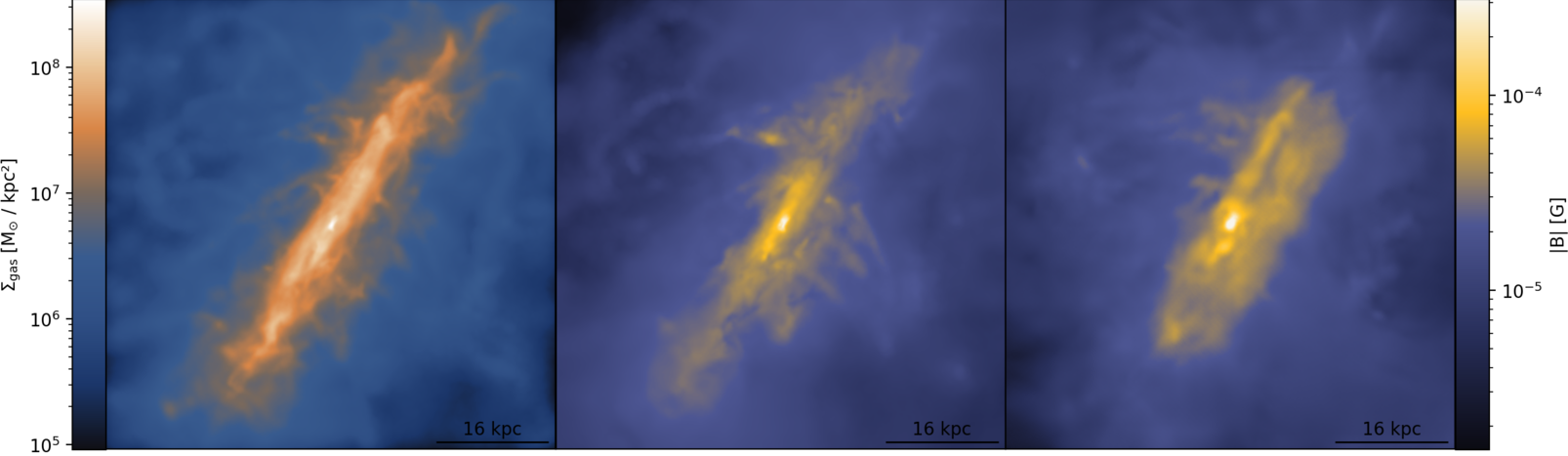}\\
      \includegraphics[width=\textwidth]{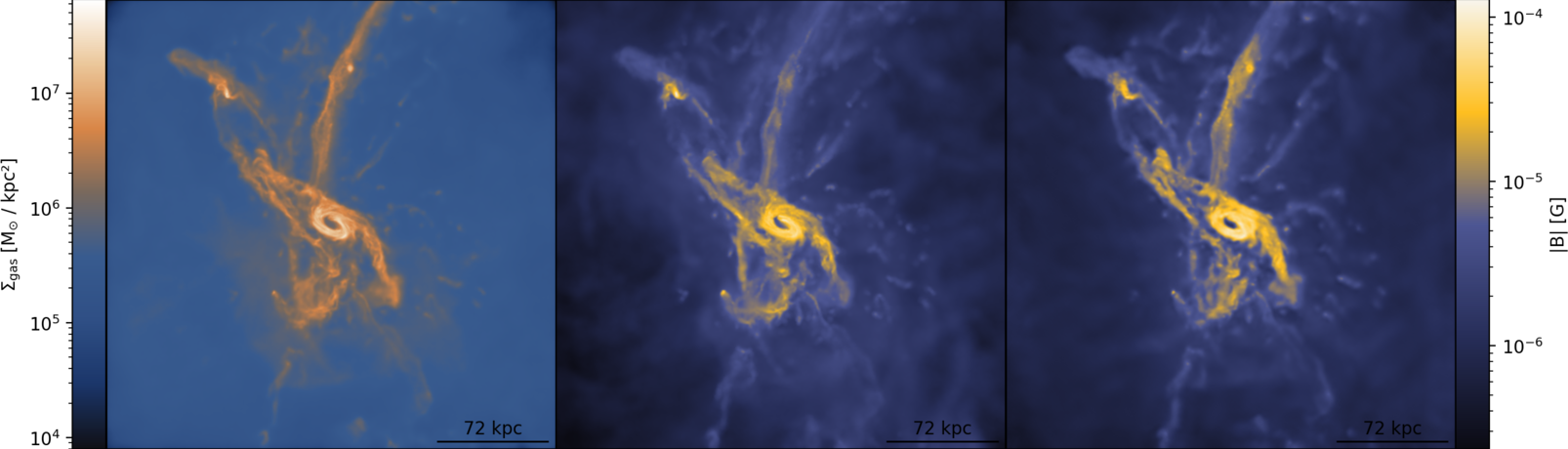}\\
      \includegraphics[width=\textwidth]{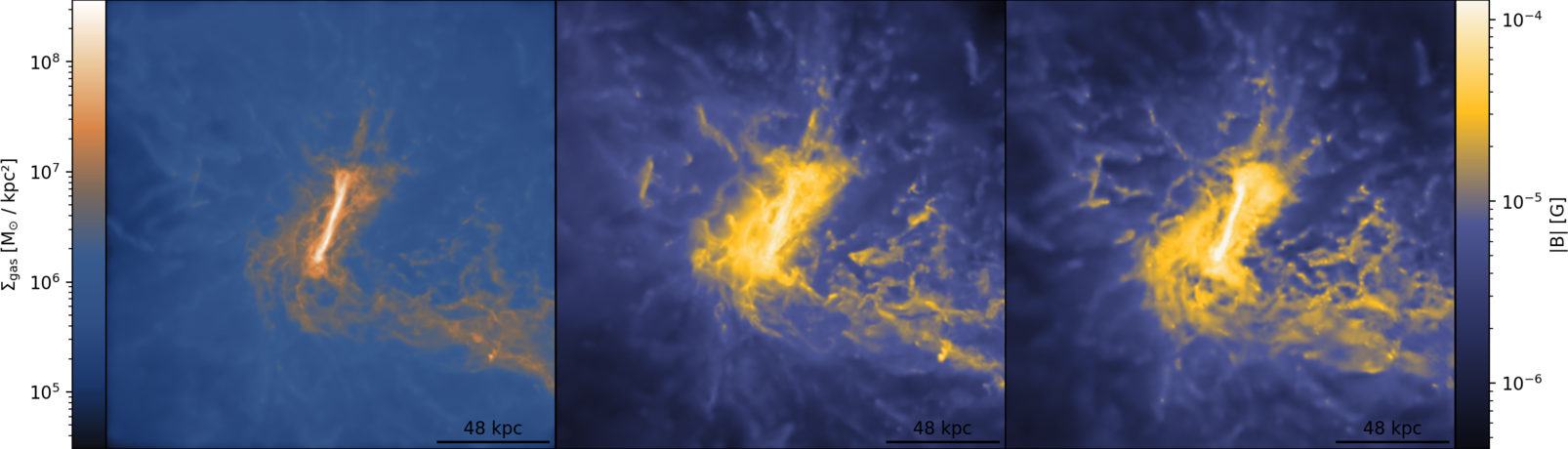}
    \end{subfigure}
    \begin{subfigure}[t]{0.48\textwidth}
      \centering
      \caption{\centering\gas{}$\rightarrow$\bfield{}: DDPM-inferred samples.\\
        \hspace{4.2em} input \hfill ground truth \hfill prediction \hspace{3.8em}
      }\label{fig:gas_bfield_ddpm_samples}
      \includegraphics[width=\textwidth]{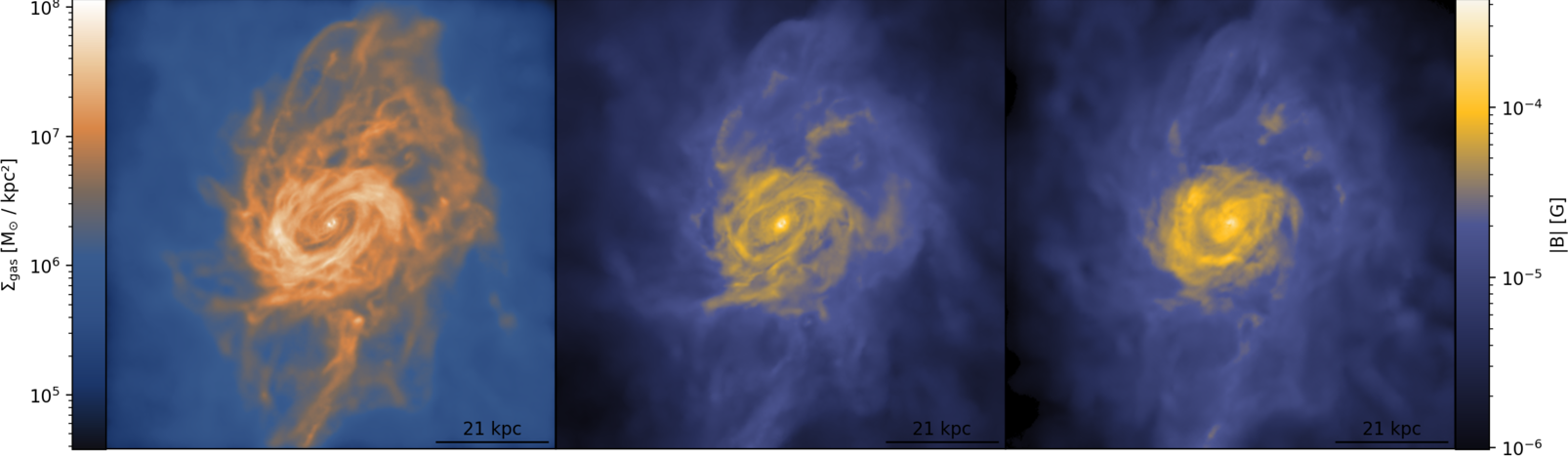}\\
      \includegraphics[width=\textwidth]{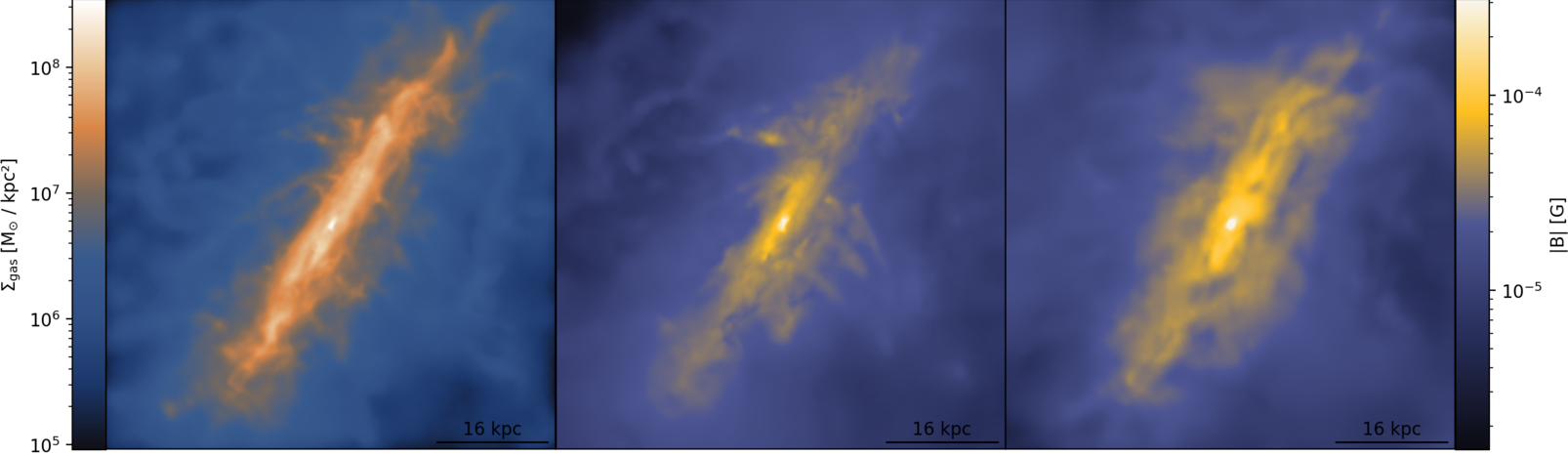}\\
      \includegraphics[width=\textwidth]{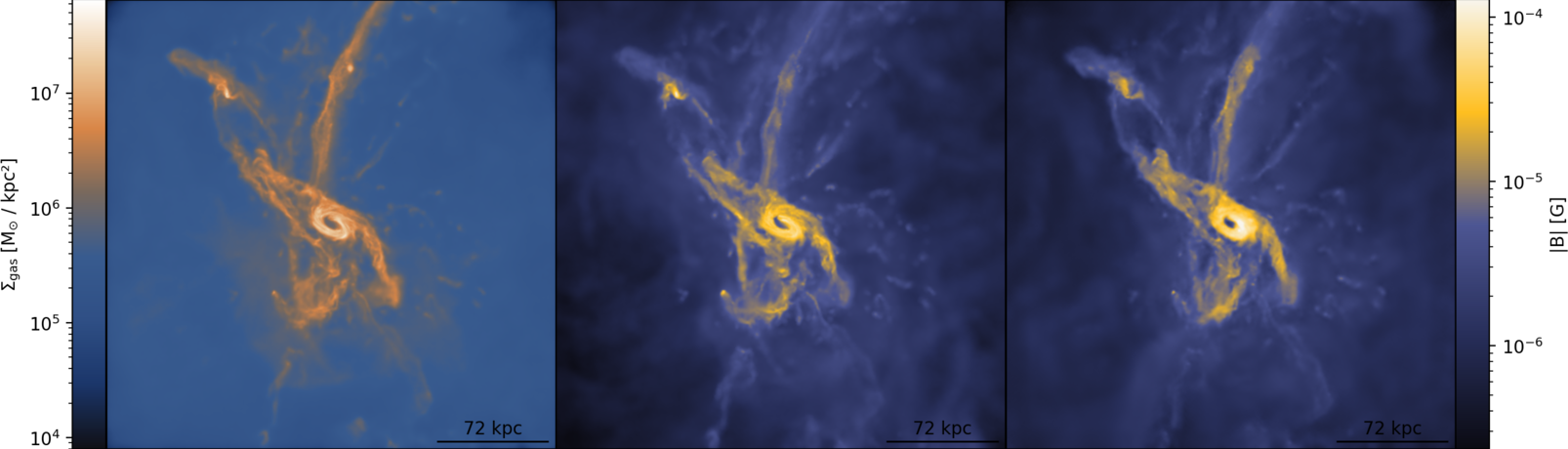}\\
      \includegraphics[width=\textwidth]{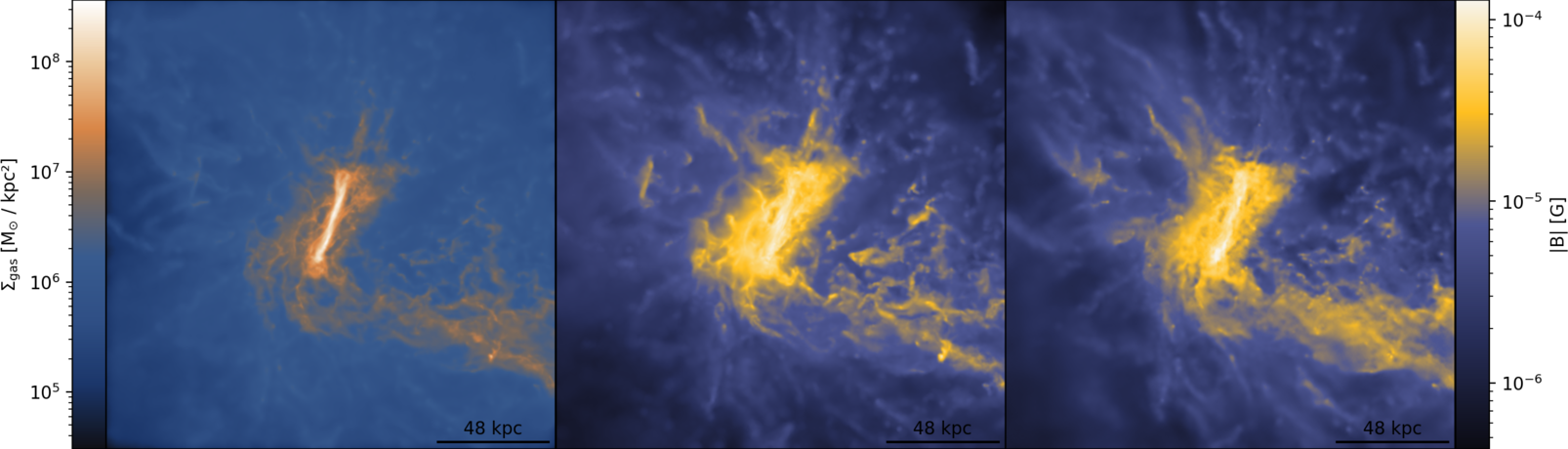}
    \end{subfigure}
    
  \end{center}
\end{figure*}

\begin{figure*}
  \ContinuedFloat
  \begin{center}

    \begin{subfigure}[t]{0.48\textwidth}
      \centering
      \caption{\centering\dm{}$\rightarrow$\gas{}: GAN-inferred samples.\\
        \hspace{4.2em} input \hfill ground truth \hfill prediction \hspace{3.8em}
      }\label{fig:dm_gas_gan_samples}
      \includegraphics[width=\textwidth]{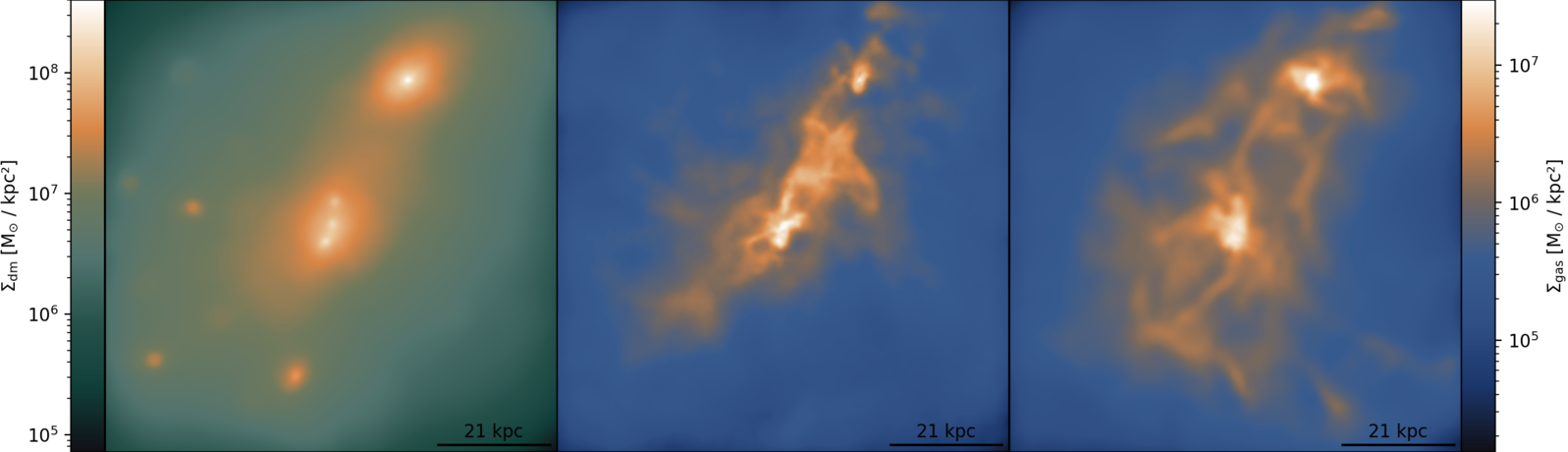}\\
      \includegraphics[width=\textwidth]{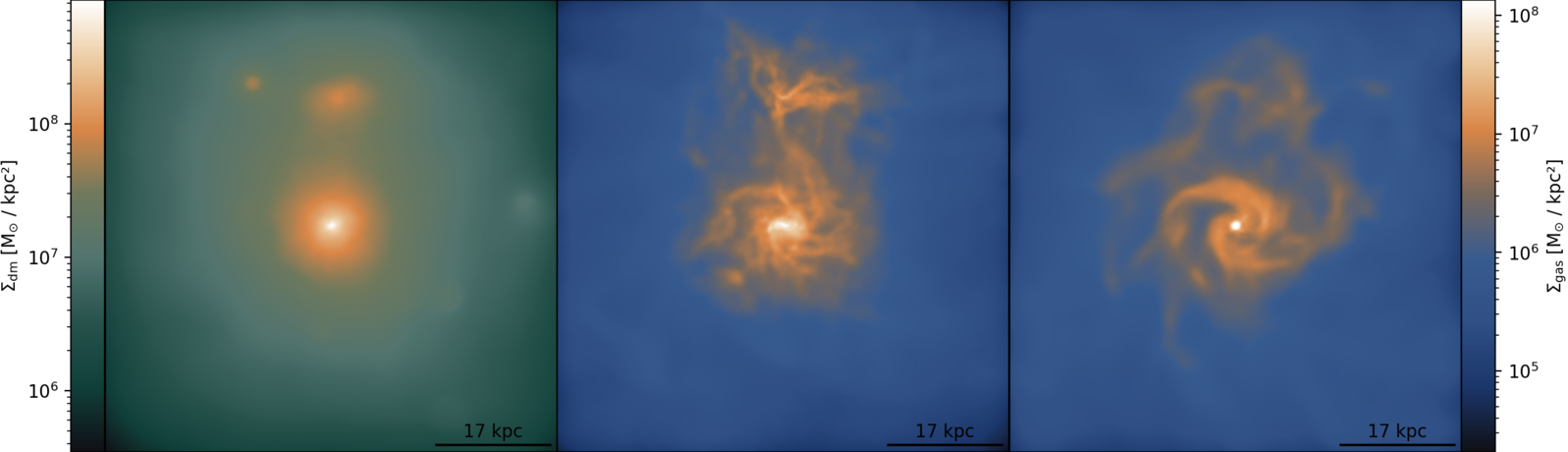}\\
      \includegraphics[width=\textwidth]{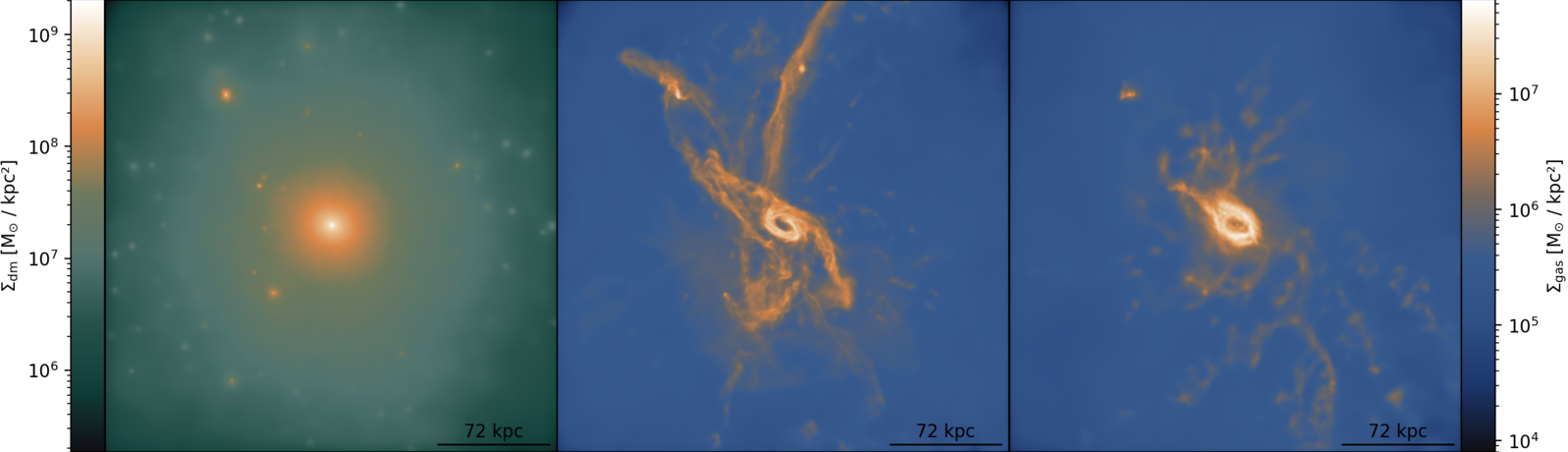}\\
      \includegraphics[width=\textwidth]{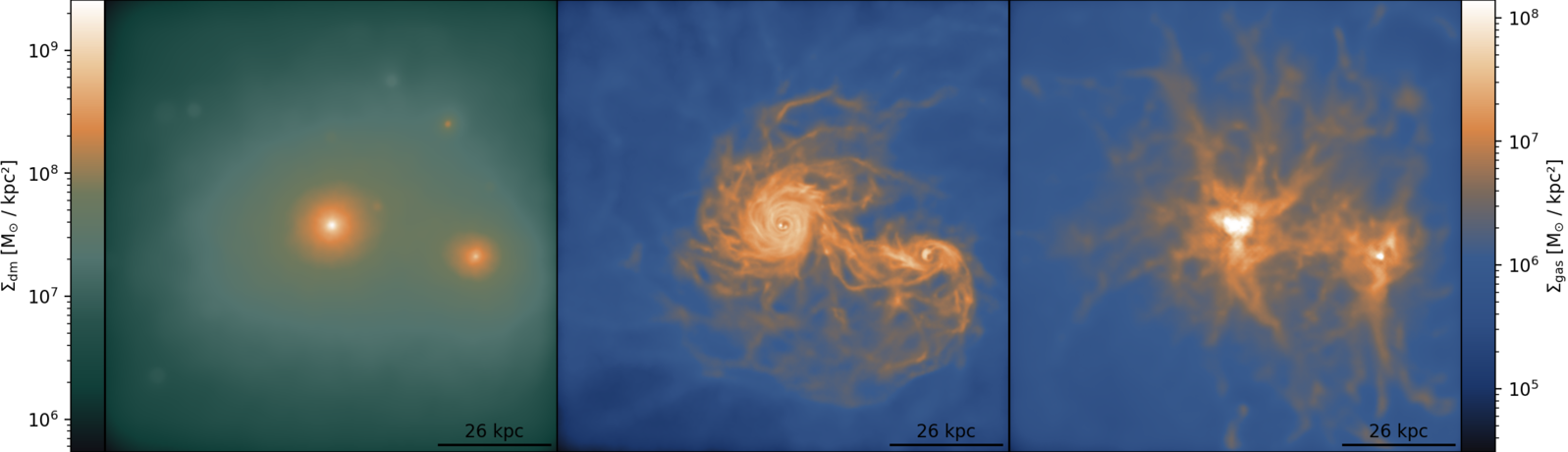}
    \end{subfigure}
    \begin{subfigure}[t]{0.48\textwidth}
      \centering
      \caption{\centering\dm{}$\rightarrow$\gas{}: DDPM-inferred samples.\\
        \hspace{4.2em} input \hfill ground truth \hfill prediction \hspace{3.8em}
      }\label{fig:dm_gas_ddpm_samples}
      \includegraphics[width=\textwidth]{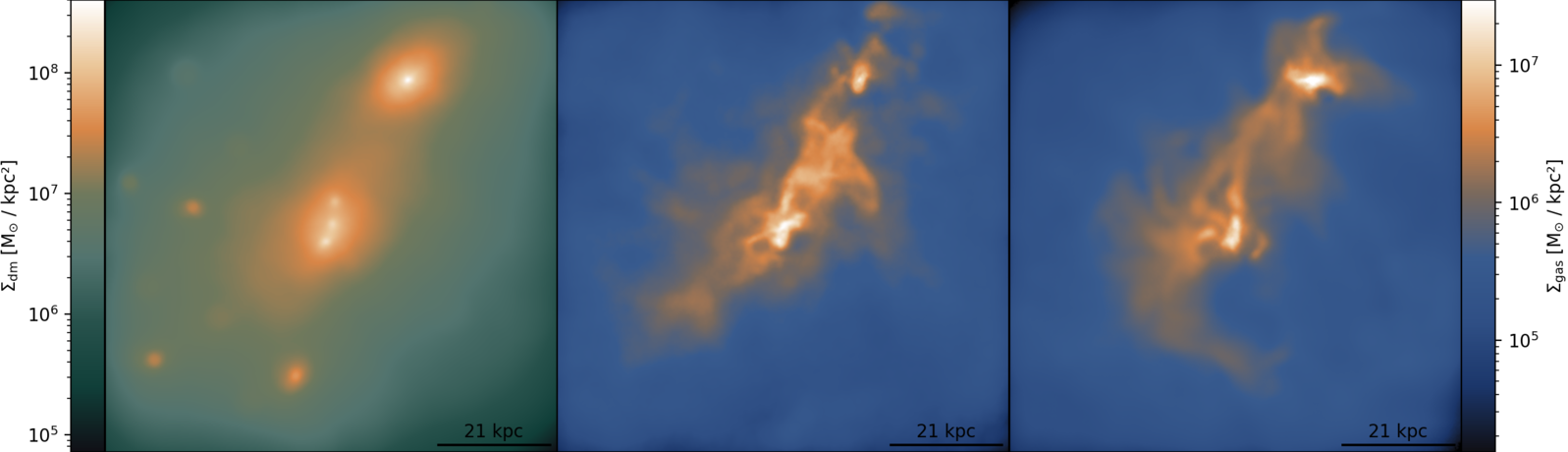}\\
      \includegraphics[width=\textwidth]{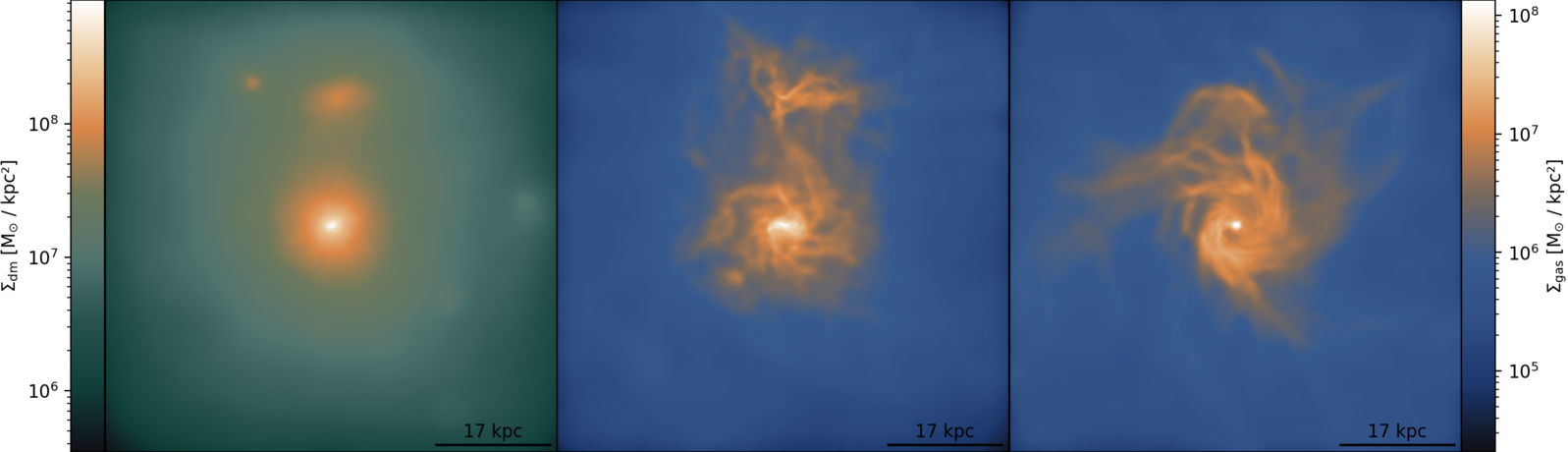}\\
      \includegraphics[width=\textwidth]{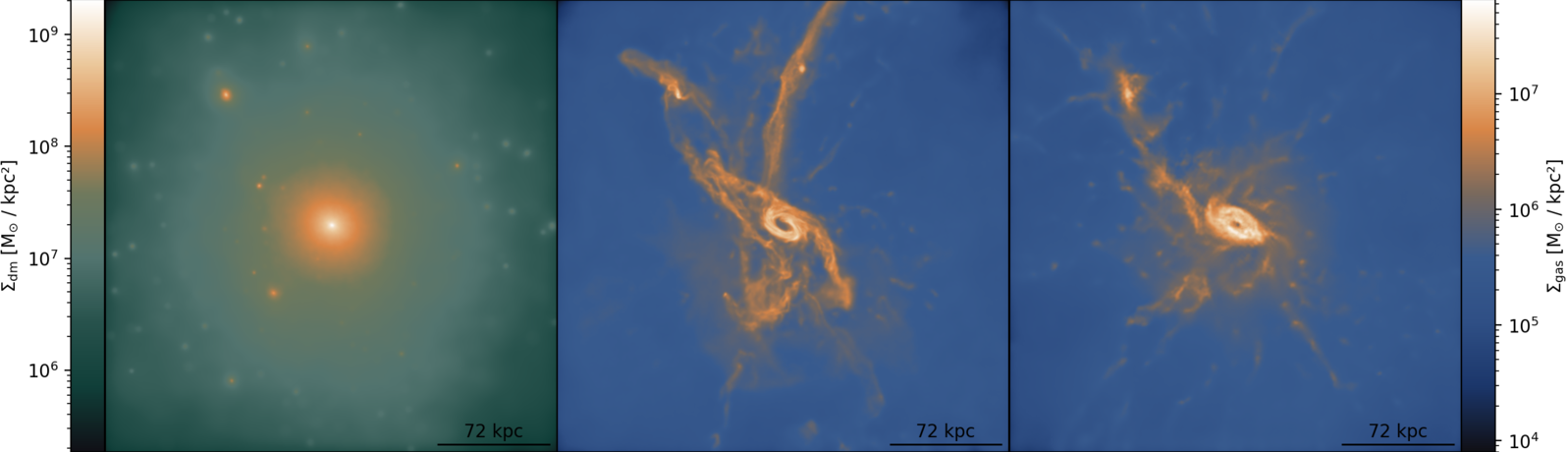}\\
      \includegraphics[width=\textwidth]{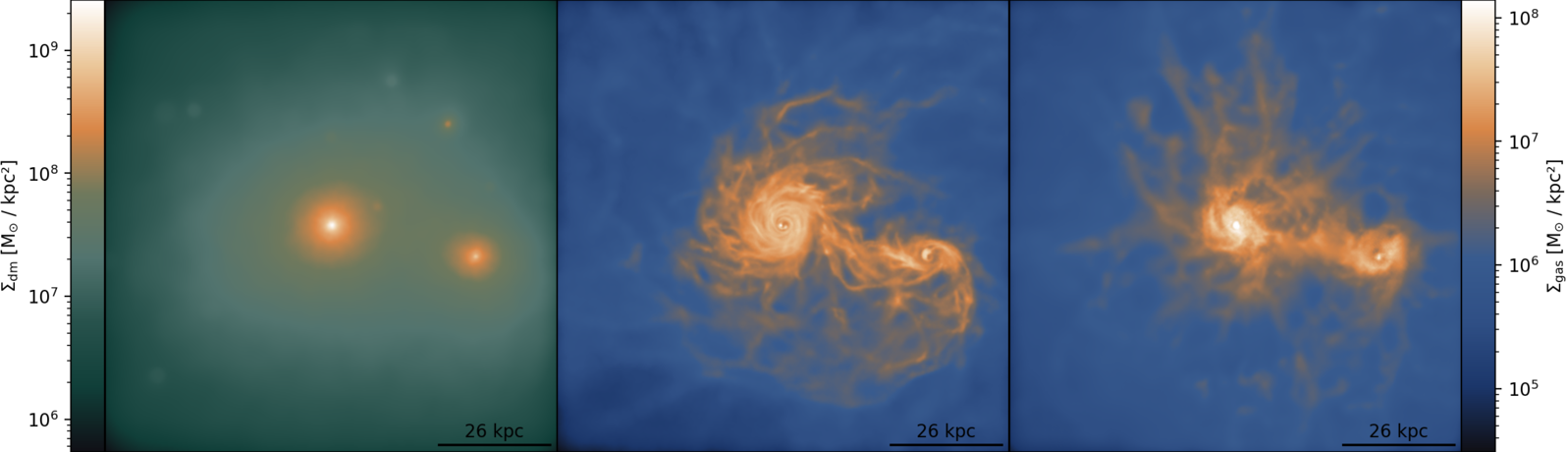}
    \end{subfigure}

    \vspace{2em}

    \begin{subfigure}[t]{0.48\textwidth}
      \centering
      \caption{\centering\hicm{}$\rightarrow$\gas{}: GAN-inferred samples.\\
        \hspace{4.2em} input \hfill ground truth \hfill prediction \hspace{3.8em}
      }\label{fig:hicm_gas_gan_samples}
      \includegraphics[width=\textwidth]{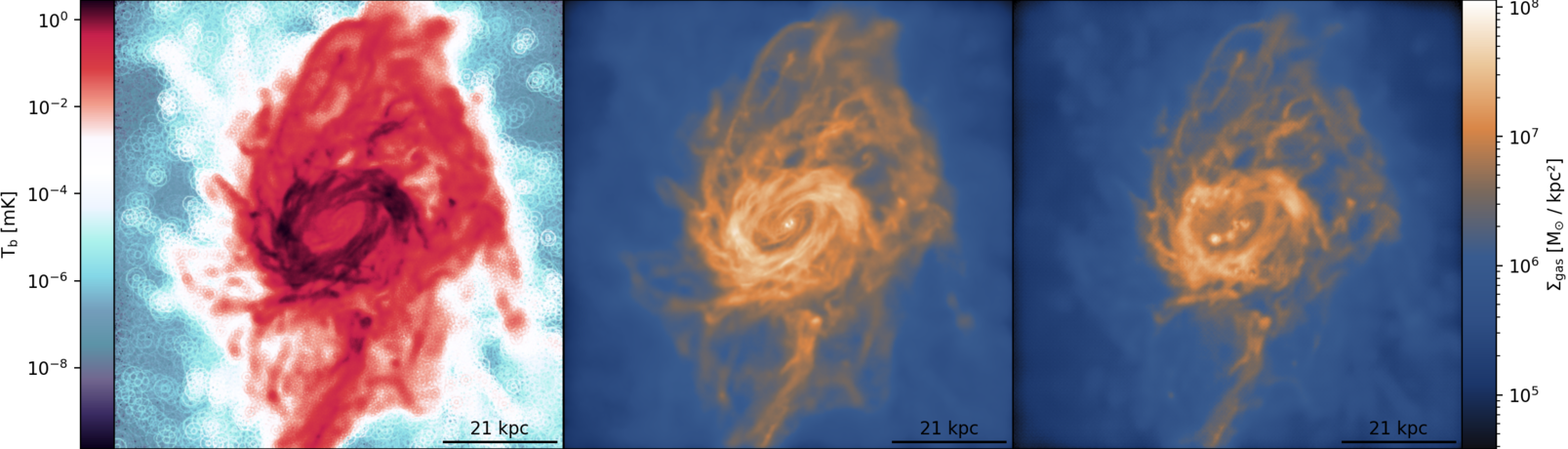}\\
      \includegraphics[width=\textwidth]{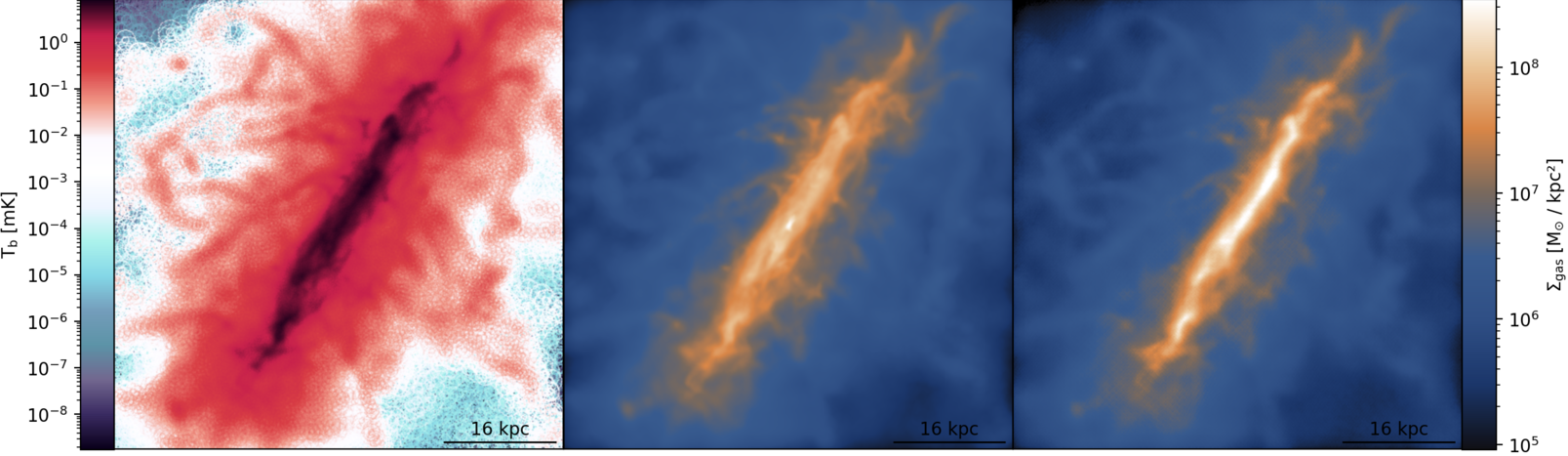}\\
      \includegraphics[width=\textwidth]{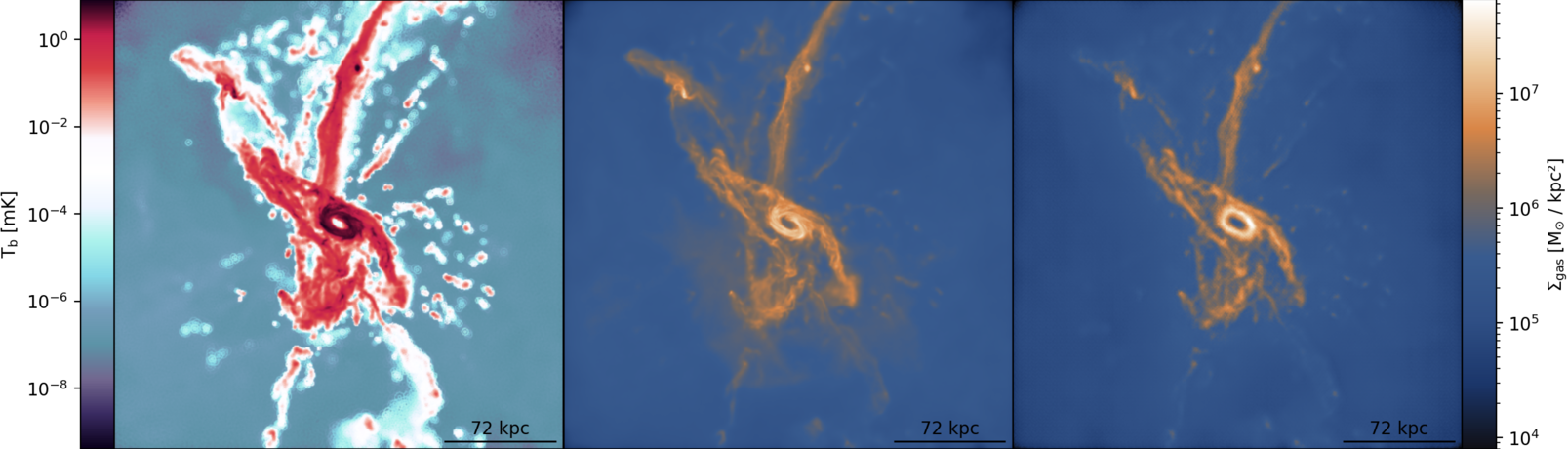}\\
      \includegraphics[width=\textwidth]{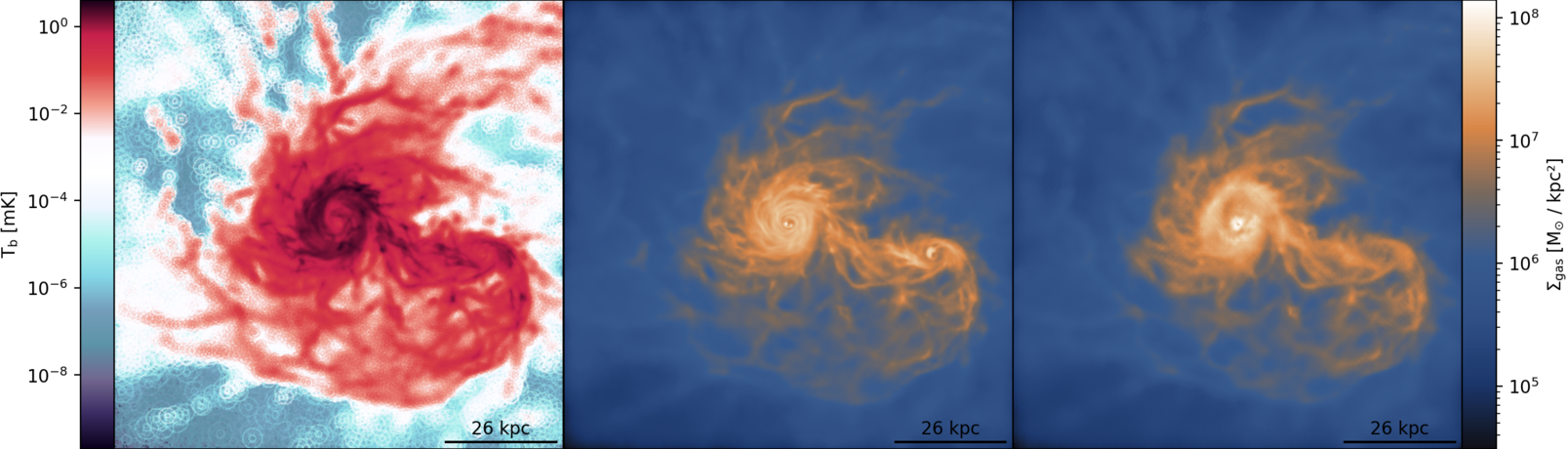}
    \end{subfigure}
    \begin{subfigure}[t]{0.48\textwidth}
      \centering
      \caption{\centering\hicm{}$\rightarrow$\gas{}: DDPM-inferred samples.\\
        \hspace{4.2em} input \hfill ground truth \hfill prediction \hspace{3.8em}
      }\label{fig:hicm_gas_ddpm_samples}
      \includegraphics[width=\textwidth]{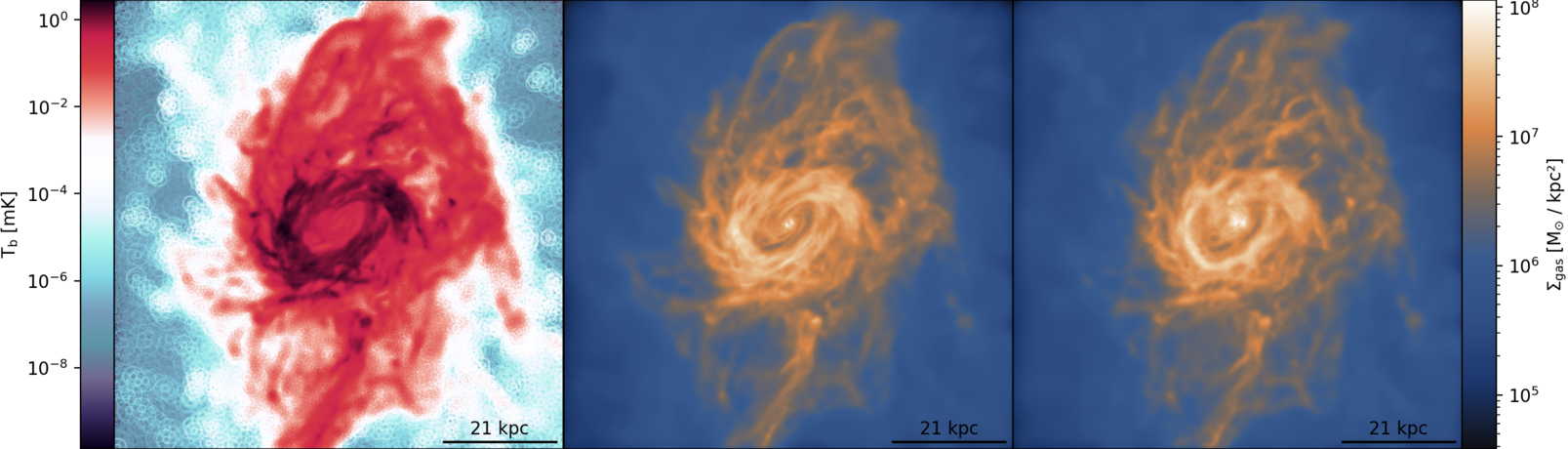}\\
      \includegraphics[width=\textwidth]{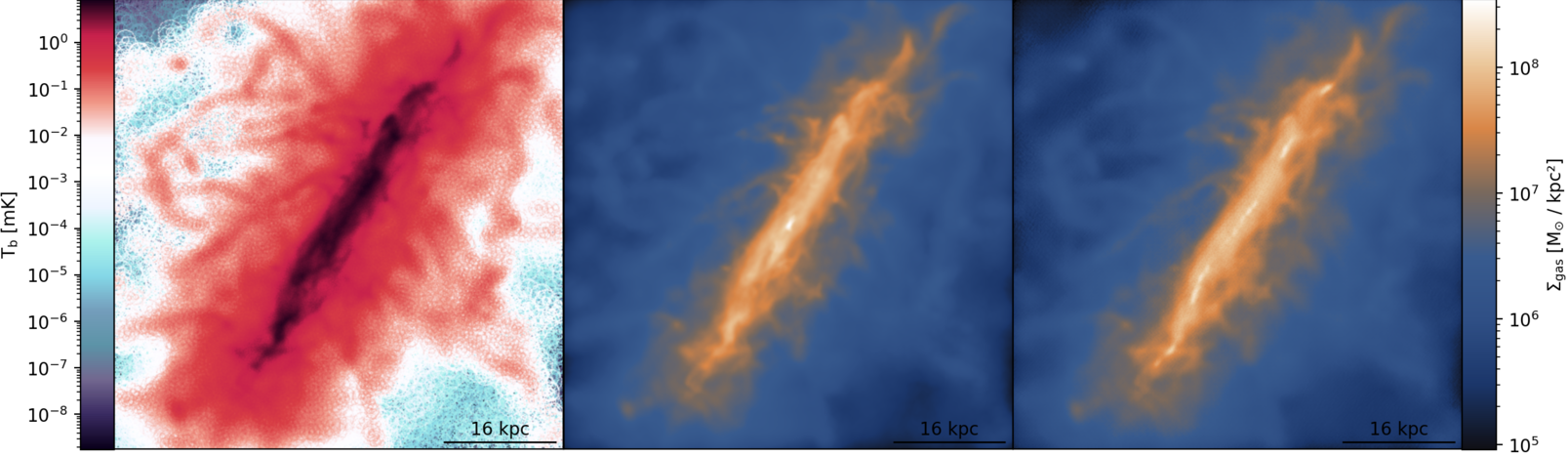}\\
      \includegraphics[width=\textwidth]{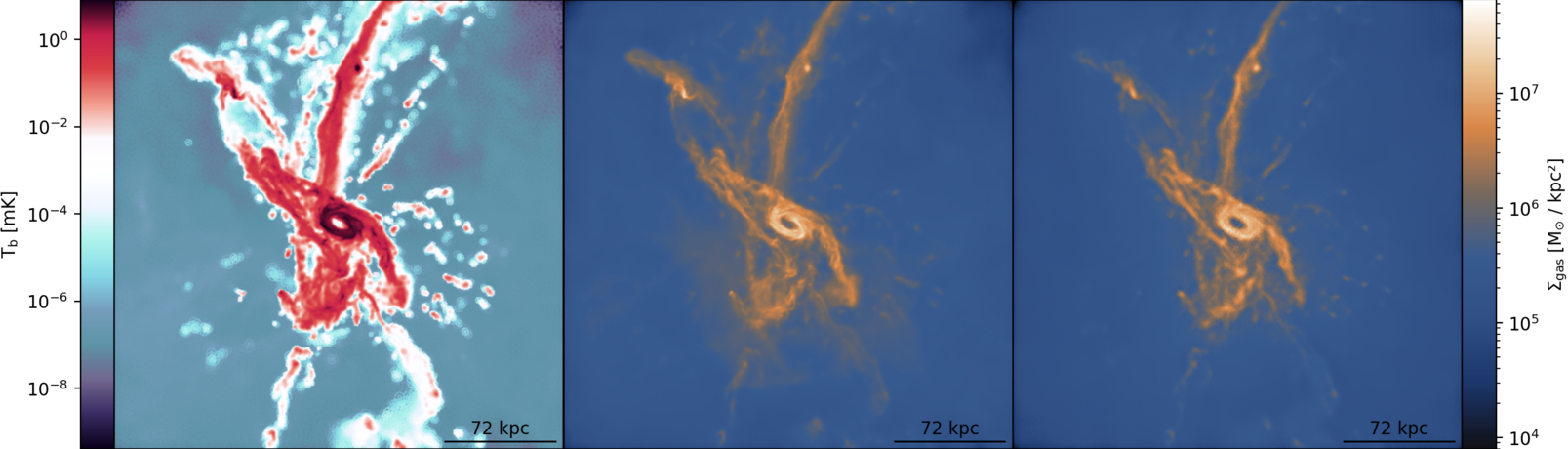}\\
      \includegraphics[width=\textwidth]{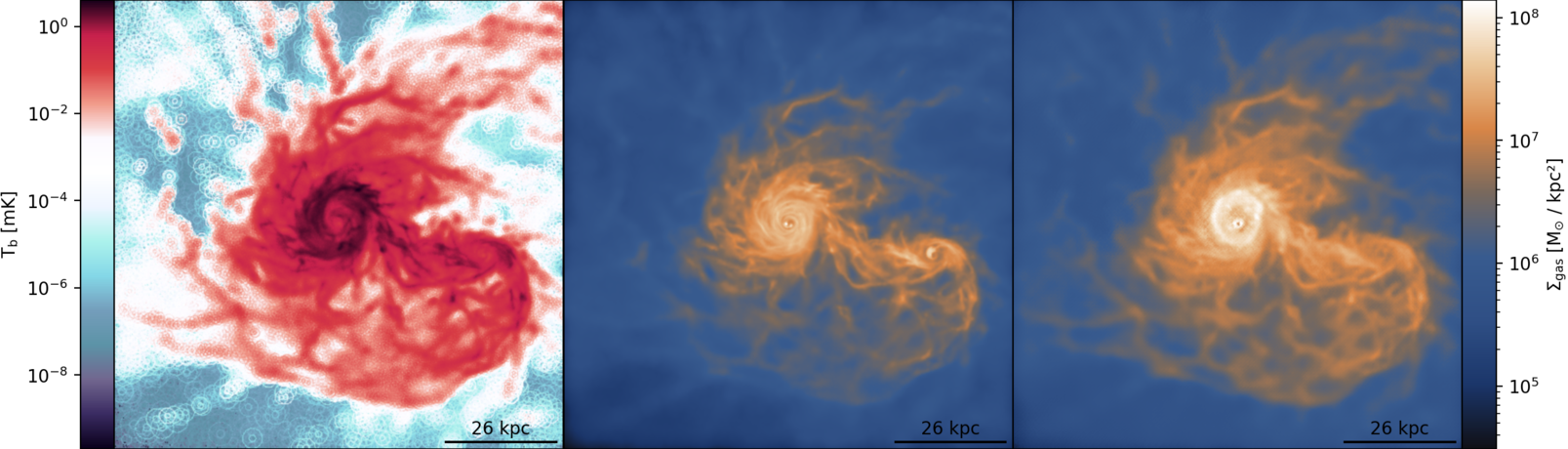}
    \end{subfigure}
    
  \end{center}
\end{figure*}

\subsection{Overall performance across translation tasks}\label{sec:overall_performance}
The measured model performance varies systematically with the physical coupling between source and target domains (Tables~\ref{tab:map_to_map_results_pixelwise}~\&~\ref{tab:map_to_map_results_astroph}).
As in previous experiments, the SSIM metric saturates quickly during training and is less discriminative than the other image-based metrics.

Table~\ref{tab:map_to_map_results_pixelwise} lists image-based (traditional CV) metric evaluations for all domain translation tasks, grouped in pairs of GAN and DDPM. The best mean value of each metric across all tasks and models is listed in bold. Similarly, Table~\ref{tab:map_to_map_results_astroph} shows the set astrophysical metric evaluations in the same order and grouping.

Among all tested translations, \gas{}$\rightarrow$\dm{} attains the highest overall fidelity: GAN and DDPM models reach best FID scores of 1.56 $\pm$ 0.36 and 2.03 $\pm$ 0.08, respectively, with PSNR values above 35 dB and SSIM $\gtrapprox$ 0.997 (see Table~\ref{tab:map_to_map_results_pixelwise}).
The astrophysical metric evaluations listed in Table~\ref{tab:map_to_map_results_astroph} confirm this trend for \gas{}$\rightarrow$\dm{}: asymmetry and clumpiness errors are among the smallest, COM offsets are negligible, and cumulative total mass deviations (evaluated at R$_{50}$) and power-spectrum errors remain modest.

Translations within the baryonic sector also perform strongly when the target is closely tied to the gas morphology.
\gas{}$\rightarrow$\hi{} and \gas{}$\rightarrow$\hicm{} achieve low FID values of 4–6 and competitive PSNR/MSE.
These models show low morphological errors (AE and SCE), minimal COM drift, excellent recovery of the radial profiles, and reproduce the expected near-monotonic relations between the total gas mass, neutral hydrogen mass, and 21-cm brightness temperature.

Moreover, the inverse mapping \hicm{}$\rightarrow$\gas{} remains tractable with similar FID values up to $7.6$, competitive ranges for the other pixel-wise metrics. The aligned performance with its counterpart across all astrophysical metrics suggests that reconstructing gas maps from observational 21-cm inputs is feasible.

In contrast, mapping \dm{}$\rightarrow$\gas{} is substantially harder, only scoring within an FID range between 22 and 45, and slightly but consistently worse results across all astrophysical metrics.

However, the most challenging mapping is clearly \gas{}$\rightarrow$\stars{}, which yields FID scores above well above 50, PSNR below 20 dB, and SSIM values well below the normal saturation levels. The large morphological errors, especially in asymmetry, reflect the models' inability to capture the alignment and ellipticity of the mass distributions, and the clumpiness errors indicate the models' difficulty to cope with the high non-locality of the stellar components.

We note that hyper-parameters (U-Net size/attention and PatchGAN settings) were primarily optimized on the \gas{}$\rightarrow$\dm{} task (Section~\ref{sec:experiments}); this may confer a slight advantage to \gas{}$\rightarrow$\dm{} in cross-task comparisons.
Thus, we repeated a small hyper-parameter sweep for a balanced set of tasks (\gas{}$\rightarrow$\stars{}, \gas{}$\rightarrow$\hi{}, and \dm{}$\rightarrow$\gas{}) to assess possible task-selection bias and performed a regret analysis based on the average FID score.
The resulting task ranking was unchanged and the regret of the optimal configuration (as in Section \ref{sec:experiments}) remained small across tasks, indicating that the results reflect intrinsic task difficulty rather than tuning alone.

\subsection{Metric interpretation}\label{sec:metric_interpretation}

Direct comparison of PSNR and MSE across models require care because of the different preprocessing ranges (see Equation~\ref{eq:hep_scaling}):
GAN inputs/outputs are mapped $[0,1]$, while DDPMs use $[-1,1]$.
The DDPM pixel value range is a factor of 2 larger, which biases PSNR by roughly 6.02 dB for the same MSE.
Thus, throughout the cross-model PSNR comparisons in this Section~\ref{sec:results} we implicitly remove this bias.

On these data domains, pixel distortion metrics PSNR, MSE, and especially SSIM are near-ceiling for several tasks (e.g. \gas{}$\rightarrow$\dm{}) and can under-discriminate subtle morphological differences in smooth, high-resolution simulation maps. Conversely, the suite of astrophysically motivated metrics (AE, SCE, COMD, CRCE, and PSE) remains sensitive to structural realism and are model-agnostic.

\begin{figure}
  \centering
  \includegraphics[width=0.8\linewidth]{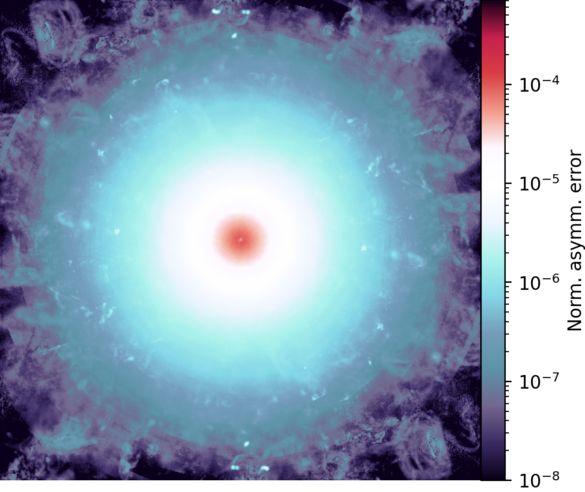}
  \includegraphics[width=0.8\linewidth]{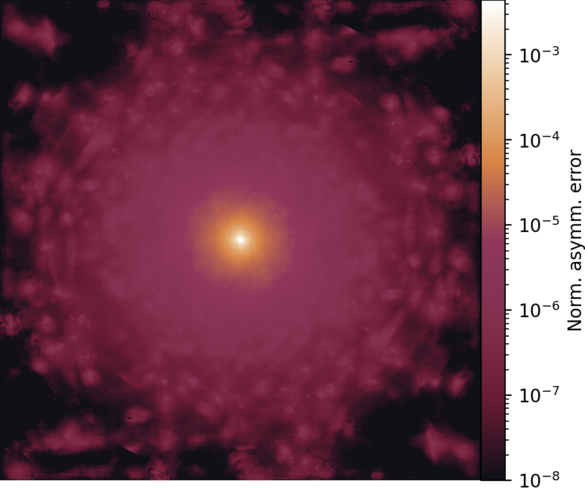}
  \caption{
    Examples of normalized asymmetry error maps for the mappings \gas{}$\rightarrow$\hicm{} (top) and \gas{}$\rightarrow$\stars{} (bottom) in the test set, inferred by GANs.
    The overall mean error is around an order of magnitude larger for \gas{}$\rightarrow$\stars{} and relatively uniform but exhibits a slight chequerboard pattern, indicating the difficulty to model the fine-grained structure of the stellar mass distribution.
    \gas{}$\rightarrow$\hicm{} exhibits smaller irregular errors which are noticeable due to overall lower average error.
  }\label{fig:asymmetry}
\end{figure}

\begin{figure}
  \centering
  \includegraphics[width=0.8\linewidth]{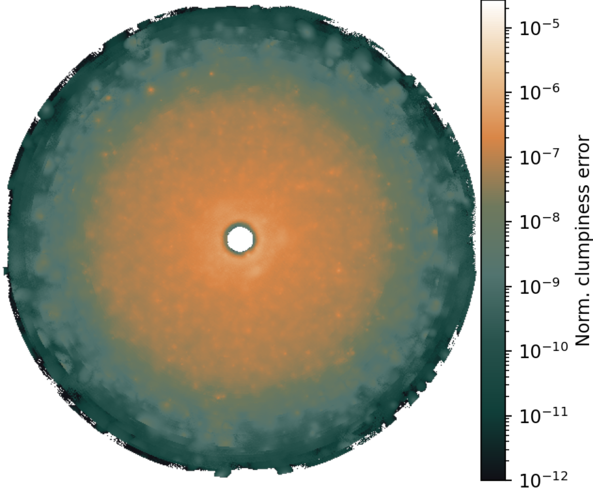}
  \includegraphics[width=0.8\linewidth]{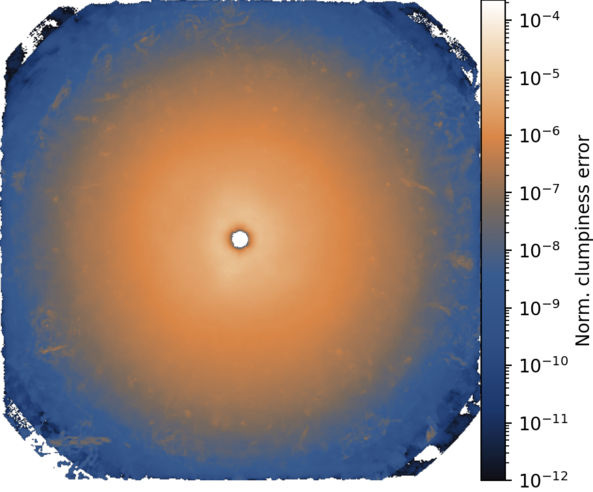}
  \caption{
    Examples of normalized clumpiness error maps for the mappings \gas{}$\rightarrow$\dm{} (top) and \dm{}$\rightarrow$\gas{} (bottom) in the test set, inferred by GANs.
    To keep numerical stability, the inner regions of the error maps have been masked to 5\% of the map's respective half-mass radius.
    The overall mean error is around an order of magnitude larger for \dm{}$\rightarrow$\gas{}, indicating the increased difficulty of predicting baryonic components from DM compared to the inverse mapping.
    Moreover, due to the collisionless nature of DM, its distributions tend to be smoother, which also contributes to the lower mean error.
    For \gas{}$\rightarrow$\dm{}, errors mainly arise due to the wrong estimate of DM substructure in the haloes, whereas errors for \dm{}$\rightarrow$\gas{} indicate unrealistic fragmentation in small-scale structures.
  }\label{fig:clumpiness}
\end{figure}

Figures~\ref{fig:asymmetry} and \ref{fig:clumpiness} illustrate how global errors manifest spatially.
Both metrics measure important features (e.g., structural symmetry and fine-structure resolution) indicative of morphological realism and overall plausibility of generated samples.
Figure~\ref{fig:asymmetry} shows the average asymmetry error map for \gas{}$\rightarrow$\hicm{} versus \gas{}$\rightarrow$\stars{}.
Errors of the latter task are roughly an order of magnitude larger with a slight chequerboard pattern, indicating unresolved fine-structure and adversarial artefacts.
For \gas{}$\rightarrow$\dm{} and \dm{}$\rightarrow$\gas{} (Figure~\ref{fig:clumpiness}) the harder inverse mapping (latter) exhibits higher small-scale residuals consistent with unrealistic fragmentation.

\begin{figure}
  \centering
  \includegraphics[width=0.85\linewidth]{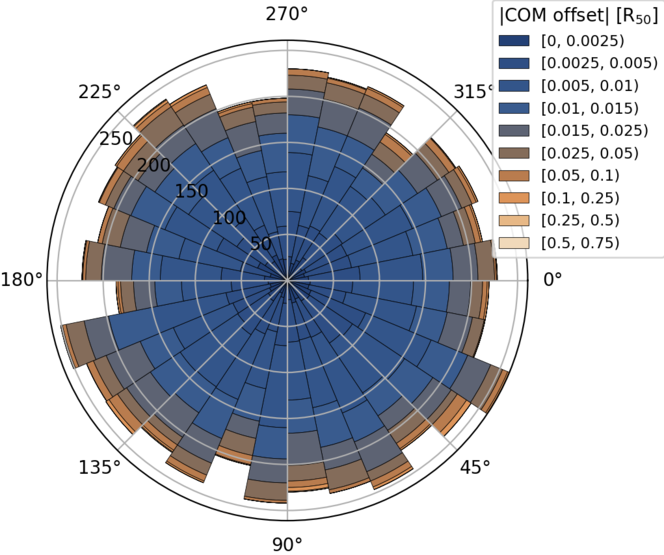}
  \includegraphics[width=0.85\linewidth]{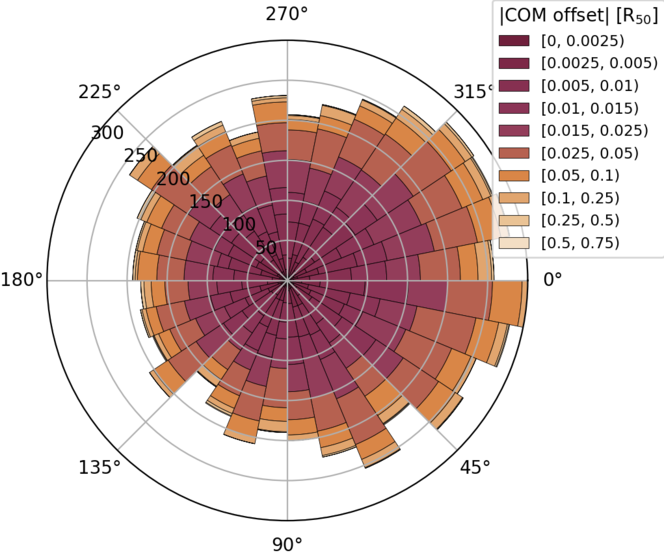}
  \caption{
    Examples of the angular distribution of COM drifts for the mappings \gas{}$\rightarrow$\hi{} (top) and \gas{}$\rightarrow$\stars{} (bottom) in the test set, inferred by DDPMs.
    While the upper wind rose diagram shows a uniform distribution for \gas{}$\rightarrow$\hi{} COM drifts, \gas{}$\rightarrow$\stars{} exhibits an angular bias towards 0$\textdegree$.
    The concentration of these errors in lower offset bins (in units of R$_{50}$), as shown for \gas{}$\rightarrow$\hi{}, indicates low overall drift and typically good agreement with the ground truth.
  }\label{fig:com}
\end{figure}

Centre-of-mass drift errors can also be decomposed in more detail (Figure~\ref{fig:com}).
While the COMD only measures the scalar global drift, higher values may have different causes: the upper panel shows a near-uniform distribution of COM drifts (good positional agreement), whereas the lower panel exhibits a noticeable angular bias, signalling a systematic vectorial drift of the inferred mass centroid.

\subsection{Model types: performance and trade-offs}

There is no universal winner between GANs and DDPMs across all tasks.
GANs tend to achieve lower FIDs when the target is tightly tied to the gas morphology (e.g., \gas{}$\rightarrow$\dm{}, \gas{}$\rightarrow$\hi{}, and \gas{}$\rightarrow$\hicm{}), while DDPMs often deliver more favourable astrophysical fidelity (lower AE, SCE, and COMD) for less strongly related quantities such as \gas{}$\rightarrow$\temp{}, or \gas{}$\rightarrow$\bfield{}.
Moreover, GANs inherently exhibit more quality fluctuations even long into training due to the adversarial nature of their objective; this is evidenced by the typically higher standard deviations of the metric results from the last five epochs.
These complementary behaviours suggest that adversarial training sharpens structural realism in strongly coupled mappings, whereas diffusion-based modelling better preserves global morphology for thermodynamic and field-like targets.
From a resource perspective, the GAN models in this work required $\sim140$ kWh training energy versus $\sim520$ kWh for DDPMs in our setup (summarized read-outs from the GPU monitoring system for all runs, not including ablation tests).
Both approaches are orders of magnitude more energy-efficient than re-running comparable hydrodynamical simulations $\mathcal{O}(\text{GWh})$ \citep[cf. Table 1 in][]{nelson_2019_first_results}, but the $\sim4\times$ advantage of GANs can be decisive when many map-to-map translation models need to be trained.

\subsection{Global consistency of inferred quantities}

\begin{figure*}
  \caption{
    Global statistics of inferred vs true integrated quantities.
    The colour scheme qualitatively indicates histogram density and matches the task assignment analogous to Figure~\ref{fig:samples}.
    In general, GANs and DDPMs show no biases and minimal scatter of integrated quantities (except for \gas{}$\rightarrow$\stars{}).
  }\label{fig:map_total}
  \begin{center}

  \begin{subfigure}[t]{0.48\textwidth}
    \centering
    \caption{\centering\gas{}$\rightarrow$\dm{}: Inferred total DM mass of all galaxies\\ in the test set.}\label{fig:mass_gas_dm}
    \includegraphics[width=\textwidth]{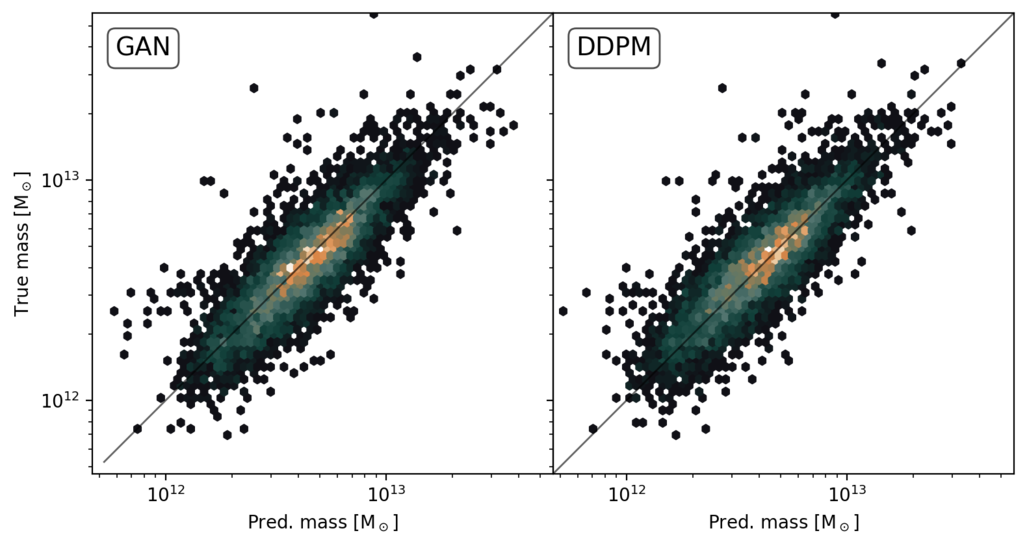}
  \end{subfigure}
  \begin{subfigure}[t]{0.48\textwidth}
    \centering
    \caption{\centering\gas{}$\rightarrow$\stars{}: Inferred total stellar mass of all galaxies\\ in the test set.}\label{fig:mass_gas_star}
    \includegraphics[width=\textwidth]{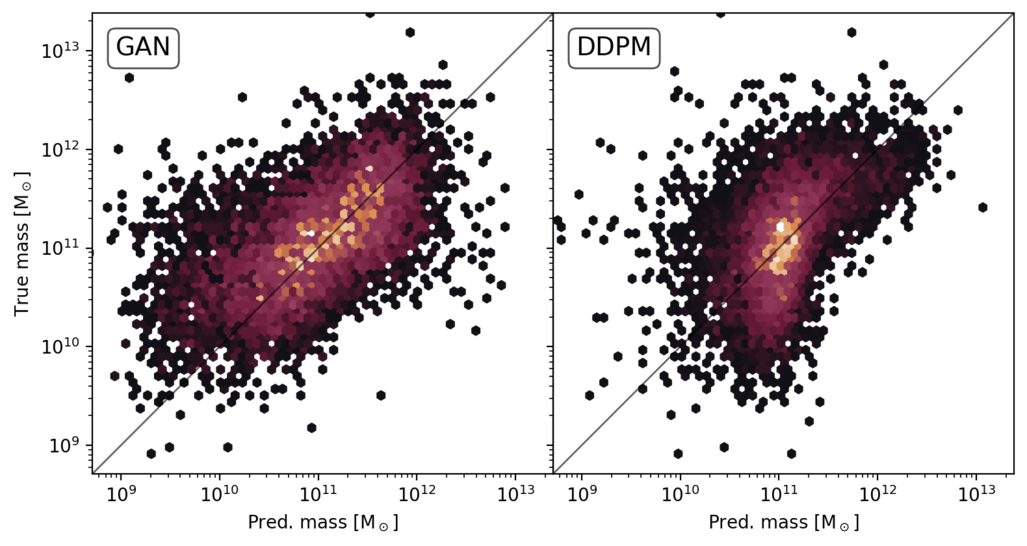}
  \end{subfigure}

  \begin{subfigure}[t]{0.48\textwidth}
    \centering
    \caption{\centering\gas{}$\rightarrow$\hi{}: Inferred total neutral hydrogen gas mass of all galaxies in the test set.}\label{fig:mass_gas_hi}
    \includegraphics[width=\textwidth]{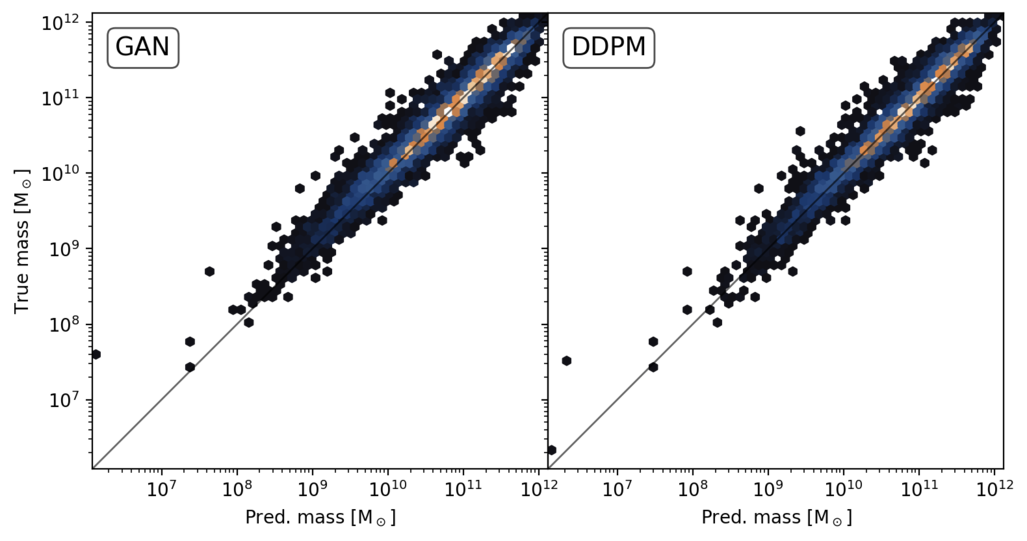}
  \end{subfigure}
  \begin{subfigure}[t]{0.48\textwidth}
    \centering
    \caption{\centering\gas{}$\rightarrow$\hicm{}: Inferred average 21-cm brightness temperature of all galaxies in the test set.}\label{fig:mass_gas_hicm}
    \includegraphics[width=\textwidth]{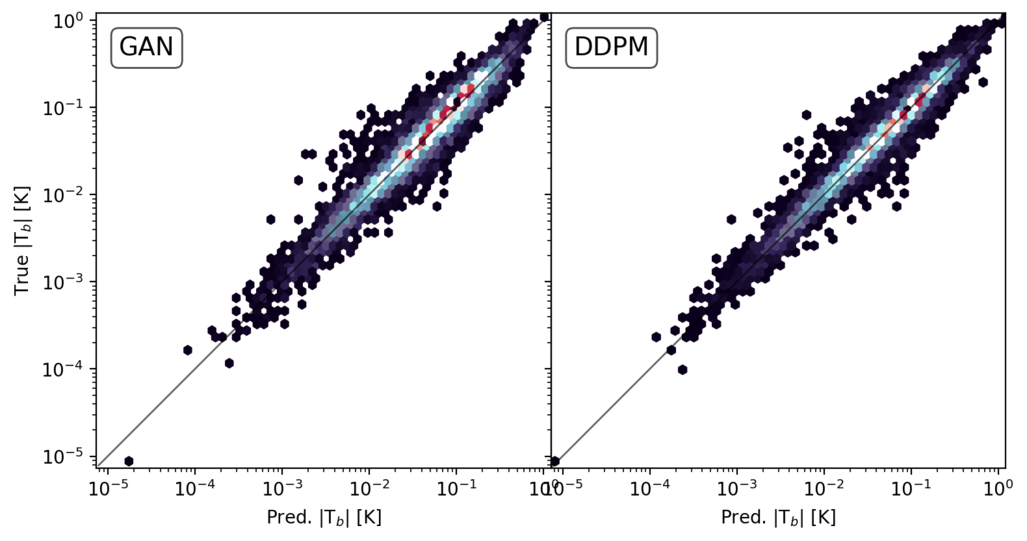}
  \end{subfigure}

  \begin{subfigure}[t]{0.48\textwidth}
    \centering
    \caption{\centering\gas{}$\rightarrow$\temp{}: Inferred average temperature of all galaxies in the test set.}\label{fig:mass_gas_temp}
    \includegraphics[width=\textwidth]{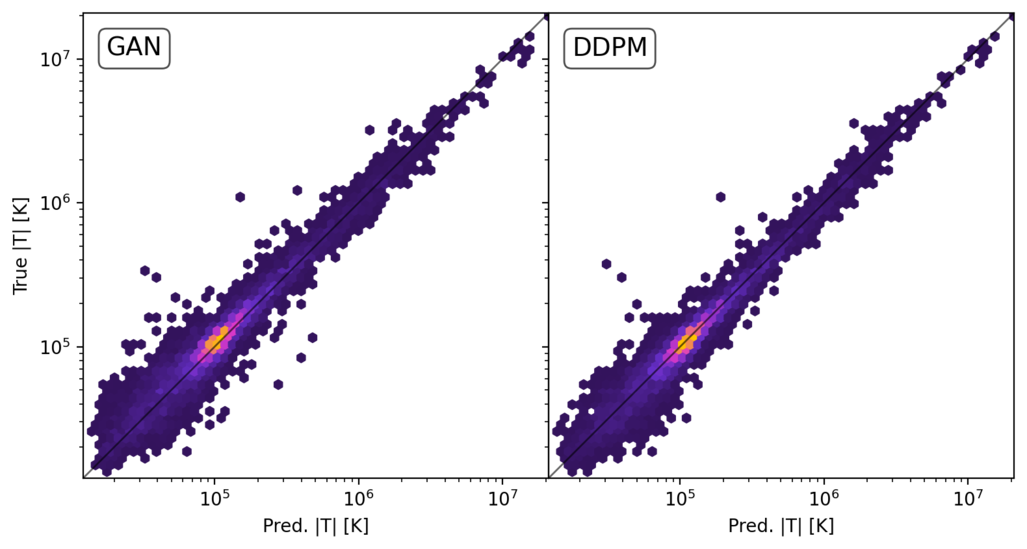}
  \end{subfigure}
  \begin{subfigure}[t]{0.48\textwidth}
    \centering
    \caption{\centering\gas{}$\rightarrow$\bfield{}: Inferred average magnetic field strength of all galaxies in the test set.}\label{fig:mass_gas_bfield}
    \includegraphics[width=\textwidth]{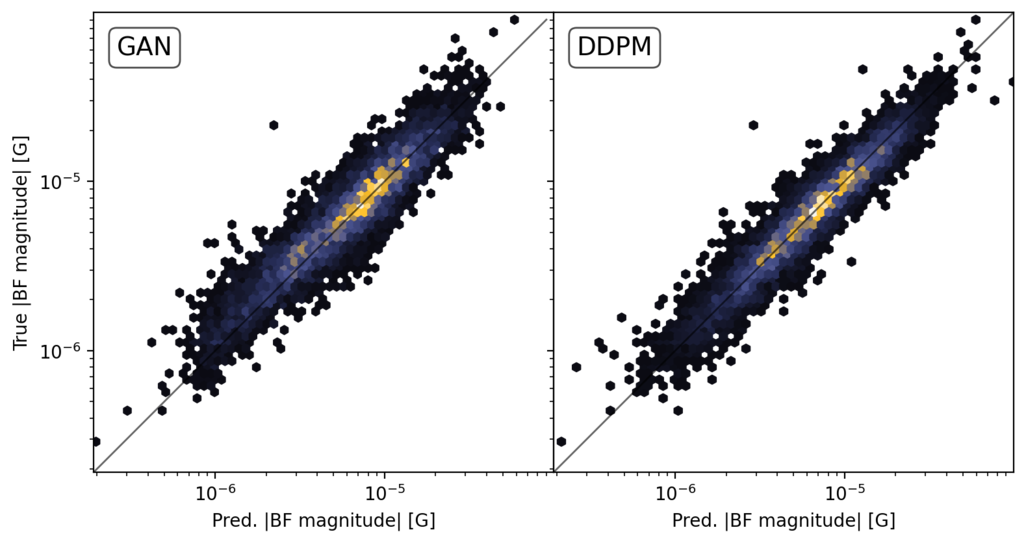}
  \end{subfigure}

  \begin{subfigure}[t]{0.48\textwidth}
    \centering
    \caption{\centering\dm{}$\rightarrow$\gas{}: Inferred total gas mass of all galaxies in the test set (from DM maps).}\label{fig:mass_dm_gas}
    \includegraphics[width=\textwidth]{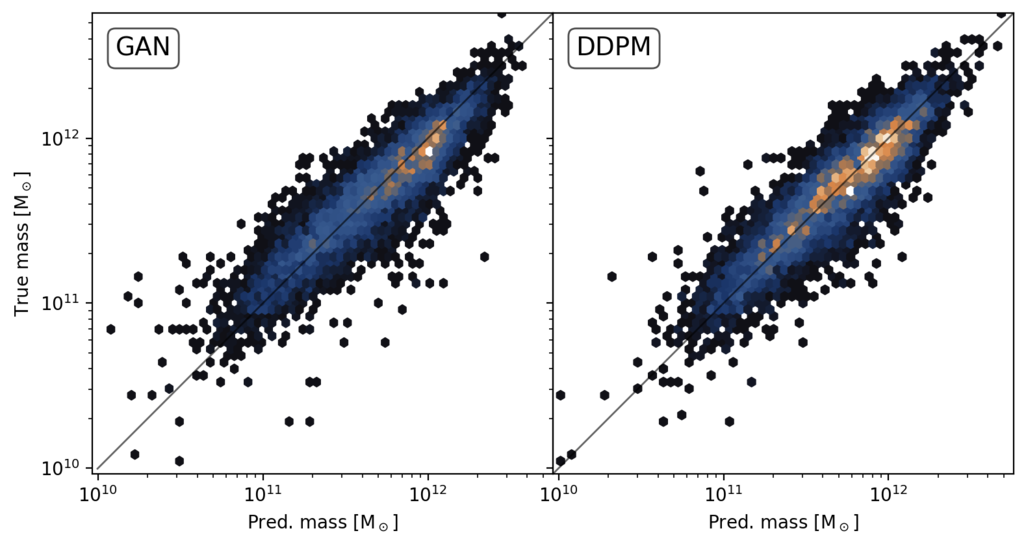}
  \end{subfigure}
  \begin{subfigure}[t]{0.48\textwidth}
    \centering
    \caption{\centering\hicm{}$\rightarrow$\gas{}: Inferred total gas mass of all galaxies in the test set (from 21-cm maps).}\label{fig:mass_hicm_gas}
    \includegraphics[width=\textwidth]{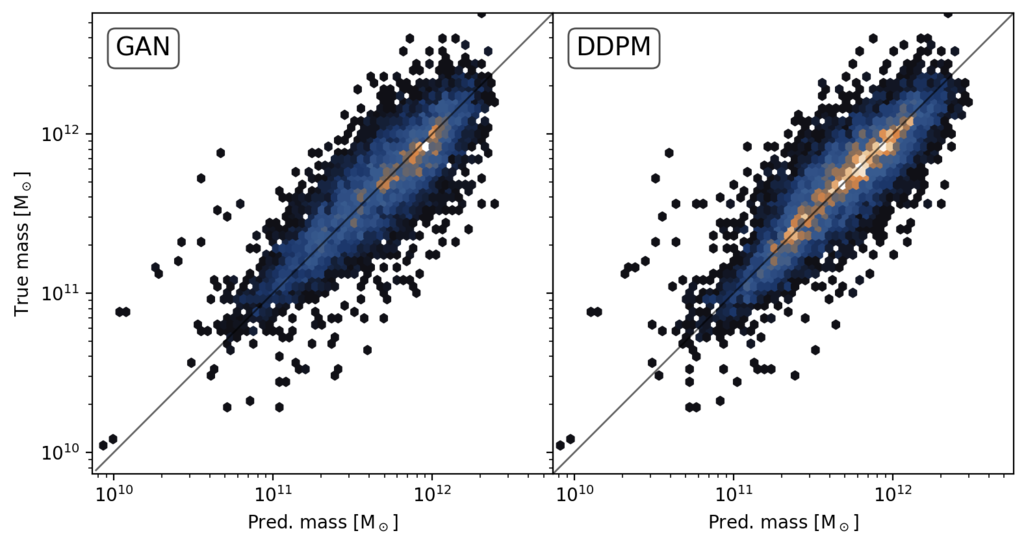}
  \end{subfigure}
  
  \end{center}
\end{figure*}

Figure~\ref{fig:map_total} compares integrated inferred properties against ground truth for the unseen test population.
For strongly coupled mappings such as \gas{}$\rightarrow$\hi{} and \gas{}$\rightarrow$\hicm{}, both GAN and DDPM models recover total masses with minimal bias and scatter, indicating robust conservation of global properties.
Notably, 99.9\% of errors for all listed mappings, including \gas{}$\rightarrow$\dm{}, \hicm{}$\rightarrow$\gas{}, \gas{}$\rightarrow$\temp{}, \gas{}$\rightarrow$\bfield{}, are within a factor of 10 (see also Table~\ref{tab:map_to_map_results_astroph}).
In contrast, the mapping \gas{}$\rightarrow$\stars{} is exceptionally challenging for both models and presents large scatter and bias patterns (Figure~\ref{fig:mass_gas_star}):
DDPMs over-predict at low masses and under-predict at the high-mass end, while GANs exhibit smaller mean bias but extreme scatter reaching beyond two orders of magnitude.
These outcomes mirror the expected entropy and non-local differences among target domains and underline the task difficulty ordering observed in the other astrophysical metrics.

\section{Conclusion}\label{sec:conclusion}


We presented the first systematic study of multi-domain map-to-map translations for galaxy formation simulations, introducing deep generative models as scalable data-driven alternatives that map between seven physical domains (DM, stellar mass, gas mass, neural hydrogen mass, 21-cm mock brightness, temperature, and magnetic field strength), comparing adversarial (GAN) and diffusion (DDPM) deep learning approaches under unified preprocessing and evaluation.
Both approaches are able to learn physically plausible solutions to these domain translations, demonstrated on a dataset of galaxy maps extracted from the \tng{} suite (\siml{TNG50--1}).
Across extensive ablations and metrics -- distortion (MSE, PSNR, SSIM), perceptual (FID) and astrophysical metrics (asymmetry, clumpiness, centre-of-mass drift, radial/cumulative curves, power spectra) -- we find that translation difficulty strongly correlates with the physical coupling of source and target:
\gas{}$\rightarrow$\dm{} achieves the best fidelity measured by image-based metrics (FID $\approx$ 2.0), \gas{}$\rightarrow$\hi{}, \gas{}$\rightarrow$\hicm{}, and \hicm{}$\rightarrow$\gas{} are likewise strong and conserve integrated quantities, while \dm{}$\rightarrow$\gas{} is substantially harder but still produces plausible results.
\gas{}$\rightarrow$\stars{} remains the most challenging across all measures.
GANs tend to excel for tightly coupled targets with sharper structure and lower FID, whereas DDPMs better preserve global morphology and thermodynamic or field-like structure;
this complementarity comes with a $\sim4\times$ difference in training energy in our setup ($\sim140$ kWh vs $\sim520$ kWh).
These results demonstrate the feasibility of learnt representations that encapsulate aspects of a simulation's formation scenario $\Phi$ from different observationally motivated inputs, while underscoring the need for domain-ware metrics and physics-informed inductive biases to tackle weakly constrained mappings.
Notably, despite the controversy around the use of FID in scientific domains (cf. \ref{sec:cv_metrics}), it correlated surprisingly strongly with the astrophysical metrics which capture structural realism, suggesting it is an appropriate discriminator for our use case.

\paragraph{Physical couplings.}
The empirical task ordering we observe follows the expected information coupling among galaxy components.
Gas traces the gravitational potential well and interacts collisionally, so \gas{}$\rightarrow$\dm{} is comparatively well-posed:
large-scale morphology and substructure are strongly correlated, enabling excellent astrophysical veracity (low FID and morphological errors; see Tables~\ref{tab:map_to_map_results_pixelwise}~and~\ref{tab:map_to_map_results_astroph} and Figure~\ref{fig:clumpiness}).
In contrast, the inverse mapping \dm{}$\rightarrow$\gas{} is under-constrained:
while the DM halo delineates the potential, baryon distributions are additionally set by feedback, heating, and cooling;
our models thus exhibit more clumpiness residuals and centre-of-mass drift (Figure~\ref{fig:clumpiness}), consistent with fragmentation artefacts.
The most difficult case, \gas{}$\rightarrow$\stars{}, reflects the intrinsically non-local nature and higher entropy of stellar mass assembly:
star formation depends on history and feedback cycles only weakly encoded in a single gas snapshot, leading to large asymmetry and clumpiness errors, poor FID, and strong biases in integrated stellar mass (see Tables~\ref{tab:map_to_map_results_pixelwise}~and~\ref{tab:map_to_map_results_astroph} and Figures~\ref{fig:asymmetry}~and~\ref{fig:com}).
Altogether, the results corroborate the conceptual view in Equations~\ref{eq:galaxy_population}~and~\ref{eq:nuisance_margin}:
learning conditional terms is easier when nuisance parameters are few and the conditional entropy of the target given the source is low.

Integrated quantities provide an orthogonal check of global physical plausibility. We find minimal bias and scatter for most mappings.
The outlier is \gas{}$\rightarrow$\stars{}, which shows systematic bias and large scatter for both model types (Figure~\ref{fig:mass_gas_star}).
These findings imply that for a subset of domains with strong coupling, chaining of models (e.g. \hicm{}$\rightarrow$\gas{}$\rightarrow$\dm{}) may be feasible without much loss of information (without explicitly training for cycle-consistency).

\paragraph{Model choice guidance.}
No single model type dominates across all translations.
Adversarial training yields good high-frequency results (at times mildly exaggerated), especially for targets strongly coupled to the gas morphology, but exhibits larger epoch-to-epoch variability -- a hallmark of the minimax optimization game (cf. Tables~\ref{tab:map_to_map_results_pixelwise}, \ref{tab:map_to_map_results_astroph}, and Section~\ref{sec:gans}, Equation~\ref{eq:minimax}).
Diffusion models tend to preserve global gradients and in more complex couplings and often improve astrophysical plausibility at the cost of slower sampling and higher training time and energy.
From a practitioner's standpoint:
\begin{itemize}[leftmargin=*,topsep=1pt]
\item choose GANs for tight, morphology-driven mappings where fast inference is worth the small trade-off in accuracy;
\item choose DDPMs when the target encodes smoother or more complex fields, when robustness in astrophysical accuracy has highest priority.
\end{itemize}

Importantly, through targeted architectural and training optimizations, including U-Net depth/width tuning, attention placement near the bottleneck, and discriminator sizing (Section~\ref{sec:experiments}), we demonstrate that GAN-based models can achieve performance on par with state-of-the-art DDPMs for most mappings.
This parity, combined with GANs' lower training energy and single-pass inference, positions them as a competitive and computationally sustainable alternative for large-scale deployment.
Moreover, a hybrid approach which draws from each methods advantages while mitigating their disadvantages, could be an promising avenue for future work.

\paragraph{Implications for observations.}
This work offers a direct path to observational validation by incorporating domains that are measurable in practice, such as 21cm brightness and neutral hydrogen, into the translation process.
This capability is particularly critical for the SKA, which will probe the distribution of \HI{} in nearby galaxies to unprecedented precision.
Here, two practical applications of our models emerge:
\begin{itemize}[leftmargin=*,topsep=1pt]
\item \textit{Forward modelling}: predicting 21-cm brightness from simulated gas maps and pass through an instrument response pipeline \citep[such as Karabo;][]{sharma_2025_karabo} for high-realism mock observations including SKA-like systematics.
\item \textit{Reconstruction}: inferring gas distributions and related galactic properties from observed 21-cm maps of nearby galaxies to support feedback and morphological studies.
\end{itemize}

By embedding observational proxies and incorporating, e.g., beam smoothing, thermal noise, and foreground residuals into the generative framework (during training or via data augmentation), domain-shift robustness is increased;
our astrophysical metrics are naturally suited to quantify degradation after instrumental effects.
This provides a scalable pathway to interpret SKA data within the context of galaxy formation scenarios.

\paragraph{Limitations.}
Our models learn by design conditional slices of a simulation's formation scenario $\Phi$.
Because $\Phi$ depends on sub-grid physics and calibration, generalization across suites (e.g. \tng{}, \siml{SIMBA}, \siml{FIRE}, or \siml{EAGLE}) and redshift evolution must be demonstrated rather than assumed.

Furthermore, perceptual metrics such as FID carry domain-mismatch assumptions; fine-tuning feature extractors on domain-specific (astrophysical) data could provide an even better measure for astrophysical veracity. More flexible alternatives to FID such as LPIPS~\citep{zhang_2018_unreasonable_effectiveness} could improve evaluation fidelity even further.

Translation with weak couplings could be improved with additional constraints.
Models for the \gas{}$\rightarrow$\stars{} mapping lack sufficient mutual information between input and target domains, making the task particularly challenging.

\paragraph{Outlook.}
Future work will focus on addressing these limitations.

Weakly constrained mappings could be improved by further extending the dataset domains with intermediates.
For instance, since \Hmolecular{} is more closely tied to star formation, it should provide better constraints for the stellar mass prediction via \gas{}$\rightarrow$\Hmolecular{}$\rightarrow$\stars{}.

Alternatively, various inductive biases could also provide stronger constraints during training:
\begin{itemize}[leftmargin=*,topsep=1pt]
\item \textit{Regularization of the objective function}: directly physics-informed networks through, e.g., constraining mass within aperture, or penalties on radial-profile mismatch.
\item \textit{Structure-aware discriminators}: adversarial heads operating on radial profiles, power spectra, or multi-scale losses.
\item \textit{Equivariant architecture}: SO(2)-aware U-Nets can reduce sample complexity, and inherently enforce symmetries, thereby explicitly handling nuisance parameters.
\item \textit{Multi-domain training}: predicting several targets at once in multiple channels would increase cross-domain robustness but increase processing time.
\item \textit{Cross-suite transfer learning}: cross-suite transfer learning and domain adaptation avoids re-training models on other simulation suites from scratch, requiring only a small amount of fine-tuning on the target simulation.
\item \textit{Redshift conditioning}: redshift introduces temporal information to models and helps capture the true galaxy evolution through cosmic time.
\end{itemize}

Our findings demonstrate that learnt generative surrogates can transform galaxy formation research by bridging simulations and observations, reducing reliance on costly, repeated hydrodynamical runs.
By coupling our blueprint for domain-aware assessment of physical realism with computational scalability, this work marks a significant step towards efficient, next-generation modelling pipelines, automated survey interpretation, and managing the ensuing data deluge in the SKA era.

\section*{Acknowledgements}


PD, FS, and EG acknowledge support from SERI as part of the SKACH consortium.
We would also like to thank the ZHAW Centre for Artificial Intelligence's science cluster admin M. Stadelmann for his support and management of the high resources this project demanded.

\section*{Data Availability}

The original source of the dataset is publicly released by the \href{https://www.tng-project.org/data/}{\tng{} project}.
The extracted dataset and model weights can be shared upon reasonable request.
Our \Code{PyTorch}-based code used for the training of the presented deep learning models is publicly released on GitHub under a GPLv3 license (\href{https://github.com/CAIIVS/chuchichaestli}{\faIcon{github}~\Code{chuchichaestli}}) and (\href{https://github.com/phdenzel/skais-mapper}{\faIcon{github}~\Code{skais-mapper}}), including scripts and \emph{hydra} configurations \citep{yadan_2019_hydra} to re-create the results in this work.



\bibliographystyle{mnras}
\bibliography{refs}








\bsp
\label{lastpage}
\end{document}